%% file: main.tex
\documentclass[%
nofootinbib,
twocolumn
,tightenlines
,amsmath
,amssymb
,pra,
,tikz,letterpaper,eqsecnum]{revtex4-2}
\usepackage[]{graphicx}%
\usepackage{dcolumn}%
\usepackage{mathtools}
\usepackage{bm}%
\usepackage[colorlinks=true]{hyperref}%
\usepackage{tikz}
\usepackage{multirow}
\usepackage{qcircuit}
\usepackage{braket}
\usepackage{xcolor}
\usepackage{units}
\usepackage{adjustbox}
\let\arrow\vec

\input{Menicucci_preamble.tex}
\let\originalleft\left
\let\originalright\right
\renewcommand{\left}{\mathopen{}\mathclose\bgroup\originalleft}
\renewcommand{\right}{\aftergroup\egroup\originalright}

\newcommand{\Rtheta}{\theta}

\newcommand{\withchars}[2]{\begin{bmatrix} {#1} \\ {#2} \end{bmatrix}}
\newcommand{\withperiodandchars}[3][{}]{_{#1} \withchars {#2} {#3}}

\newcommand{\thetafxn}[3][{}]{\theta \withperiodandchars [{#1}] {#2} {#3}}
\newcommand{\Rthetafxn}[3][{}]{\Rtheta \withperiodandchars [{#1}] {#2} {#3}}
\newcommand{\Shafxn}[3][{}]{\Sha \withperiodandchars [{#1}] {#2} {#3}}

\newcommand{\qwax}[1][-1]{\ar @{->} [#1,0]}
\newcommand{\bsbal}[1]{\qwax[{#1}] \qw}

\newcommand{\erf}[1]{Eq.~\eqref{#1}}

\newcommand{\sqrpi}{{\sqrt \pi}}
\newcommand{\qql}{{\frac \sqrpi 2}}

\usepackage[OT2,T1]{fontenc}
\DeclareSymbolFont{cyrletters}{OT2}{wncyr}{m}{n}
\DeclareMathSymbol{\Sha}{\mathalpha}{cyrletters}{"58}

\begin{document}

\title{Phase-space methods for representing, manipulating, and correcting Gottesman-Kitaev-Preskill qubits}

\author{Lucas J. Mensen}
\email{lucasjmensen@gmail.com}
\author{Ben Q. Baragiola}%
\author{Nicolas C. Menicucci}%
\affiliation{%
	Centre for Quantum Computation and Communication Technology, School of Science, RMIT University, Melbourne, Victoria 3000, Australia\\
}
\date{\today}%

\begin{abstract}
The Gottesman-Kitaev-Preskill~(GKP) encoding of a qubit into a bosonic mode is a promising bosonic code for quantum computation due to its tolerance for noise and all-Gaussian gate set. 
We present a toolkit for phase-space description and manipulation of GKP encodings that includes Wigner functions for ideal and approximate GKP states, for various types of mixed GKP states, and for GKP-encoded operators. One advantage of a phase-space approach is that Gaussian unitaries, required for computation with GKP codes, correspond to simple transformations on the arguments of Wigner functions. We use this fact and our toolkit to describe GKP error correction, including magic-state preparation, entirely in phase space using operations on Wigner functions. While our focus here is on the square-lattice GKP code, we provide a general framework for GKP codes defined on any lattice.

\end{abstract}

\pacs{Valid PACS appear here}%
\maketitle

\section{Introduction}
\label{sec:intro}

In recent decades, there has been an effort to adapt the scalable features of continuous-variable (CV) quantum systems into a viable platform for 
fault-tolerant quantum computation.
In 2001, Gottesman, Kitaev, and Preskill~(GKP) proposed a novel means of encoding a qubit within a quantum oscillator~\cite{GKP}. One key feature of such an encoding is protection from small phase-space displacements, which is desirable for reliable quantum computing~(QC). Crucially, using a qubit encoding 
makes CV systems compatible
with quantum error correction~\cite{Shor1995,Steane2003}, which is the process of using redundantly encoded quantum information 
to identify errors and implement appropriately targeted 
recovery operations  to resolve them. \blk

GKP proposed syndrome-extraction and recovery
techniques that restore a corrupted encoded qubit state, with some potential to incur a qubit-level error in the process. However, provided that the resulting error rate is sufficiently low, these qubit-level errors can be managed by concatenating with a qubit-level error-correcting code within a larger  fault-tolerant
architecture~\cite{Nick2014,wuQuantumComputingMultidimensional2020,yamasakiPolylogoverheadHighlyFaulttolerant2020}. Concatenation also enables the use of topological error correcting schemes~\cite{kosukeprx2018,vuillotQuantumErrorCorrection2019,nohFaulttolerantBosonicQuantum2020,terhalScalableBosonicQuantum2020} and noise-biased codes~\cite{hangglienchancednoise2020}.

GKP codes are among a larger set of discrete encodings into bosonic modes, known as bosonic codes ~\cite{cochraneMacroscopicallyDistinctQuantumsuperposition1999,michaelNewClassQuantum2016,ralphQuantumComputationOptical2003,knillSchemeEfficientQuantum2001,mirrahimiDynamicallyProtectedCatqubits2014,grimsmoQuantumComputingRotationSymmetric2020,chuangSimpleQuantumComputer1995,leghtasHardwareEfficientAutonomousQuantum2013}.
Useful bosonic codes
protect against some form of error, which may include phase-space displacements~\cite{GKP} and\slash or other types of errors such as loss or dephasing~\cite{michaelNewClassQuantum2016}. 
Interestingly, codes designed for one type of error may also perform well against other types. As a key example,
the GKP encoding turns out to afford the best protection against pure loss 
when compared to other bosonic codes
~\cite{noh2019}, outperforming other codes specifically designed for this type of noise. 
This motivates
further work using the GKP encoding in particular.

A practical  
drawback in using the GKP encoding
is that the 
highly nonclassical code states are difficult to produce experimentally. 
Recent breakthrough experiments 
have risen to the challenge and produced GKP states in two physical architectures: 
the mechanical oscillations of a trapped ion~\cite{Fluhmann:2019aa} 
and a microwave cavity in superconducting circuits~\cite{campagne2020quantum}. \blk
Numerous proposals also exist to generate optical GKP states~\cite{daknaGenerationArbitraryQuantum1999a,fiurasekConditionalGenerationArbitrary2005,sabapathyProductionPhotonicUniversal2019,suConversionGaussianStates2019,quesadaSimulatingRealisticNonGaussian2019,tzitrinProgressPracticalQubit2020,vasconcelosAllopticalGenerationStates2010,weigandGeneratingGridStates2018,eatonNonGaussianGottesmanKitaev2019,shuklaSqueezedCombStates2020}, although they have not yet been demonstrated experimentally.
Optical implementations have the advantage of room-temperature operation and can be miniaturized, for increased stability,
using integrated photonic platforms~\cite{slussarenko2019photonic}. Furthermore, recent results have shown the potential of using optical GKP states directly in measurement-based schemes~\cite{kosukeprx2018,fukuiHighthresholdFaulttolerantQuantum2019,bourassaBlueprintScalablePhotonic2020} and with
highly scalable CV cluster states~\cite{menicucci2006universal,Nick2014,walshe2020continuous,wuQuantumComputingMultidimensional2020,asavanant2019generation,larsen2019deterministic}.\blk
 
Another distinct advantage of the GKP encoding 
over other bosonic codes
is that
the entire Clifford group of encoded-qubit operations 
can be implemented 
using Gaussian unitaries~\cite{GKP}. This means the most common operations in QC~\cite{nielsen2002quantum} can be performed easily and reliably. This property makes the GKP encoding dovetail seamlessly with CV measurement-based quantum computing schemes~\cite{Nick2014,pantaleoni2020modular} since all Gaussian operations can be implemented using just homodyne detection on a CV cluster state~\cite{menicucci2006universal,Gu2009}.
There are a variety of methods for modelling GKP codewords~\cite{GKP,nohFaulttolerantBosonicQuantum2020,pantaleoni2020modular,matsuura2019equivalence}. Phase-space techniques, however, have the versatility to model both pure and mixed states and to treat them on equal footing~\cite{fabre2020}. The action of Gaussian unitaries (required for GKP Clifford QC) are represented simply in phase space 
by
linear transformations on the arguments of the input Wigner function~\cite{simonrwigner1987,arvindRealSymplecticGroups1995}. In this way, computational operations on pure or mixed state are simple transformations on quasiprobability distributions. Additionally, many of the bosonic noise channels expected in a laboratory have well-known effects on phase space~\cite{albertPerformanceStructureSinglemode2018,Gu2009,michaelNewClassQuantum2016}. 
Such an approach also provides the ability to visualize the features and properties of codewords, operations, measurements, and environmental noise channels. This helps one develop intuition for these concepts and for how they relate to each other.

Although universal QC is not possible with Gaussian resources and measurements alone~\cite{bartlett2002efficient}, supplementing Gaussian elements with a source of just a single type of non-Gaussian state, measurement, or operation enables universal QC~\cite{Lloyd2003}. In fact, the non-Gaussian features of GKP-encoded states themselves can be used as the non-Gaussian resource required to achieve universal QC~\cite{allGaussianPRL,yamasakicostreducedaps2020}.  This means that, at a conceptual level, one can model the unitary part of universal GKP computation in terms of linear (symplectic) transformations acting on GKP and Gaussian Wigner functions. Simulation of full universality (including measurements), however, will not generally be efficient unless restricted to Clifford operations~\cite{Garcia-Alvarez2020}.

The phase-space picture is particularly useful for the GKP encoding, and this work provides the tools and concepts to use it effectively.
Some of these properties have been reported previously (\emph{e.g.},~\cite{matsuura2019equivalence}), and we extend this to a full suite of such tools, along with additional properties and relations that make them easy to use.

We begin in Sec.~\ref{section:theta} by defining and giving the key properties of the quasiperiodic functions that
will be employed extensively throughout this work. In Sec.~\ref{section:gkpcode}, we introduce the GKP encoding and give wave-function representations for
ideal
and approximate GKP codewords using these tools. 
In Secs.~\ref{section:phasespacegkp} and~\ref{section:approxgkpphasespace}, we construct phase-space descriptions of pure and mixed GKP states (both ideal and approximate), as well as a basis of GKP-encoded Pauli operators that transform amongst themselves under the Gaussian unitaries that implement the Clifford group. \blk
Then, in Sec.~\ref{section:gkpec} we consider GKP error correction in phase space. And in Sec.~\ref{section:magicstate}, we illustrate this result using the example of heterodyne detection on a GKP Bell pair to produce GKP-encoded magic states~\cite{allGaussianPRL}. Finally, we conclude with a discussion in Sec.~\ref{sec:discussion}.

\subsection*{Summary of key results}
\begin{itemize}

    \item In Sec.~\ref{subsection:mvlatticethetas} we define multidimensional Jacobi $\theta$-functions on a lattice and their asymptotic limits as $\Sha$-functions, which represent quasiperiodic Dirac combs. These functions can be used to describe various GKP codes including square lattice and hexagonal lattice.
    \blk
   
    \item We outline 
    the periodic features of GKP codewords (both wavefunctions and in phase-space) in terms of quasiperiodic functions. We consider ideal and approximate codewords within this framework in Secs.~\ref{subsection:idealgkp} and~\ref{subsection:pureapproxwf}, respectively.
    
   \item We construct a phase-space representation of GKP codewords in Sec.~\ref{subsection:idealgkpoperators} that is preserved 
   Gaussian blurring, such that it can generally describe both pure and mixed states under additive Gaussian noise both prior to and after GKP-qubit Clifford operations as shown in Sec.~\ref{subsection:idealgkpstatesmixed}. 
   
   \item In Sec.~\ref{subsection:minenvapproxgkp}, we provide sufficient conditions for phase-space descriptions of finite-energy approximate GKP codewords---both pure and impure.
   These conditions enable the direct construction of physical GKP codewords in phase space,
   including circumstances where the states have experienced additive Gaussian noise. 
   This allows theorists to work directly in phase space---bypassing the wave function or density operator entirely---with a unified representation of both pure and mixed approximations of both pure or mixed GKP-encoded states, with a guarantee that the constructed Wigner function represents a physical approximate GKP state of a given quality.
   
   \item We analyze these phase-space representations in the limit of high-quality GKP states. In this limit, they simplify to a form similar to perfect codewords transformed under an additive Gaussian noise channel, as shown in Sec.~\ref{subsection:twirling}. We also provide approximate normalization factors for the associated pure states in these circumstances in both Sec.~\ref{subsection:minenvapproxgkp} and the Appendices~\ref{appendix:approxnormalization} and~\ref{appendix:exactnormalization}. 
   
   \item 
   In Sec.~\ref{subsection:gkpecphasespace}, we derive the map for GKP error correction
   directly in phase space, which is given by two successive convolutions with the Wigner functions for the GKP ancilla states used in the process.   
   Each convolution periodically ``rakes'' the input Wigner function 
   in the quadrature to be corrected and simultaneously replicates it in the other quadrature (to induce quasiperiodicity). 
   We illustrate and visualize this procedure in Sec.~\ref{section:magicstate} by using GKP error correction to generate encoded magic states from the vacuum~\cite{allGaussianPRL}.  
   \blk

\end{itemize}

\section{Periodic and Quasiperiodic Functions} \label{section:theta}

\subsection{Jacobi and Siegel $\theta$-functions}

A class of elliptic functions---Jacobi $\theta$-functions and their multivariate form, Siegel $\Rtheta$-functions~\cite{igusaThetaFunctions1972b,mumfordIntroductionMotivationTheta1983,mumfordBasicResultsTheta1983} (henceforth referred to collectively as $\theta$-functions)---exhibit periodic structure that makes them useful for modelling GKP codewords
in phase spase. As we extensively employ Jacobi 
and Siegel $\theta$-functions throughout this work, we provide a brief review of their forms and properties, and we show their connections to Dirac combs.

A univariate Jacobi $\theta$-function of the third kind is defined as~\cite{mumfordIntroductionMotivationTheta1983}
    \begin{align} \label{classic1dtheta}
        \thetafxn{v_1}{v_2}(z,\tau)
        \coloneqq \sum_{n \in \mathbb{Z}} e^{2\pi i \big[\frac{1}{2} (n+v_{1})^{2}\tau + (n+{v_{1}})(z+v_{2}) \big]}\, ,
    \end{align}
where $z, \tau \in \mathbb{C}$ with $\Im(\tau)>0$, and the parameters $v_{1},v_{2} \in \mathbb{Q}$ are known as the \emph{characteristics} of the $\theta$-function. An alternate, recursive form is found by factoring out terms independent of the sum,
\begin{align}
    \thetafxn{v_1}{v_2}(z,\tau) = e^{2 \pi i (\frac{1}{2} \tau v^{2}_{1} + v_{1}(z + v_{2}))} \theta (z + v_{1}\tau + v_{2},\tau).
\end{align}
Multivariate $\theta$-functions, known as \emph{Siegel} (or \emph{Riemann}) $\theta$-functions, 
are defined analogously to (\ref{classic1dtheta})~\cite{mumfordBasicResultsTheta1983}:
\begin{equation} \label{classicSiegelTheta}
\Rthetafxn{\vec{v}_{1}}{\vec{v}_{2}}
(\vec{z},\mat{\tau}) \coloneqq \sum_{\vec{n} \in \mathbb{Z}^{d}} e^{2\pi i (\frac{1}{2}(\vec{n}+\vec{v}_1)^\tp \mat{\tau}(\vec{n}+\vec{v}_1) + (\vec{n}+\vec{v}_1)^\tp(\vec{z}+  \vec{v}_2)) }.
\end{equation}
Here, $\vec{z} \in \mathbb{C}^d$ is a $d$-dimensional vector of complex arguments, and $\mat{\tau} = \mat{\tau}^\tp$ is now a $d$-dimensional, complex, square matrix in the \emph{Siegel upper half space}---\emph{i.e.},~$\Im(\mat{\tau}) > \vec{0}$, where this notation indicates that $\mat \tau$ is a positive-definite matrix.
The rational characteristics are analogously extended to column vectors $\vec{v}_{1},\vec{v}_{2} \in \mathbb{Q}^{d}$. An expanded multivariate form can be written as
\begin{align}
    \Rthetafxn{\vec{v}_{1}}{\vec{v}_{2}} (\vec{z},\mat{\tau}) = e^{2\pi i 
    (\frac{1}{2}\vec{v}_{1}^{\tp}\mat{\tau}\vec{v}_{1} + \vec{v}_{1}^{\tp} (\vec{z} + \vec{v}_{2})) } \theta (\vec{z} + \mat{\tau}\vec{v}_{1} + \vec{v}_{2} , \mat{\tau}).
\end{align}

\subsection{$T$-periodic univariate $\theta$-functions}

\blk We define a convenient variation of Jacobi $\theta$-functions, \erf{classic1dtheta},
\begin{align} \label{1dthetaT}
        \thetafxn [T] {v_1} {v_2}
         (z, \tau ) \coloneqq \frac{1}{\sqrt{\abs{T}}}
         \thetafxn {v_1} {v_2}
         \bigg(\frac{z}{T}, \frac{\tau}{T^{2}}\bigg)\, . %
 \end{align}
    for real, non-zero period $T \in \mathbb{R}_{\neq 0}$. Including the factor $1/\sqrt{\abs T}$ preserves the $L^2$ norm of the function for all~$T$ and simplifies many of the relations that follow. This form is useful for describing Gaussian pulse trains of period $T$ (and thus features of the GKP encoding), and we refer to it henceforth as a $\theta$-\emph{function} in this work. 
We refer to the case of $[0,0]^\tp$ characteristics as a \emph{canonical $\theta$-function}, and
for notational convenience denote it simply as
    \begin{align} \label{canonicalthetaT}
        \theta_T(z,\tau) := 
        \thetafxn [T] 0 0 (z, \tau)
    \end{align}
$\theta$-functions (both \erf{classic1dtheta} and our new definition \erf{1dthetaT}; we focus on the latter) exhibit several useful properties. 
The expanded form of this function is given as:
  \begin{align} \label{1dthetaTalt}
        &\thetafxn[T]{v_1}{v_2} (z,\tau) = \nonumber \\
          & \qquad e^{2 \pi i (\frac{1}{2} v_1^{2}\frac{\tau}{T^{2}} + v_1 \frac{z}{T} + v_{1}v_{2})}\theta_T (z+ (  v_1 \tfrac{\tau}{T^2}  + v_2)T, \tau). 
    \end{align}
Here, the characteristics for our $T$-periodic function $\theta_{T}$ describe two distinct transformations: the first characteristic gives is a shift by the quasiperiod $v_1 \frac{\tau}{T}$ and introduces an exponential factor, and the second characteristic gives a shift by a factor proportional to the period, $v_2 T$. Canonical $\theta$-functions, \erf{canonicalthetaT}, are integer-periodic in $z$ and quasiperiodic for integer displacements of $\tau$. These relations, referred to as \emph{quasiperiodicity}, are described by the formula
\begin{align} \label{quasiperiodicity}
    \theta_T[z + (m_1\tfrac{\tau}{T^{2}} + m_2)T,\tau] = e^{-2 \pi i (\frac{1}{2} m_1^{2}\frac{\tau}{T^{2}} + m_1 \frac{z}{T})}\theta _T(z, \tau) \, ,
\end{align}
for integer characteristics $m_1, m_2 \in \mathbb{Z}$. 
Together, these relations reveal that for fixed $\tau$, $\theta$-functions are invariant under integer-valued characteristics:
    \begin{align} \label{thetaTinvariance}
        \thetafxn[T]{m_1}{m_2} (z,\tau)
        &= \theta_T (z,\tau) \, .
    \end{align}
\blk
Integer increments $m_{1},m_{2}$ on the characteristics give the following identity for our $T$-periodic functions
\begin{align}
    \thetafxn[T]{v_{1} + m _{1}}{v_{2} + m_{2}} (z,\tau) = e^{2\pi i v_{1} m _{2}} \thetafxn[T]{v_{1}}{v_{2}} (z,\tau).
\end{align}
A useful feature of half-period characteristics is that they yield identities that relate $\theta$-functions of double and half frequency; for binary $j,k \in \{0,1\}$, these are given as
~\cite{ruzziJacobiThFunctions2006},
\begin{subequations} \label{periodsplitting}
\begin{align}
    \thetafxn[T]{\nicefrac{j}{2}}{0} (z, \tau)
    &= \frac{1}{\sqrt{2}} \bigg( \theta_{2T}(z,\tau) + (-1)^j \thetafxn[2T]{0}{\nicefrac{1}{2}} (z,\tau)  \bigg), 
    \\
    \thetafxn[T]{0}{\nicefrac{k}{2}} (z, \tau)
    &= \frac{1}{\sqrt{2}} \bigg( \theta_{T/2} (z , \tau ) + (-1)^{k} \thetafxn[T/2]{\nicefrac{1}{2}}{0} (z, \tau) \bigg) .
\end{align}
\end{subequations}
For $j,k = \{0,0\}$, we recover the canonical $\theta$-function, \erf{canonicalthetaT}. The collection of $\theta$-functions associated with the remaining three pairs, $\{0,1 \}$, $\{1,0 \}$, and $\{1, 1 \}$, are sometimes referred to as \emph{auxiliary functions}~\cite{thetacharref}. These are shown in the right-hand column of Fig.~\ref{shahthetatable}.

\subsubsection{Connection to Gaussian pulse trains}
A normalized, univariate Gaussian distribution on $x \in \mathbb{R}$ with variance $\sigma^{2}$ and mean $x_{0}$ is given by
    \begin{align} \label{normalizedGaussian}
        G_{\sigma^{2}} (x - x_{0}) &\coloneqq \frac{1}{\sqrt{2\pi \sigma^2}} e^{-\frac{(x-x_{0})^{2}}{2\sigma^{2}}}.
    \end{align}
A pulse train consists of a sum of identical such Gaussians whose means are centered at integer multiples of the period $T$. 
Pulse trains are a special case of $\theta$-functions for real argument $x$ and purely imaginary quasiperiod $\tau= 2\pi i \sigma^2$.
The $\theta$-function characteristics in \erf{1dthetaT} allow us to describe the more general case of a pulse train of period $T$ that is translated and phased,
    \begin{equation} \label{Gaussianpulsetrain}
        \sum_{n \in \mathbb{Z}} e^{ -2 \pi i n v_{1}} G_{\sigma^2}  (x + (n+v_2)T) =  \frac{1}{\sqrt{\abss{T}}} \thetafxn[T]{v_1}{v_2}(x, 2\pi i \sigma^2)\, .
    \end{equation} 
The first characteristic $v_1$ describes periodic phasing along the pulse train (with fractional first characteristics applying sequential roots of unity), and the second characteristic describes shifts by units of the period $T$.
These features are useful for representing the GKP code, where logical states are encoded---for both wavefunctions and Wigner functions---in amplitudes and phases of alternating peaks of a Gaussian pulse train (up to a large envelope). In \erf{Gaussianpulsetrain}, this corresponds to half-integer characteristics---\emph{i.e.}~with $\{v_1, v_2\} = \{ \tfrac{m_1}{2}, \tfrac{m_2}{2} \}$ for integers $m_1, m_2 \in \mathbb{Z}$.
With this fact and Eqs.~\eqref{periodsplitting}, we can also compose and decompose Gaussian pulse trains of different periods, which is useful for describing GKP states.

\subsubsection{Connection to Dirac combs}
The Dirac $\delta$-function can be described as the weak limit (\emph{i.e.},~as a distribution) of
a normalized Gaussian, \erf{normalizedGaussian},
as the variance tends to zero. 
\blk
We take an analogous weak limit of a Gaussian pulse train, through its description as a  $\theta$-function in \erf{Gaussianpulsetrain}, to define a variation of a Dirac comb of period $T$ with characteristics $v_1,v_2$ analogous to \erf{1dthetaTalt}:\footnote{Note that the quasiperiod $v_{1}\tau$ vanishes in the limit $\tau\to i0^{+}$  function such that the function becomes simply periodic.}
\begin{align} \label{shadefin}
        \Shafxn[T]{v_1}{v_2} (x) &\coloneqq \lim_{\sigma^2 \to 0^+} \thetafxn[T]{v_1}{v_2}(x,2\pi i \sigma^2) \\
        & = e^{ 2 \pi i v_1(\frac{x}{T} + v_2)} \sqrt{\abss{T}} \sum_{n=-\infty}^\infty  \delta (x + (n + \blk v_2) T) 
        \, .
    \end{align}
The ``shah function'' $\Sha_T(x)$ is a Dirac comb of period $T$ multiplied by an additional scaling factor of $\sqrt{\abss{T}}$. 
This definition is convenient because, under the Fourier transform $\mathcal{F}$,
    \begin{align}
        \mathcal{F}[f](y) \coloneqq \frac{1}{\sqrt{2\pi}}\int dx\; e^{- i x y } f(x), \label{foudefin}
    \end{align}
a $\Sha$-function with period $T$ transforms to another $\Sha$-function with period $\frac{2\pi}{T}$:
    \begin{align} \label{Fouriershah}
        \mathcal{F} [\Sha_{T}](y) = \mathcal{F}^{-1} [\Sha_{T}](y) = \Sha_{\frac{2\pi}{T}} (y) \, . 
    \end{align}
(Without the $\sqrt{\abs T}$ prefactor, there would be additional scaling factors that would complicate this and other relations.) 
Just as above, we have defined a canonical $\Sha$-function with suppressed characteristics in our notation:
    \begin{align}
        \Sha_T(x) := 
        \Shafxn [T] 0 0 (x)\, .
    \end{align}

\begin{figure}[t]
	\includegraphics[width=0.9\linewidth]{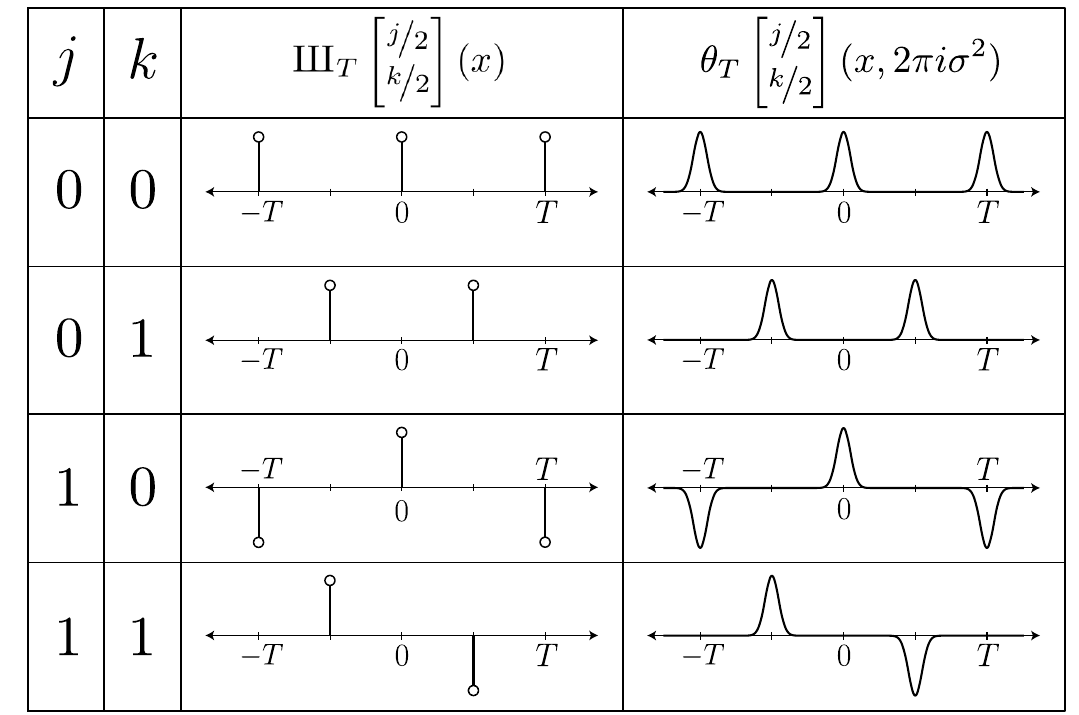}
	\caption{\label{shahthetatable}  Half-period (auxiliary) $\Sha$-functions, Eqs.~\eqref{shadefin},  and $\theta$-functions, Eqs.~\eqref{1dthetaT}. Lines capped with circles indicate $\delta$-functions. The first characteristic $j$ introduces phases (alternating positivity and negativity), and the the second characteristic $k$ shifts by half a period. 
}
\end{figure}

Since $\Sha$-functions, \erf{shadefin}, are (weak) limits of $\theta$-functions, they also have the alternate form
    \begin{equation} \label{diraccombcharacteristics}
        \Shafxn[T]{v_1}{v_2}(x)  
        = e^{ 2 \pi i v_1(\frac{x}{T} + v_2)} \Sha_{T} (x + v_2T) \, ,
    \end{equation}  
showing that a nonzero $v_1$ introduces a phase and $v_2$ a shift. Moreover, integer increments of the characteristics give an analogous identity
\begin{align}
     \Shafxn[T]{v_1 + m_{1}}{v_2 + m_{2}}(x) = e^{2\pi i v_{1}m_{2}} \Shafxn[T]{v_1 }{v_2} (x)  .
\end{align}
The $\Sha$-functions inherit integer-characteristic invariance [\erf{thetaTinvariance}] and the relations between auxiliary functions [\erf{periodsplitting}],
\begin{subequations} \label{periodsplittingShah}
\begin{align}
    \Shafxn[T]{\nicefrac{j}{2}}{0} (x)
    &= \frac{1}{\sqrt{2}} \bigg( \Sha_{2T}(x) + (-1)^j \Shafxn[2T]{0}{\nicefrac{1}{2}} (x)  \bigg), 
    \\
    \Shafxn[T]{0}{\nicefrac{k}{2}} (x)
    &= \frac{1}{\sqrt{2}} \bigg( \Sha_{T/2} (x ) + (-1)^k \Shafxn[T/2]{\nicefrac{1}{2}}{0} (x) \bigg) ,
\end{align}
\end{subequations}
for binary $j,k \in \{0,1\}$.
These are shown in the left-hand column of Fig.~\ref{shahthetatable}.

To summarize, our definition of a $\Sha$-function, \erf{shadefin}, differs from a typical Dirac comb in two distinct ways. First, we define $\Sha$-function as weak limits of real-valued $\theta$-functions, from which they inherit characteristics, \erf{diraccombcharacteristics}.
Second, we scale the Dirac comb by a factor $\sqrt{\abss{T}}$ to ensure the Fourier relations in \erf{Fouriershah} (as well as others) remain as clean as possible.

\subsection{Multivariate $\theta$-functions on a lattice}
\label{subsection:mvlatticethetas}
We generalize the $\theta$-functions defined above to the multivariate case, where the period is generalized to a lattice. We define a scaled multivariate $\theta$-function as 
   \begin{align} \label{SiegelThetaT}
        &\Rthetafxn[\mat A]{\vec{v}_1}{\vec{v}_2}  (\vec z, \mat\tau) \coloneqq
        \frac{1}{\sqrt{\abs{\det \mat A}}}
       \Rthetafxn{\vec{v}_1}{\vec{v}_2} (\mat A^{-1}\vec z, \mat A^{-1} \mat\tau \mat A^{-\tp}),
    \end{align}
where $\mat{A} \in GL_{d}(\mathbb{R})$ is a linear transformation from the integer lattice~$\mathbb Z^d$ to the new lattice---\emph{i.e.},~the columns of~$\mat A$ are the desired lattice vectors. Again, we include a normalizing factor, this time $1/\sqrt{\abs{\det{\mat A}}}$, to preserve the $L^2$ norm and simplify important relations. For the real, normalized, $d$-dimensional, multivariate Gaussian with covariance matrix $\mat{\Sigma}$ and mean $\vec{x}_{0} \in  \mathbb{R}^{d}$,
    \begin{align} \label{2DGaussian}
        G_{\mat{\Sigma}} (\vec{x} - \vec{x}_{0}) \coloneqq \frac{1}{\sqrt{\det(2\pi \mat{\Sigma})}} e^{-\frac{1}{2} (\vec{x}-\vec{x}_{0})^{\tp}\mat{\Sigma}^{-1}(\vec{x} - \vec{x}_{0})},
    \end{align}
we may write the corresponding $\theta$-function associated with sums of this Gaussian over the lattice $\mat{A}$ with analogous constraints $\vec{x} \in \mathbb{R}^{d}$ and $\Re(\mat{\tau}) =0$:
    \begin{align} \label{pulsetrainA}
        &\sum_{\vec n \in \mathbb{Z}^d} e^{-2\pi i \vec{v}^{\tp}_{1}\vec{n}} G_{\mat \Sigma} (\vec x + \mat A (\vec n + \vec{v}_{2})) \nonumber\\
    &\quad\quad\quad=
         \frac{1}{\sqrt{\abs{\det \mat A}}} 
        \Rthetafxn [\mat A] {\vec{v}_{1}} {\vec{v}_{2}} (\vec x, 2\pi i \mat\Sigma).
    \end{align}
 This function can be expressed in terms of the general characteristic vectors $\vec{v}_{1},\vec{v}_{2}$ as 
 \begin{align}
     \Rthetafxn[\mat A]{\vec{v}_1}{\vec{v}_2}  (\vec z, \mat\tau) &= e^{2 \pi i (\frac{1}{2}\vec{v}^{\tp}_{1}\mat{A}^{-1}\mat{\tau}\mat{A}^{-\tp}\vec{v}_{1} + \vec{v}_{1}^{\tp}(\mat{A}^{-1}\vec{z} + \vec{v}_{2}))} \nonumber \\
     &\quad\times \Rthetafxn[\mat A]{\vec{v}_1}{\vec{v}_2}  (\vec z + \mat{A}(\mat{A}^{-1}\mat{\tau}\mat{A}^{-\tp}\vec{v}_{1} + \vec{v}_{2}), \mat\tau).
 \end{align}
 In this form, the integers added to the characteristics give an analogous identity,
     \begin{align} \label{Thetaquasiperiod}
        \Rthetafxn [\mat{A}] {\vec{v}_{1} + \vec{m}_{1}} {\vec{v}_{2} + \vec{m}_{2}} (\vec{z}, \mat{\tau}) = e ^{2\pi i \vec{v}_{1}^{\tp}\vec{m}_{2}} \Rthetafxn [\mat{A}] {\vec{v}_{1}} {\vec{v}_{2}} (\vec{z}, \mat{\tau}).
     \end{align}
Taking the (weak) limit of infinitesimal, purely imaginary~$\mat{\tau}$ forms a $d$-dimensional Dirac comb, which we define as the multivariate $\Sha$-function
    \begin{subequations} \label{ShaA}
    \begin{align} 
        \Sha_{\mat{A}}
        \begin{bmatrix}
            \vec{v}_{1} \\ 
            \vec{v}_{2}
        \end{bmatrix}  (\vec{x}) & \coloneqq   \lim_{\mat{\Sigma} \to \mat 0^+} \theta_{\mat{A}} \begin{bmatrix}
            \vec{v}_{1} \\ 
            \vec{v}_{2}
        \end{bmatrix}(\vec{x},2 \pi i \mat{\Sigma}) \\
        &=\sqrt{|\det \mat{A}|} \sum_{\vec{n} \in \mathbb{Z}^{d}}  e^{-2\pi i \vec{v}^{\tp}_{1}\vec{n}} \delta (\vec{x} + \mat{A}(\vec{n}+\vec{v}_{2})) \\
           & =
        \frac{1}{\sqrt{\abs{\det \mat A}}}
       \Shafxn{\vec{v}_1}{\vec{v}_2} (\mat A^{-1}\vec x), 
\end{align}
\end{subequations}
where $\mat 0^+$ indicates an infinitesimal positive-definite matrix. Lastly, we can also express the periodicity of the multivariate $\Sha$-functions in terms of the characteristics
\begin{align}
\Sha_{\mat{A}} \begin{bmatrix}
\vec{v}_{1} \\ 
\vec{v}_{2}
\end{bmatrix}(\vec{x}) &= e^{2 \pi i (\vec{v}_{1}^{\tp} (\mat{A}^{-1}\vec{x} + \vec{v}_{2}))} \Sha_{\mat{A}} (\vec{x} + \mat{A}\vec{v}_{2}),
\end{align}
with integer increments on each characteristic having an analogous property,
     \begin{align} 
        \Shafxn [\mat{A}] {\vec{v}_{1} + \vec{m}_{1}} {\vec{v}_{2} + \vec{m}_{2}} (\vec{x}) = e ^{2\pi i \vec{v}_{1}^{\tp}\vec{m}_{2}} \Shafxn [\mat{A}] {\vec{v}_{1}} {\vec{v}_{2}} (\vec{x}). \label{integerid}
     \end{align}

\subsection{Blurring and deblurring} \label{Sec:Convolutions}
The convolution of two functions $f(\vec{x})$ and $g(\vec{x})$ in $\mathbb{R}^{n}$ is given by
\begin{align}
    [f * g] (\vec{x}) \coloneqq \int d^{n}\vec{x}'\;f(\vec{x} - \vec{x}')g(\vec{x}')\, .
    \label{conolutiondef}
\end{align}
This describes a ``blurring'' of $f(\vec{x})$ by $g(\vec{x})$, or vice versa since convolutions are symmetric. 
We consider the case when $g(\vec{x})$ is a normalized, $d$-dimensional Gaussian distribution $G_{\mat \Sigma}(\vec{x})$ with $\mat{\Sigma} \in \mathbb{R}^{d \times d}$. Using Fourier representations of the convolution and its inverse, deconvolution, we define the action of a \emph{blurring} and a \emph{deblurring} operator, $\mathcal{D}_{\mat{\Sigma}}$ and  $\mathcal{D}^{-1}_{\mat{\Sigma}}$, on $f(\vec{x})$:
\begin{subequations} \label{bludefin}
\begin{align} 
    \mathcal{D}_{\mat{\Sigma}} f(\vec{x}) &\coloneqq \mathcal{F}^{-1} \big[\mathcal{F}[f]\cdot \mathcal{F}[G_{\mat{\Sigma}}]\big] (\vec{x}), \\
    \mathcal{D}^{-1}_{\mat{\Sigma}} f(\vec{x})&\coloneqq \mathcal{F}^{-1} \bigg[\frac{\mathcal{F}[f]}{\mathcal{F}[G_{\mat{\Sigma}}]}\bigg](\vec{x}),
\end{align}
\end{subequations}
where the Fourier transform $\mathcal{F}$ is defined in (\ref{foudefin}). These operators describe a generalized forward and backward Weierstrass transforms~\cite{widderConvolutionTransform1954}.
The blurring and deblurring operators are equivalently described by the convenient Gaussian form~\cite{kempfNewDiracDelta2014a}
    \begin{subequations} \label{blurringops}
    \begin{align}
	    \mathcal{D}_{\mat{\Sigma}} &= e^{\frac{1}{2}\vec{\nabla}^{\tp}\mat{\Sigma}\vec{\nabla}} \\
	    \mathcal{D}^{-1}_{\mat{\Sigma}} &= e^{-\frac{1}{2}\vec{\nabla}^{\tp}\mat{\Sigma}\vec{\nabla}},
    \end{align} 
    \end{subequations}
where $\vec{\nabla} \coloneqq [\partial_{1}, \dots, \partial_{d} ]^{\tp}$. In this form, blurring and deblurring can be applied directly to a function.

\begin{figure}[t]
	\includegraphics[width=0.95\linewidth]{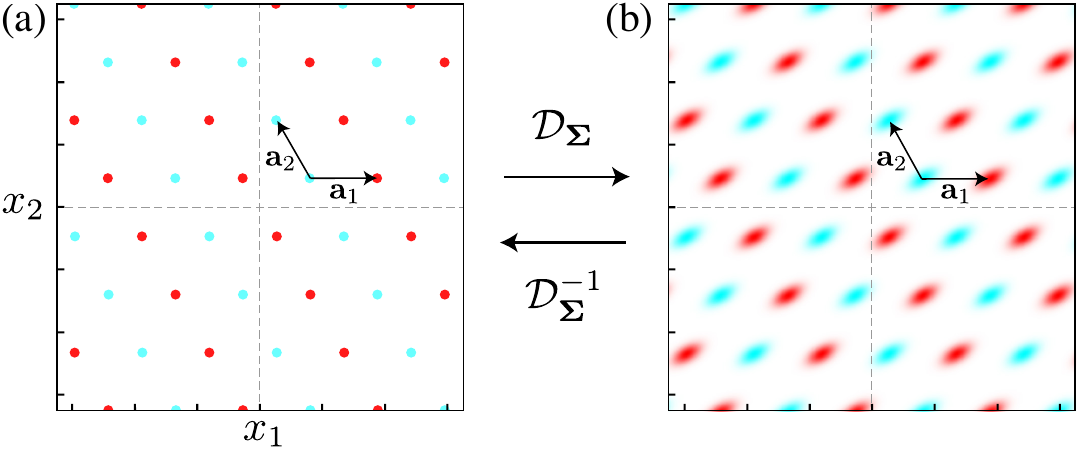}
	\caption{\label{shahtheta} Transformation between a two-dimensional (a) $\Sha$-function and (b) $\theta$-function by blurring and deblurring. 
	A blurring operator $\mathcal{D}_{\mat \Sigma}$ transforms a $\Sha$-function, Eq.~\eqref{ShaA}, into a $\theta$-function, Eq.~\eqref{SiegelThetaT}, with purely imaginary parameter $\mat \tau = 2 \pi i \mat \Sigma$; see Eq.~\eqref{ShaConvolution}. Deblurring with the same $\mat \Sigma$ undoes this operation and transforms back. Shown is
	$\mat \Sigma 
	= \tfrac{1}{2}\mat{R}^{\tp}_{\pi/3}\Big[\begin{smallmatrix}
	(0.2)^{2} & 0 \\ 
	0 & (0.4)^{2}
	\end{smallmatrix}\Big]\mat{R}_{\pi/3}$,
	where $\mat{R}_{\pi/3}$ is a rotation matrix, \erf{rotationmatrix}.
	The lattice $\mat{A} = [ \vec{a}_1 \, \vec{a}_2 ]$ for lattice vectors $\vec{a}_1 = (2\sqrt{3})^{-1/2} [2, 0]^\tp$ and $\vec{a}_2 = (2\sqrt{3})^{-1/2} [-1, \sqrt{3} ]^\tp$ corresponds to a triangular lattice.
	Characteristic $\vec{v}_1 = [\frac{1}{2}, 0]^\tp$ gives rise to the phasing along $\vec{a}_1$ (alternating positivity and negativity) and characteristic $\vec{v}_2 = [0, \frac{1}{2}]^\tp$ gives the half-lattice-period shift along $\vec{a}_2$ (from the origin). In both, red indicates positive and cyan negative, the color scales are arbitrary, tick marks are at integers, and in (a) the dots indicate two-dimensional $\delta$-functions.}
\end{figure}

Blurring and deblurring transform back and forth between $\Sha$-functions and $\theta$-functions, without disturbing the lattice $\mat{A}$ or characteristics, via the general relations,
    \begin{subequations} \label{ShaConvolution}
    \begin{align} 
	    \mathcal{D}_{\mat{\Sigma}} \Rthetafxn[\mat{A}]{\vec{v}_{1}}{\vec{v}_{2}} (\vec{x}, 2 \pi i\mat{\Sigma}') &= \Rthetafxn[\mat{A}]{\vec{v}_{1}}{\vec{v}_{2}} \big(\vec{x}, 2 \pi i (\mat{\Sigma}'+\mat{\Sigma}) \big),\\
	    \mathcal{D}^{-1}_{\mat{\Sigma}} \Rthetafxn[\mat{A}]{\vec{v}_{1}}{\vec{v}_{2}} (\vec{x}, 2 \pi i\mat{\Sigma}') &= \Rthetafxn[\mat{A}]{\vec{v}_{1}}{\vec{v}_{2}} \big(\vec{x}, 2 \pi i (\mat{\Sigma}'-\mat{\Sigma}) \big)
	    \, .
    \end{align}
    \end{subequations}
The connection to $\Sha$-functions arises from the special case of deblurring where $\mat{\Sigma}'=\mat{\Sigma}$, illustrated in Fig.~\ref{shahtheta}.

Note that deblurring may take a $\theta$-function out of the Siegel upper-half space---\emph{i.e.}, after deblurring, it may no longer hold that $\Im (\mat{\tau})>0$. 
This behavior is not unusual for highly localized phase-space distributions---for example, the $P$-distribution~\cite{cahillglauberorder1969,cahillglauberdensity1969} of a squeezed state (which is found by deconvolving a squeezed Gaussian with a Gaussian of vacuum variance) is highly singular~\cite{kim1989squeezednumber}. GKP states are more localized (periodically) than the vacuum state in all quadratures, which means their $P$-distributions are even more singular than a squeezed vacuum state. We focus on Wigner functions in this work, which are always well behaved---at worst, being as singular as a $\delta$-function. Still, the fact that deblurring can lead to a $\mat \tau$ that leaves the Siegel upper-half space is important to remember.

\section{ The GKP Encoding}
\label{section:gkpcode}
In 2001, Gottesman, Kitaev, and Preskill developed the GKP encoding of a logical, two-dimensional qubit into the continuous-variable Hilbert space of a bosonic mode~\cite{GKP}. The GKP encoding, described in detail below, is founded on translational invariance of the code words. The periodic structure of $\theta$-functions introduced in Sec.~\ref{section:theta} make these functions powerful mathematical tools for describing many aspects of the GKP encoding.

A major advantage of the GKP code over alternative bosonic codes is that the full set of encoded Pauli and Clifford operations are implemented via quadrature displacements and Gaussian unitaries, respectively. Additionally, universality and fault tolerance require no further resources beyond the Gaussian Cliffords and a single type of encoded state~\cite{allGaussianPRL, yamasakicostreducedaps2020}.
In the next subsection, we briefly review the square-lattice GKP encoding and show how univariate $\theta$-functions, \erf{1dthetaT}, and their limit as $\Sha$-functions, \erf{shadefin}, describe pure ideal and approximate GKP states and their transformations under GKP Pauli and Clifford operations. This serves as a more recognizable introduction to the main focus of this work: using multivariate $\theta$-functions and $\Sha$-functions to simplify phase-space representations of the GKP encoding. Importantly, although throughout this work we are using the square-lattice GKP code, one may easily convert to other lattices such as the hexagonal code~\cite{GKP,harrington2001achievable,albertPerformanceStructureSinglemode2018} using the appropriate transformation~$\mat A$.

\subsection{GKP states and encoded Clifford operators}
\label{subsection:idealgkp}
A GKP qubit is encoded into a bosonic mode, whose position- and momentum-quadrature operators are defined in terms of bosonic creation and annihilation operators, 
    \begin{align}
        \op{q} \coloneqq \tfrac{1}{\sqrt{2}}(\op{a} + \op{a}^\dagger), \quad \quad \op{p}\coloneqq \tfrac{-i}{\sqrt{2}}(\op{a} - \op{a}^\dagger),
    \end{align}
which satisfy the canonical commutation relation $[\op{q}, \op{p}] = i$, with $\hbar=1$. (With these conventions, the measured vacuum variance in both quadratures is $\tfrac{1}{2}$.)
Respective eigenstates of the position and momentum operators, $\qket s$ and $\pket t$, satisfying
\begin{align}
	\op q \qket s & = s  \qket s,
\quad \quad
	\op p \pket t = t \pket t \, ,
\end{align}
constitute two useful bases, indicated by subscripts~$q$ and~$p$, respectively.
An arbitrary pure state $\ket{\psi}$ is represented in either basis using the position and momentum wavefunctions, respectively,
\begin{align}
	&\psi(s) \coloneqq \inprodsubsub{s}{\psi}{q}{}, 	
	&\tilde{\psi}(t) \coloneqq \inprodsubsub{t}{\psi}{p}{},
\end{align}
as
    \begin{align}
        \ket{\psi} = \int ds \, \psi(s) \qket s = \int dt \,  \tilde{\psi}(t) \pket t \, .
    \end{align}
The canonical quadrature operators are generators of displacements through the position- and momentum-shift operators,
    \begin{align} \label{displacementops}
        \op{X} (h) \coloneqq  e^{-ih\op{p}}\, ,\quad \quad 
        \op{Z} (h) \coloneqq  e^{ih\op{q}}\, , 
    \end{align}
respectively, for $h \in \mathbb{R}$. General translations in phase space are described by
    \begin{align} \label{shiftopV}
        \op{V} (\vec h) 
        = \op{V} (h_q, h_p) 
        \coloneqq  
        e^{-i h_q \op p + i h_p \op q}
        \, ,
    \end{align}
where $\vec h = [h_q,h_p]^\tp$ is the displacement vector in phase space.
The shift operator in \erf{shiftopV} is related to the standard displacement operator
$\op{D} (\alpha) \coloneqq  e^{\alpha \op{a}^\dagger - \alpha^* \op{a} }$ (with $\alpha \in \mathbb{C}$) via the relation $\op{V} (\vec{h}) = \op{D}\left(\frac{h_q + i h_p }{\sqrt{2}}\right)$.

\begin{figure} [t]
	\includegraphics[width=1\linewidth]{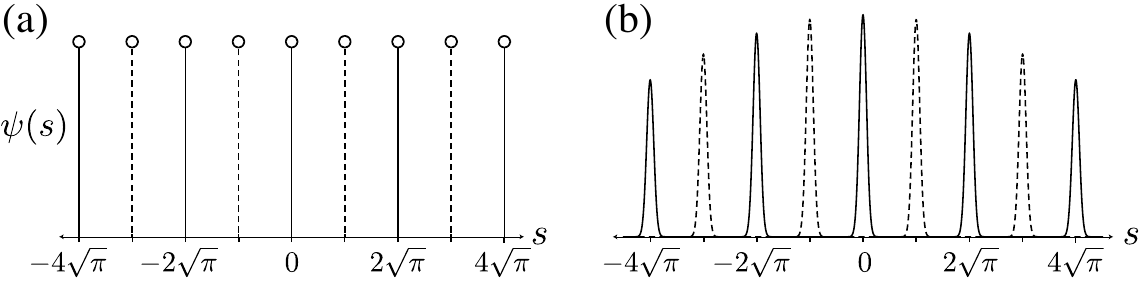}
	\caption{ Position wavefunctions for the square-lattice GKP states $\ket{0_L}$ (solid) and $\ket{1_L}$ (dashed). Ideal GKP states, \erf{GKPwavefunctions}, are shown in (a) and approximate GKP states, \erf{approxGKPlimit}, are shown in (b). Lines capped in circles are $\delta$-functions.
	} \label{idealplot}
\end{figure}

The ideal computational-basis states for the square-lattice GKP encoding are defined
    \begin{align} \label{GKPstates}
    	\ket{j_L} &\coloneqq  (2 \sqrt{\pi})^{1/2} \sum_{n=-\infty}^\infty \qket{(2n+j)\sqrt{\pi}} \\
    	          &=          (\sqrt{\pi})^{1/2}   \sum_{n=-\infty}^\infty e^{i j \pi n } \pket{n\sqrt{\pi}} 
    \end{align}
with $j \in \{0,1 \}$ indicating the logical-qubit state. Throughout, we use the subscript $L$ to label logical states and operators for the square-lattice GKP code. The states in \erf{GKPstates} are represented in the position and momentum bases as,
    \begin{subequations} \label{GKPwavefunctionsketform}
        \begin{align} 
    	\ket{j_L}  =  \int ds  \, \psi_{j,L}(s) \qket{s} 
    	           =  \int dt  \, \tilde{\psi}_{j,L}(s) \pket{t}
    \end{align}
    \end{subequations}
where the position and momentum GKP wavefunctions are, respectively,
    \begin{subequations} \label{GKPwavefunctions}
        \begin{align} 
    	\psi_{j,L}(s)         & \coloneqq \inprodsubsub{s}{\psi}{q}{} = \Shafxn[2\sqrt{\pi}]{0}{\nicefrac{j}{2}}(s)\, , \\
    	\tilde{\psi}_{j,L}(t) & \coloneqq \inprodsubsub{t}{\psi}{p}{} = \Shafxn[\sqrt{\pi}]{\nicefrac{j}{2}}{0}(t) \, ,
    \end{align}
    \end{subequations}
and the $\Sha$-functions are defined in \erf{shadefin}. 
For reference, Table~\ref{Table:GKPstates} provides the wavefunctions for the other Pauli-basis states,\footnote{The wavefunctions for the GKP Pauli-$Y$ eigenstates, $\ket{\pm i_L} \coloneqq \tfrac{1}{\sqrt{2}}(\ket{0_L} \pm i \ket{1_L})$, can be represented as single $\Sha$-functions (with characteristics) in the $\frac{1}{\sqrt{2}}(q+p)$-basis. In the $q$ or $p$ basis, they require two $\Sha$-functions.
}
    \begin{align}
        \ket{\pm_L}   &\coloneqq \tfrac{1}{\sqrt{2}}(\ket{0_L} \pm \ket{1_L}) \label{GKPpm} \, .
    \end{align}
The different periods for the position and momentum wavefunctions in \erf{GKPwavefunctions} (and those in Table~\ref{Table:GKPstates}) are a consequence of the fact that the position and momentum bases are related by a Fourier transform.
The square-lattice GKP wavefunctions, described by $\Sha$-functions of periods $2\sqrpi$ and $\sqrpi$, 
halve or double, respectively, under the action of a Fourier transform, \erf{Fouriershah}. This halving or doubling forms a new symmetric or antisymmetric superposition of $\Sha$-functions, described by the half-period transformations in \erf{periodsplittingShah}. 
A consequence is that both the position and momentum wavefunctions for an arbitrary pure GKP state 
    \begin{align} \label{approxstate}
        \ket{\psi_L} = c_0 \ket{0_L} + c_1 \ket{1_L}
    \end{align} 
($|c_0|^2 + |c_1|^2 = 1$)
have two key properties: they only have support only on multiples of $\sqrt{\pi}$, and they are $2\sqrt{\pi}$-periodic.

GKP logical Pauli operators for the square-lattice encoding, $\op{X}_L$ and $\op{Z}_L$, are position and momentum displacements, \erf{displacementops}, of magnitude $\sqrpi$,\footnote{
We focus on the square-lattice GKP encoding, for which an encoded Hadamard operator $\op{H}_{L}$ is a realized via Fourier transform $\op{F}$. 
For other GKP lattices, see Refs.~\cite{GKP, albertPerformanceStructureSinglemode2018}. } 
    \begin{subequations} \label{GKPPaulis}
    \begin{align} 
        \op{X}_L &\coloneqq \op{X} (\sqrt{\pi}) = e^{-i \sqrt{\pi} \op{p}} \, ,\\
        \op{Z}_L &\coloneqq \op{Z} (\sqrt{\pi}) = e^{i \sqrt{\pi} \op{q}}\, .
    \end{align}
    \end{subequations}    
The GKP stabilizers are position and momentum shifts by $2\sqrt{\pi}$,  
    \begin{subequations} \label{stabilizers}
    \begin{align}
        \op{S}_X &\coloneqq (\op{X}_L)^2 = \op{X}(2\sqrt{\pi})\, , \\ 
        \op{S}_Z &\coloneqq (\op{Z}_L)^2 = \op{Z}(2\sqrt{\pi}) 
        \, ,
    \end{align}
    \end{subequations}    
and satisfy $[ (\op{S}_X)^m, (\op{S}_Z)^n] = 0$ for any integers $m,n$.
GKP Hadamard $\op{H}_L$ and phase $\op{S}_L$ gates, which together generate the single-qubit Clifford group, are realized by the Fourier transform operator $\op{F}$ and unit-shear operator $\op{P}$, respectively:
     \begin{subequations} \label{GKPCliffords}
     \begin{align} 
        \op{H}_{L} &\coloneqq \op{F} = e^{i\frac{\pi}{4}(\op{q}^2+\op{p}^2)}\, ,  \\
        \op{S}_{L} &\coloneqq  \op{P} = e^{i\frac{1}{2} \op{q}^2}.
    \end{align}
    \end{subequations}    
The full set of encoded single-qubit Pauli and Clifford operators, $\{\op{X}_L, \op{Z}_L, \op{H}_L, \op{S}_L \}$, are generated by linear and quadratic combinations of $\op{q}$ and $\op{p}$. 
Importantly, when operators from this set are applied to GKP states, they generate half-period transformations of the $\Sha$-functions that comprise the square-lattice GKP wavefunctions; see Table~\ref{Table:GKPstates}.

\begingroup
\renewcommand{\arraystretch}{1.8}
\begin{table}[t]
	\begin{tabular}{ccc}
	\cline{2-3}
\multicolumn{1}{c|}{} & \multicolumn{1}{c|}{$\psi(s)$} & \multicolumn{1}{c|}{$\tilde{\psi}(t)$} \\ \hline
		\multicolumn{1}{|c|}{$\ket{0_L}$} & \multicolumn{1}{c|}{$ \Sha_{2\sqrpi}(s)$} & \multicolumn{1}{c|}{$\Sha_{\sqrpi} (t)$} \\ \hline
		\multicolumn{1}{|c|}{$\ket{1_L}$} & \multicolumn{1}{c|}{$\Sha_{2\sqrpi}\big[\begin{smallmatrix}
			0 \\ 
			\nicefrac{1}{2}
			\end{smallmatrix}\big](s)$} & \multicolumn{1}{c|}{$\Sha_{\sqrpi}\big[\begin{smallmatrix}
			\nicefrac{1}{2} \\ 
			0
			\end{smallmatrix}\big](t)$} \\ \hline
		\multicolumn{1}{|c|}{$\ket{+_L}$} & \multicolumn{1}{c|}{$\Sha_{\sqrpi} (s)$} & \multicolumn{1}{c|}{$\Sha_{2\sqrpi}(t)$} \\ \hline
		\multicolumn{1}{|c|}{$\ket{-_L}$} & \multicolumn{1}{c|}{$ \Sha_{\sqrpi}\big[\begin{smallmatrix}
			\nicefrac{1}{2} \\ 
			0
			\end{smallmatrix}\big](s)$} & \multicolumn{1}{c|}{$\Sha_{2\sqrpi}\big[\begin{smallmatrix}
			0 \\ 
			\nicefrac{1}{2}
			\end{smallmatrix}\big](t)$} \\ \hline
	\end{tabular}
\caption{\label{Table:GKPstates}Position and momentum wavefunctions for ideal GKP Pauli eigenstates. Our choice of definition for $\Sha$-functions, \erf{shadefin}, makes these forms simple while ensuring they all have the same Hilbert-space norm.
}
\end{table}
\endgroup

\subsection{Embedded-error operator} \label{subsection:embsection}
Ideal GKP states are unphysical since their wavefunctions are not $L^2$-normalizable. We consider here physical, normalized approximations to GKP states, $\ket{\bar{\psi}_L}$, that result from the application of the nonunitary embedded-error operator~\cite{GKP, motes_encoding_2017},
    \begin{align} \label{embeddederroroporig}
        \op \xi (\mat{\Xi}) \coloneqq \exp \left( - \frac{1}{2}\opvec{x}^\tp \mat{\Xi}^{-1} \opvec{x} \right)\, ,
    \end{align}
to ideal GKP states.
The real, symmetric, positive-definite, matrix $\mat{\Xi}$, which encodes the details of the (coherent) errors applied to the ideal state, is diagonalizable via a rotation matrix $\mat{R}_\varphi \in$ SO(2)
    \begin{align} \label{rotationmatrix}
    \mat{R}_{\varphi} &\coloneqq 
    \begin{bmatrix}
        &\cos \varphi &-\sin \varphi \\
        &\sin \varphi &\cos \varphi
    \end{bmatrix},
    \end{align}
giving
    \begin{align}
 \label{diagnoisematrix}
        \mat{\Xi} &= \mat{R}_\varphi^\tp \mat{\Xi}_0 \mat{R}_\varphi
        \, ,
    \end{align}
where the elements of the diagonal matrix $\mat{\Xi}_0$ describe envelopes in the two principal quadratures determined by $\varphi$ and $\varphi + \frac{\pi}{2}$. Without lack of generality, we parameterize $\mat{\Xi}_0$ using the standard notation~\cite{GKP}, 
    \begin{align} \label{GKPnoisematrix}
        \mat{\Xi}_0 = \begin{bmatrix}
        \frac{1}{\kappa^2} & 0 \\ 
        0                  & \frac{1}{\Delta^2}
        \end{bmatrix} \, ,
    \end{align}
which, for $\mat{R}_\varphi = \mat{I}$, produces envelopes in the $\op{q}$ and $\op{p}$ quadratures. As a consequence of \erf{diagnoisematrix}, $\op \xi (\mat{\Xi})$ can also be written as
    \begin{align} \label{embeddederrorop}
        \op \xi (\mat{\Xi}) = \op{R}^\dagger(\varphi) \exp \left( - \frac{1}{2}\opvec{x}^\tp \mat{\Xi}^{-1}_0 \opvec{x} \right) \op{R}(\varphi)\, ,
    \end{align}
where $\op{R}(\varphi) = e^{i \varphi \op n}$ is the phase delay (rotation) operator. For symmetric error parameters ($\Delta^{2}=\kappa^{2}=\beta$), $\mat \Xi_0 = \beta^{-1} \mat{I}$, the embedded-error operator is the \emph{damping operator}~\cite{motes_encoding_2017, walshe2020continuous},
    \begin{align}
        \op{N}(\beta) \coloneqq e^{ -\beta \op{n}} = e^{-\frac{\beta}{2}(\op{q}^2 + \op{p}^2)} = \op \xi(\beta^{-1} \mat{I} ) \, .
    \end{align}
A useful, single-parameter measure for the quality of an approximate GKP state is the \emph{squeezing}~\cite{motes_encoding_2017}, given by
    \begin{align}
	    (\text{squeezing in dB}) = -10 \log (\beta).
    \end{align}
The connection between this notion of squeezing and that in squeezed vacuum states is given in Ref.~\cite{walshe2020continuous}.

In the limit of small errors, $\mat{\Xi}^{-1}_0 \ll \mat{I}$ (\emph{i.e.},~$ \Delta^2,\kappa^2 \ll 1$), the embedded-error operator in \erf{embeddederrorop} can be written as
    \begin{subequations} \label{highqualitynoise}
    \begin{align} 
        \op{\xi}( \mat{\Xi} ) 
        &= \op{R}^\dagger(\varphi) e^{-\frac{1}{2} \left( \kappa^{2}\op{q}^2 +\Delta^{2}\op{p}^2 \right) } \op{R}(\varphi) \\*
        & \approx \op{R}^\dagger(\varphi) e^{-\frac{1}{2}\Delta^{2}\op{p}^2}e^{-\frac{1}{2}\kappa^{2}\op{q}^2} \op{R}(\varphi)
        \label{highsqapprox} \\*
        &\approx \op{R}^\dagger(\varphi) e^{-\frac{1}{2}\kappa^{2}\op{q}^2}e^{-\frac{1}{2}\Delta^{2}\op{p}^2} \op{R}(\varphi) \, ,\label{highsqapprox2}
    \end{align}
    \end{subequations}
and the exponential operators approximately commute~\cite{kimuraExplicitDescriptionZassenhaus2017,motes_encoding_2017}.\footnote{Note that this embedded-error operator is the same as that resulting from sequential single-bit teleportation through squeezed states in a canonical continuous-variable cluster-state setting~\cite{Nick2014,Alexander2014,walshe2020continuous} }

Each of the exponential operators above has a particular effect on the wavefunctions of the input state. 
In position space, $e^{-\frac{1}{2}\kappa^{2}\op{q}^2}$ applies a Gaussian envelope of variance 1/$\kappa^2$ to the position wavefunction,
    \begin{align} \label{eq:positiondampedwavefunction}
        e^{-\frac{1}{2} \kappa^2 \op{q}^2} \ket{\psi} 
        & \propto \int ds \, G_{\kappa^{-2}}(s) \psi(s) \qket{s}
        \, ,
    \end{align}
noting that the wavefunction is not square normalized.
In momentum space, it convolves the momentum wavefunction with a Gaussian of variance $\kappa^{2}$. 
\begin{align}
    e^{-\frac{1}{2} \kappa^2 \op{q}^2} \ket{\psi} 
     &\propto \int dt \left[ \int ds \, G_{\kappa^2}(s) \tilde{\psi}({ t-s})\right] \pket{t}
     \\
     &= \int dt \, [\psi * G_{\kappa^2}](t) \pket{t}
    \, ,
\end{align}
where the term in the last line in square brackets is the new momentum wavefunction (up to normalization).
This relation can be shown by inserting a complete set of momentum eigenstates in Eq.~\eqref{eq:positiondampedwavefunction} and using the convolution theorem $\mathcal{F}[f \cdot g] = \mathcal{F}[f] * \mathcal{F}[g]$. Another way to see this connection is by noting that the $\op q \to i \partial_p$ when acting on momentum wave functions, so $e^{-\frac{1}{2} \kappa^2 \op{q}^2} \to e^{\frac{1}{2} \kappa^2 \partial_ p^2} = \mathcal D_{\kappa^2}$, which is the one-dimensional version of the blurring operator from Eqs.~\eqref{blurringops}. The operator $e^{-\frac{1}{2}\Delta^{2}\op{p}^2}$ works conversely, applying a Gaussian envelope of variance 1/$\Delta^2$ on the momentum wavefunction and convolving the position wavefunction with a Gaussian of variance $\Delta^{2}$.

Applying the two exponential operators consecutively in a specified order gives both an envelope and a convolution on the position or momentum wavefunction. 
For example, $e^{-\frac{1}{2} \kappa^2 \op{p}^2} e^{-\frac{1}{2} \Delta^2 \op{q}^2} \ket{\psi}$ gives the (unnormalized) position wave function
    \begin{align} 
        \psi_\text{out}(s) &= \int dt \, G_{\Delta^2}(t) G_{\kappa^{-2}}(s-t) \psi(s - t) \\
        &= [(G_{\kappa^{-2}} \psi) * G_{\Delta^2}](s)
        \, .
    \end{align}
Applying the operators in the other order, $e^{-\frac{1}{2} \kappa^2 \op{q}^2} e^{-\frac{1}{2} \Delta^2 \op{p}^2} \ket{\psi}$, gives
    \begin{align} 
        \psi_\text{out}(s) &=  G_{\kappa^{-2}}(s) \int dt \, G_{\Delta^2}(t) \psi(s - t) \\
        &= G_{\kappa^{-2}}(s) [\psi * G_{\Delta^2}](s)
        \, .
    \end{align}
\blk
In the low-noise limit, these operations approximately commute (see Appendix~\ref{appendix:multiplyingGaussians}), and the action on position wavefunction $\psi(s)$ is described by successive applications in either order:
\begin{align} \label{whichorder}
     [ (G_{\kappa^{-2}}\psi) * G_{\Delta^{2}}](s) \approx G_{\kappa^{-2}}(s) [\psi * G_{\Delta^{2}}](s) \, .
 \end{align}
The action of the embedded error on momentum wavefunctions is identical with the roles of $\kappa^2$ and $\Delta^2$ reversed.

\subsection{Pure, approximate GKP states} \label{subsection:pureapproxwf}
Pure, approximate GKP states are generated by applying the embedded-error operator \erf{embeddederrorop}, to an arbitrary ideal GKP state, \erf{approxstate}:
    \begin{align} \label{approximateGKPstate}
        \ket{ \bar{\psi}_L} \coloneqq \frac{1}{\sqrt{ \mathcal{N}}} \op \xi (\mat{\Xi}) \ket{\psi_L},
    \end{align}
The normalization, 
$\mathcal{N} \coloneqq \bra{\psi_L} [\op \xi (\mat{\Xi})]^2 \ket{\psi_L}$,
is a function of the state itself due to the fact that the approximate GKP basis states are not perfectly orthogonal\footnote{The fact that physical codewords may not be orthogonal is common to other bosonic codes, notably cat codes~\cite{ralphQuantumComputationOptical2003}, although such codes can be made orthogonal~\cite{grimsmoQuantumComputingRotationSymmetric2020}.}.
The embedded error in approximate GKP states spoils their periodicity to a degree determined by the eigenvalues of $\mat{\Xi}_0$ ($\kappa^{-2}$ and $\Delta^{-2}$~  in the original setting~\cite{GKP}). Note that, since the states are still pure, this error is distinct from that introduced by external decoherence.\footnote{Using a subsystem decomposition~\cite{lau2017, marshall2019, pantaleoni2020modular}, one can restore the interpretation of embedded error as decoherence arising from entanglement between two \emph{virtual} subsystems.} 
In the limit of low embedded error, $\Delta^2,\kappa^2  \ll 1$, the approximate GKP states are \emph{nearly} periodic and approximately obey the discrete translational symmetries of the GKP code. (For a detailed study of the case without this approximation, see Ref.~\cite{matsuura2019equivalence}.)  This periodicity is the basis for the faithful encoding and decoding of digital quantum information. When the embedded error is low enough, approximate GKP states allow for error correction and can be used fault tolerantly~\cite{Nick2014}.

\subsubsection{$\theta$-function representations of high-quality GKP-state wavefunctions}

We consider the action of an embedded-error operator, in diagonal form as in the original GKP setting, $\mat{\Xi}_0$ in \erf{GKPnoisematrix}, on a state whose wavefunction is $\Sha$-function of period $T$. This replaces each spike in the $\Sha$-function with a Gaussian of variance $\Delta^{2}$, and introduces a broad Gaussian envelope of variance $\kappa^{-2}$; see \ref{idealplot}(b). These operations are described by convolution with a sharp Gaussian followed by multiplication by a broad Gaussian. In the high-quality limit, $\Delta^2, \kappa^2 \ll 1$, these operations can be performed in either order, see Eqs.~\eqref{whichorder}. 

Normalized position wavefunctions for approximate GKP states $\ket{\bar{j}_L} = \mathcal{N}^{-1/2} \op \xi(\kappa, \Delta) \ket{j_L}$ for $j \in \{0,1\}$ in the high-quality limit---that is, using \erf{highqualitynoise}---are derived in Appendix~\ref{appendix:approxnormalization}. The computational-basis position wavefunctions are~\cite{GKP,matsuura2019equivalence}
    \begin{align}
	    \bar{\psi}_{j,L} (s) &\approx \frac{1}{\sqrt{\mathcal{N}}}G_{\kappa^{-2}} (s) \left[G_{\Delta^{2}}  * \psi_{j,L}   \right] (s) \\
	    &= \sqrt{4\pi} \sqrt{ \frac{\Delta}{ \kappa}} \; G_{\kappa^{-2}}(s) \thetafxn[2\sqrt{\pi}]{0}{ \nicefrac{j}{2} } (s, 2 \pi i \Delta^2 )\, ,
	    \label{approxGKPlimit}
    	\blk
    \end{align}
where the ideal wavefunctions, $\psi_{j,L}(s)$, are given in \erf{GKPwavefunctions}, and we have used the normalization
    \begin{align}
        \mathcal{N} \approx  \frac{1}{4\pi} \frac{\kappa}{\Delta}
        \, . 
    \end{align}
These states are approximately square-normalized, $\int ds \, | \bar{\psi}_{j,L}(s)|^2 \approx 1.$\footnote{Small deviations from 1 arise in the derivation of the normalization, which assumes that neighboring spikes have vanishing overlap---as is the case for small $\Delta^2$.}
For momentum wavefunctions, $\tilde{\psi}_{j,L}(t)$, the roles of $\Delta$ and $\kappa$ in \erf{approxGKPlimit} are swapped: $\Delta^{-2}$ is the envelope variance and $\kappa^2$ is the spike variance.
The wavefunctions in both bases have narrow spikes and broad envelopes, with the result being that homodyne measurements of both position and momentum yield outcomes that are tightly clustered about multiples of $\sqrt{\pi}$ (with the frequency of even or odd multiples depending on the encoded state). 
Similar expressions can be found for wavefunctions of the other GKP Pauli eigenstates in Table~\ref{Table:GKPstates}, each of which is described by a single $\theta$-function. In the limit of vanishing error, the normalized computational-basis states, \erf{approxGKPlimit}, approach
\begin{align}
	\lim_{\Delta^2,\kappa^2 \to 0}\bar{\psi}_{j,L} (s) &\sim  \sqrt{2\Delta\kappa}\; 
	\Shafxn[2\sqrpi]{0}{\nicefrac{j}{2}}(s) \, .
\end{align}
These are indeed the ideal GKP wavefunctions in \erf{GKPwavefunctions} with an additional ``normalization constant'' that serves to connect representations of asymptotic physical GKP states to their ideal counterparts.

\section{Phase-space representations of the GKP encoding}
\label{section:phasespacegkp}

Wigner functions are phase-space representations of quantum-mechanical states and operators. For any operator $\op{O}$, the associated Wigner function $W_{\op{O}}$ is given by the Wigner-Weyl transform:%
    \begin{align} \label{wignerdef}
    	W_{\op{O}}(q,p)&=\frac{1}{\pi}\int dy \; \brasub{q + y}{q} \op{O} \ketsub{q-y}{q} e^{2ipy}.
    \end{align}
(The $q$ as subscript merely indicates the position basis.)
The Wigner function for a normalized state $\op{\rho}$ satisfying $\op{\rho}^\dagger = \op{\rho}$, and $\Tr[\op{\rho}] = 1$ is itself normalized,
    \begin{align}
        \int dq\, dp \, W_{\op{\rho}}(q,p) = 1 \, ,
    \end{align}
and bounded,
\begin{align}
    &|W_{\op{\rho}} (q,p)| \leq \frac{1}{\pi}\, . 
\end{align}    
Additionally, $W_{\op{\rho}}(q,p)$ is real valued (as is the Wigner function for any Hermitian operator), although it is not strictly nonnegative. For this reason, Wigner functions of states are called \emph{quasiprobability} distributions, whose marginal distributions over $q$ or $p$ (or any other axis in phase space) describe bonafide probability distributions for quadrature measurements (homodyne detection) along that axis. Henceforth, we will occasionally use a vector $\vec{x} = [q_1, \dotsc, q_N, p_1, \dotsc, p_N]^\tp$ as the argument of an $N$-mode Wigner function (in most cases here, $N=1$). This means that Wigner functions may appear as $W(q_1,\dotsc,q_N,p_1, \dotsc, p_N)$ or as $W(\vec{x})$.

A useful feature of Wigner functions is their simple transformations under Gaussian unitaries. The Heisenberg action of an $N$-mode Gaussian unitary operator $\op{U}_{G}$ on a $(2N \times 1)$-vector of quadrature operators, $\opvec{x} \coloneqq [\op{q}_1, \dotsc \op{q}_N, \op{p}_1, \dotsc, \op{p}_N]^\tp$ 
corresponds to an affine symplectic transformation~\cite{simonrwigner1987,arvindRealSymplecticGroups1995,weedbrook2012}:
    \begin{align}
        \opvec{x} \mapsto \hat{U}^{\dagger}_{G}\; &\vec{\hat{x}}\; \hat{U}_{G} =
        \mat{S}_{\op{U}} \vec{\hat{x}} + \vec{c},
    \end{align}
where $\mat{S}_{\op{U}} \in \mathbb{R}^{2N \times 2N}$ is a symplectic matrix associated with $\op{U}_G$, 
and $\vec{c}\in \mathbb{R}^{2N}$ is a displacement term. The matrix $\mat{S}_{\op{U}}$ allows us to describe a Schr\"{o}dinger-type transformation of operator $\op{O}$ (with the understanding that we will later be considering transformations on quantum states),
    \begin{align}
        \op{O} \mapsto \op{U}_G \op{O} \op{U}_G^\dagger,
    \end{align}
in terms of simple, phase-space transformations on the arguments of its Wigner function:
    \begin{align} \label{GaussianWignertransformation}
        W_{\op{U}_{G}\op{O}\op{U}_{G}^\dagger}(\vec{x}) = W_{\op{O}} \big(\mat{S}_{\op{U}}^{-1}(\vec{x}-\vec{c})\big) \, .
    \end{align}    
    
Consider, in particular, an operator~$\op{O}_\Sha$ whose Wigner function is a two-dimensional $\Sha$-function,
     \begin{align} \label{Osha}
         W_{\op{O}_\Sha }(\vec{x}) 
         &= \Shafxn[\mat{A}]{\vec{v}_1}{\vec{v}_2}(\vec{x})
         = \frac{1}{\sqrt{ |\det \mat{A}|}} \, \Shafxn[]{\vec v_1}{\vec v_2}(\mat{A}^{-1}\vec{x})%
         \, . 
     \end{align}
Such an operator could represent a GKP state or operator (for a square lattice or some other lattice configuration).
From \erf{GaussianWignertransformation}, it follows that the Gaussian unitary transformation $\op{U}_G \op{O}_\Sha \op{U}^\dagger_G$ proceeds in phase space by replacing $\vec x \mapsto \mat{S}_{\op{U}}^{-1}(\vec{x}-\vec{c})$ in the Wigner function, \erf{Osha}: 
    \begin{align}
        W_{ \op{U}_G \op{O}_\Sha \op{U}^\dagger_G }(\vec{x}) 
      &= \frac{1}{\sqrt{ |\det \mat{A}|}} \Shafxn[]{\vec v_1}{\vec v_2}(\mat{A}^{-1} \mat{S}_{\op{U}}^{-1} (\vec{x}-\vec{c})) \\
        &= \frac{1}{\sqrt{ |\det \mat{A}|}} \Shafxn[]{\vec v_1}{\vec v_2}( (\mat{S}_{\op{U}} \mat{A})^{-1}(\vec{x}-\vec{c})) \\
        &= \Shafxn[\mat{S}_{\op{U}}\mat{A}]{\vec{v}_1}{\vec{v}_2 - (\mat{S}_{\op{U}}\mat{A})^{-1}\vec{c}}(\vec{x})\, \\
        &=\Shafxn[\mat{A}]{\bar{\mat{S}}^{-\tp}\vec{v}_{1}}{\bar{\mat{S}}\vec{v}_{2} - \mat{A}^{-1}\vec{c}}(\vec{x}),\label{OShatransformed}
    \end{align}
where $\bar{\mat{S}} \coloneqq  \mat{A}^{-1}\mat{S}_{\op{U}}\mat{A}$. Equation~ \eqref{OShatransformed} is useful for describing transformations on GKP-encoded states and operators since the Clifford group can be realized with Gaussian operations~\cite{GKP}. On a single mode, the full set of single-qubit square-lattice GKP Clifford operations can be generated by the Fourier transform and unit-shear operations, \erf{GKPCliffords}, whose symplectic matrices are given by:
    \begin{align}
        \mat{S}_{\op{F}} &= \begin{bmatrix}
        0 & -1 \\ 
        1 & 0
        \end{bmatrix}\, , \label{SFourier} \\
        \mat{S}_{\op{P}} &= \begin{bmatrix}
        1 & 0 \\ 
        1 & 1
        \end{bmatrix}\, .
    \end{align}
Moreover, displacement terms of the form $\vec{c} = \mat{S}\mat{A}\vec{m}$, where $\vec{m}\in \mathbb{Z}^{2}$, give integer displacements on the characteristics, which lead to transformations of the form given in \erf{integerid}.
Other GKP codes, such as the hexagonal-GKP code~\cite{albertPerformanceStructureSinglemode2018,allGaussianPRL,grimsmoQuantumComputingRotationSymmetric2020}, also have Cliffords generated entirely by Gaussian operations.

\subsection{Ideal GKP-encoded operators and states}
\label{subsection:idealgkpoperators}
Any operator on a qubit, $\op{A}_L$, can be written in a basis of Hermitian Pauli operators $\op{\sigma}_\mu$,
    \begin{align} \label{qubitdecomp}
        \op{A}_L = \tfrac{1}{2} \left( r_0 \op{\sigma}_0 + r_1\op{\sigma}_1 + r_2\op{\sigma}_2 + r_3\op{\sigma}_3\right)\, ,
    \end{align}
where the coefficients, $r_\mu$ for $\mu \in \{0,1,2,3\}$, comprise the 4-component Bloch vector
    \begin{align}
        \vec{r} \coloneqq [r_0 , r_1, r_2, r_3]^\tp \, ,
    \end{align}
and $\op \sigma_0 = \op \id$.
For a normalized state within the qubit subspace, $\op{\rho}_L$, $r_0 = 1$, and the state is determined by the standard 3-component Bloch vector $\arrow{r} \coloneqq [r_1, r_2, r_3]^\tp$.

We now focus on the GKP encoding. In this case, the qubit is encoded in a two-dimensional subspace of the much larger CV Hilbert space. Within this two-dimensional subspace, the GKP Pauli operators are
    \begin{align} \label{GKPPaulioperators}
        \op{\sigma}^L_\mu \coloneqq \sum_{j,k} \sigma_{jk}^\mu \outprod{j_L}{k_L}\, ,
    \end{align}
where $j,k \in \{0,1\}$ label computational-basis states, and $\sigma_{jk}^\mu$ are matrix elements of the Pauli matrices,
    \begin{align}
        \mat{\sigma}^0 &= \begin{bmatrix} 1 & 0 \\ 0 & 1 \end{bmatrix} \quad \quad
        \mat{\sigma}^1 = \begin{bmatrix} 0 & 1 \\ 1 & 0 \end{bmatrix} \\
        \mat{\sigma}^2 &= \begin{bmatrix} 0 & -i \\ i & 0 \end{bmatrix} \quad \quad
        \mat{\sigma}^3 = \begin{bmatrix} 1 & 0 \\ 0 & -1 \end{bmatrix} \, ,
    \end{align}
with $\mat{\sigma}^0$ being the two-dimensional identity matrix.
For a discrete encodings in a mode, there is an important distinction to be made here. The GKP Pauli operators, \erf{GKPPaulioperators}, have support only within the two-dimensional GKP subspace, while the Gaussian operators $\op{X}_L$ and $\op{Z}_L$ in \erf{GKPPaulis}---as well as other operators, such as those in \erf{GKPCliffords}---act both inside and outside of the GKP subspace. In other words, not only do those operators generate the specified logical transformations within the GKP subspace, they also generate nontrivial transformations outside the subspace.\footnote{Using the modular subsystem decomposition~\cite{pantaleoni2020modular}, this difference manifests solely as different actions on the gauge subsystem, with identical action on the logical one. This difference vanishes entirely when the states acted upon are ideal GKP states.}

Since the Wigner transform, \erf{wignerdef}, is linear, we can use the decomposition in \erf{qubitdecomp} and associated Bloch 4-vector $\vec{r}$ to write the Wigner function for any operator in the GKP subspace as
    \begin{align} \label{AnyGKPWigner}
        W_{\op{A}_L} (\vec{x}) = \frac{1}{2} \sum_{\mu = 0}^3 r_\mu W^L_{\mu}(\vec{x}) \, ,
    \end{align}
where we use the shorthand
\begin{equation}
W_{\mu}^L (\vec{x}) \coloneqq W^L_{\op{\sigma}_\mu^L}(\vec{x})\,.
\end{equation}
The components of the Bloch 4-vector are found by taking the trace of the operator $\op{A}_L$ with the GKP Pauli operators,\footnote{This procedure can also be used to find the \emph{projection} of some general CV operator into the GKP subspace, although this is not our focus here.}
    \begin{align} \label{traceforr}
        r_\mu &= \Tr[\op{A}_L \op{\sigma}_\mu^L]
            = \iint d^2\vec{x} \, W_{\op{A}_L}(\vec{x}) W_\mu^L (\vec{x})
            \, ,
    \end{align}
which is performed in phase space by taking the total integral of the product of the Wigner functions.

\begin{figure*} [t]
	\includegraphics[width=0.95\linewidth]{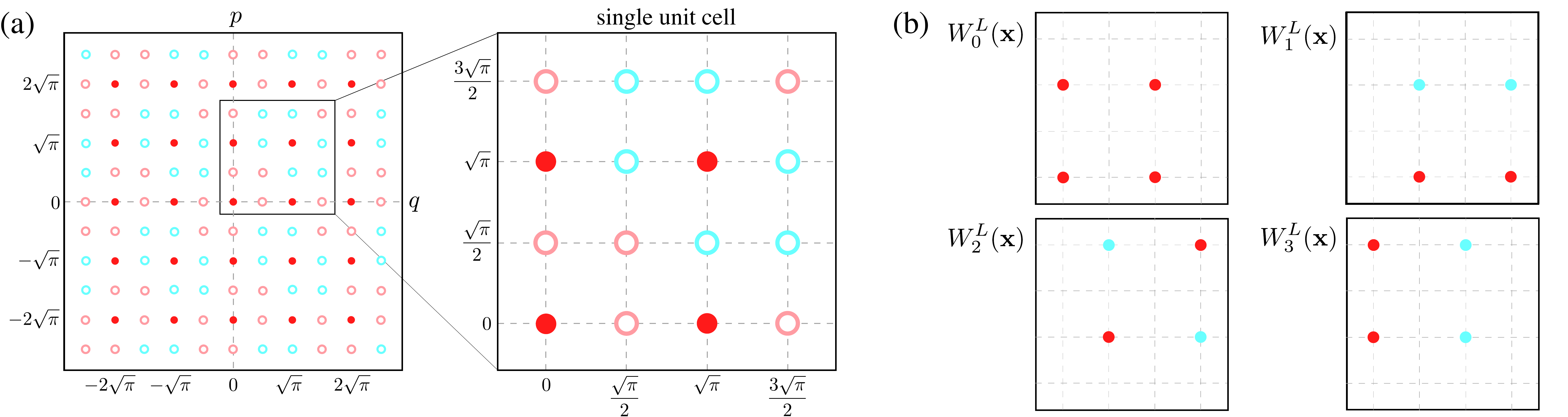} 
	\caption{\label{Fig:IdealStates}Wigner functions for the ideal square-lattice GKP encoding. These Wigner functions are periodic in phase space, and thus can be faithfully represented in a single unit cell of area $4\pi$.  (a) Wigner function for  a $\ket{+T}$ GKP state, which has $\vec{r} = [1,\frac{1}{\sqrt{3}},\frac{1}{\sqrt{3}},\frac{1}{\sqrt{3}}]^{\tp}$. (b) Wigner functions of the GKP Pauli operators (including identity), Eqs.~\eqref{WigfuncPaulis}, in the single unit cell shown in~(a). 
	Dots represent $\delta$-functions, with different styles giving relative weightings: red dot = $\frac{1}{2}$, blue dot = $-\frac{1}{2}$, open red dot = $\frac{1}{2\sqrt{3}}$, open blue dot = $-\frac{1}{2\sqrt{3}}$.
	For visual clarity, each unit cell is depicted with an additional $\frac{\sqrpi}{4}$ in each positive and negative direction for $q$ and $p$. 
	\label{Fig:idealGKPPauli} }
\end{figure*}

Thus, the Wigner function for any single-mode GKP state or operator can be written as a linear combination of the Wigner functions for the four GKP Pauli operators in \erf{GKPPaulioperators}. 
These Wigner functions are expressed as two-dimensional $\Sha$-functions, \erf{ShaA}:\footnote{Note that the additional term in the phase is omitted as $\vec{\ell}^{\tp}_{\mu}\mat{\Omega}\vec{\ell}_{\mu}=0, \forall \vec{\ell}_{\mu}$.}
    \begin{subequations} \label{WigfuncPaulis}
    \begin{align}
    	W^{L}_{\mu} (\vec{x}) &:\propto e^{i\sqrpi \vec{\ell}^\tp_{\mu}\mat{\Omega}^{-1}\vec{x} }\; \Sha_{\sqrpi \mat I}\bigg(\vec{x} + \vec{\ell}_{\mu}\frac{\sqrpi}{2}\bigg) \\
    	&=  \Shafxn[\sqrpi  \mat I]{\nicefrac{\mat{\Omega}\vec{\ell}_{\mu}}{2}}{\nicefrac{\vec{\ell}_{\mu}}{2}} (\vec{x})\, , \label{2DSha}
    \end{align}
    \end{subequations}
with $\mat{A} = \sqrt{\pi} \mat{I}$ describing a rectangular lattice of period $\sqrt{\pi}$, and $:\propto$ indicating definition up to an overall constant.
Here, $\mat{\Omega}$ is the two-dimensional symplectic form~\cite{Alexander2014},
\begin{align}
	\mat{\Omega} \coloneqq \begin{bmatrix}
	0 & 1 \\ 
	-1 & 0
	\end{bmatrix} \, , \label{sympform}
\end{align}
and each of the four Pauli operators is characterized by a two-dimensional vector $\vec{\ell}_{\mu}$:
\begin{align} \label{shiftfvectors}
\begin{split}
	\vec{\ell}_{0} \coloneqq \begin{bmatrix}
	0 \\
	0
	\end{bmatrix}, \quad
	\vec{\ell}_{1} \coloneqq  \begin{bmatrix}
	1 \\
	0
	\end{bmatrix}, \quad 
	\vec{\ell}_{2} \coloneqq  \begin{bmatrix}
	1 \\
	1
	\end{bmatrix}, \quad
	\vec{\ell}_{3} \coloneqq  \begin{bmatrix}
	0 \\
	1
	\end{bmatrix}.
 \end{split}
 \end{align}
The Wigner functions for the GKP Pauli operators, \erf{2DSha}, are shown in Fig.~\ref{Fig:IdealStates}(b). 
Each is characterized by four two-dimensional $\delta$-functions within a unit cell of size $2\sqrt{\pi} \times 2 \sqrt{\pi}$. The pattern for each Wigner function then repeats in all other unit cells and tiles the entirety of phase space.

Note that each Pauli Wigner function occupies its own, unique set of points in phase space. These set of points (and the alternating signs) are compactly described by the half-period characteristics of the $\Sha$-function in \erf{2DSha}. The second characteristic shifts the points by $\frac{\sqrt{\pi}}{2}$ in both $q$ and $p$ according to the vectors in \erf{shiftfvectors}, while the first characteristic determines their phase (sign). The action of the square-lattice Clifford operators can be compactly expressed in the form of \erf{OShatransformed} as unit symplectic matrices (\emph{i.e.}, symplectic with binary elements) and displacement terms of the form $\vec{c} = \vec{m}\sqrpi , \vec{m} \in \mathbb{Z}^{2}$,
as these displacements correspond with the GKP qubit operations $\op{X}_{L}$ and $\op{Z}_{L}$. Thus, 
perfect codewords under Clifford computation evolve as
\begin{align}
    W_{\op{U}_{G}\op{A}_{L} \op{U}^{\dagger}_{G}} (\vec{x})
    &= \frac{1}{2} \sum^{3}_{\mu=0} r_{\mu} \Shafxn[\sqrpi\mat{I}]{ \nicefrac{ \mat{\Omega} \mat{S}\vec{\ell}_{\mu} }{2} }{\nicefrac{\mat{S}\vec{\ell}_{\mu}}{2} - \vec{m}}(\vec{x})\\
    &= \frac{1}{2} \sum^{3}_{\mu=0} r_{\mu} e^{ i\pi  ( \mat{S} \vec{\ell}_{\mu})^{\tp} \mat{\Omega} \vec{m} } \Shafxn[\sqrpi\mat{I}]{\nicefrac{ \mat{\Omega} \mat{S}\vec{\ell}_{\mu} }{2}}{\nicefrac{\mat{S}\vec{\ell}_{\mu}}{2} } (\vec{x}),
\end{align}
where $\op{U}_{G}$ and $\mat{S}$ are the Gaussian unitaries and symplectic matrices associated with the square-lattice Clifford operators. Simply, increments along the lattice transform the encoded Pauli operators as per Eq.~(\ref{integerid}) and can induce a phasing of $-1$ on the encoded operators depending on $\vec{m}$ and the first characteristic. This corresponds to rotations on the encoded Bloch sphere. Two important features to consider in the context of GKP Clifford computation are the transformed vector $\mat S \vec \ell_\mu$ (where $\mat{S}$ is a unit symplectic matrix) and its symplectic inner product with the displacement term $\vec{m}$. These operations concisely describe how each GKP encoded Pauli operator transforms under general GKP logical operations within the Clifford group.

\subsection{Ideal GKP states: pure and mixed}
\label{subsection:idealgkpstatesmixed}

An ideal square-lattice GKP state, parameterized by Bloch 4-vector $\mathbf{r} = [r_0,\arrow{r}\,]^{\tp}$ is given by 
    \begin{align} \label{GKPidealgeneralstate}
        \op{\rho}_L = \frac{1}{2} \bigg( r_0 \op{\sigma}_0^L + \sum_{\mu = 1}^3 r_\mu \op{\sigma}_\mu^L \bigg)\, ,
    \end{align}
where $\op{\sigma}_i^L$ are GKP Pauli operators, \erf{GKPPaulioperators}, and $r_0 = 1$ for a normalized state.
The associated Wigner function is thus represented just as in \erf{AnyGKPWigner},
    \begin{align} \label{wigneridealGKP}
        W_{\vec{r}}(\vec{x}) \coloneqq W_{\op{\rho}_L}( \vec{x} ) = \frac{1}{2}\bigg( r_0 W^L_{0}(\vec{x}) + \sum_{\mu = 1}^3 r_\mu W^L_{\mu}(\vec{x}) \bigg) \, ,
    \end{align}
with the usual constraints for quantum states: the Bloch 3-vector for pure states satisfies $|\arrow{r}|^2 = 1$, while for mixed states it satisfies $|\arrow{r}|^2 < 1$.

Owing to the periodicity of $\Sha$-functions, Wigner functions for ideal GKP states appear simply as a convex sum of the four phased lattices in the unit cell shown in Fig.~\ref{Fig:IdealStates}, each of which itself contains four points. 
GKP Pauli eigenstates are particularly simple: each is characterized by a single non-zero element in $\arrow{r}$ taking a value $\pm 1$---giving Wigner functions that are a sum of the four positive points corresponding to the GKP identity operator, and four points (two positive, two negative) associated with the given GKP Pauli operator. The discrete translational invariance of the GKP Pauli-operator Wigner functions allows us to visualize any ideal state within a single unit cell; an example is shown in Fig.~\ref{Fig:IdealStates}(a).
Importantly, both pure and mixed ideal GKP states (described below) can be represented by GKP Wigner functions in a single unit cell. 

\subsubsection{Mixtures of ideal GKP states}
We consider a random displacement channel
    \begin{align} \label{displacementchannel}
        \mathcal{E}_D = \int d^{2} \vec{y} \, p(\vec{y}) \op{V} (\vec{y}) \odot \op{V}^\dagger(\vec{y})    ,
    \end{align}
where $\op{V}(\vec{y})$ is the shift operator in \erf{shiftopV}. For a physical channel, $p(\vec{y})$ is a two-dimensional probability distribution satisfying $\int d^{2} \vec{y} \, p(\vec{y}) = 1$. In phase space, a displacement generates a shift in the argument of the Wigner function as in \erf{GaussianWignertransformation}, giving a general relation for the displacement channel acting on operator $\op{O}$, namely
 \begin{align}
        W_{ \mathcal{E}_D (\op{O}) } ( \vec{x} ) 
        &= \int d^{2} \vec{y} \, p( \vec{y} ) W_{ \op{V}(\vec{y}) \op{O}  \op{V}^\dagger(\vec{y}) } (  \vec{x} ) \\
        &= \int d^{2} \vec{y} \, p( \vec{y} ) W_{ \op{O} } (  \vec{x} - \vec{y} ) \\ 
        &= [ p * W_{ \op{O} } ] (\vec{x})\,  \label{Wignerdisplacement1} \, ,
    \end{align}
where the convolution~$*$ is defined in \erf{conolutiondef}.
For a Gaussian probability distribution, $p(\vec{y}) = G_{\mat{\Sigma}} (\vec{y})$ from \erf{2DGaussian}, $\mathcal{E}_D$ is a Gaussian displacement channel. The associated Gaussian convolution in phase space, \erf{Wignerdisplacement1}, generates the same transformation as the blurring operation defined in \erf{bludefin}. When this channel acts on an operator $\op{O}_\Sha$ whose Wigner function is a two-dimensional $\Sha$-function, \erf{Osha}, the resulting Wigner function is
     \begin{align} \label{Gaussiandispsha}
         W_{ \mathcal{E}_{D}(\op{O}_\Sha) }(\vec{x})
         = \Rthetafxn[\mat{A}]{\vec{v}_{1}}{\vec{v}_{2}} (\vec{x}, 2 \pi i\mat{\Sigma}) \, . 
     \end{align}
The blurring operation broadens each $\delta$-function into a Gaussian parameterized by $\mat{\Sigma}$, producing a Siegel $\Rtheta$-function; see \erf{ShaConvolution}. 
From \erf{Gaussiandispsha}, we find that a Gaussian random displacement channel on an ideal square-lattice GKP state, $\mathcal{E}_{D} (\op{\rho}_L)$, gives the Wigner function
    \begin{align}
        \tilde{W}_{\vec{r}} (\vec{x}) =
         \sum^{3}_{\mu=0} r_{\mu}  
        \Rthetafxn[\sqrpi \mat \id]{\nicefrac{\mat{\Omega}\vec{\ell}_\mu}{2}}{\nicefrac{\vec{\ell}_\mu}{2}} (\vec{x},2\pi i \mat{\Sigma}), \label{mixedideal}
    \end{align}
where the tilde indicates an approximate version of~$W_{\vec r}$. This blurred ideal GKP Wigner function retains its periodicity and thus is still unphysical. However, the periodic marginal distributions (within a unit cell) have been smeared out and are no longer pointlike. Finally, we remark that $\op{\rho}_L$ may be mixed before the Gaussian displacement channel is applied---thus, in general, there are two potential types of mixing involved: logical-GKP level and CV level.

\section{Approximate GKP states in phase space: effects of an envelope}
\label{section:approxgkpphasespace}
GKP states can be approximated in many ways, some of which are laid out, analyzed, and compared in Matsuura \emph{et al.}~\cite{matsuura2019equivalence}. A feature of useful approximations is that in some parameter limit, the spikes of the Wigner function approach Dirac $\delta$-functions, thus periodically localizing the Wigner support near appropriate multiples of $\frac{\sqrt{\pi}}{2}$; this is the case for the Gaussian mixtures of ideal states in \erf{mixedideal} for $\lim_{\mat{\Sigma} \rightarrow \mat{0}} G_{\mat{\Sigma}}(\vec{y})$. An important additional requirement for physical approximations is that they are normalizable: this requires an envelope function, which ensures that the spike amplitdues get smaller with distance from the origin.  We call states with both of these properties \emph{approximate GKP states}. 

We focus on several useful approximate GKP states and their phase-space representations, with attention to the high-quality limit where such states are useful for quantum computing and error correction. We first consider phase-space descriptions of the approximate GKP states considered in Sec.~\ref{subsection:embsection}---those generated by acting the embedded-error operator on ideal GKP states. A Wigner-function approach allows us to represent not only pure states of this type but also mixed states, which arise when the embedded-error operator is applied to mixed ideal states. A feature of these states is that the spike and envelope covariances saturate a minimum uncertainty relation: for fixed spikes, smaller envelopes than the minimum are unphysical. As such, we refer to this type of state as a \emph{minimum-envelope approximate GKP state}. (We cannot generically call these states ``pure approximate GKP states'' because the encoded logical information may still be mixed.) In the subsection below, we derive the conditions for physical envelopes from the embedded-error operator in the limit of low noise.

\subsection{Embedded-error operator}

In Sec.~\ref{subsection:embsection}, we introduced the embedded-error operator~$\op \xi(\mat{\Xi})$ and examined the approximate GKP states that result when appling it to a pure ideal GKP state. In phase space, the effects of this operator is twofold: It broadens the spikes of a GKP Wigner function and also applies the minimum envelope commensurate with those spikes. A phase-space treatment allows us to widen its application significantly: $\op{\xi}(\mat{\Xi})$ can also be applied to mixed ideal GKP states to produce mixed, approximate GKP states, again with a minimum envelope that normalizes the state. These are examples of minimum-envelope approximate GKP states. Generally, the (Hermitian, nonunitary) embedded-error operator $\op{\xi}(\mat{\Xi})$ can be applied to any CV operator $\op{O}$ via
    \begin{align} \label{noisyop}
        \op{\bar{O}} \coloneqq \op{\xi}(\mat{\Xi}) \op{O}  \op{\xi}(\mat{\Xi})\, .
    \end{align}
This nonunitary transformation can be performed directly in phase space, where the composition of two operators $\op{A}\op{B}$ is described by the \emph{Moyal product} \cite{curtrightQuantumMechanicsPhase2012,curtrightConciseTreatiseQuantum2014} (or \emph{star product}) between their respective Wigner functions (found individually using \erf{wignerdef}):
    \begin{align} \label{MoyalProductdef}
        W_{\op{A}\op{B}}(q,p)
        &= \big[W_{\op{A}} \star W_{\op{B}}\big](q,p)  \\
        & \coloneqq \frac{2}{\pi} \int dq'dq''dp'dp'' 
         W_{\op{A}} (q+q',p+p')  \nonumber\\ 
         & \quad \times W_{\op{B}} (q+q'',p+p'') e^{2i(q'p''-p'q'')}\, .
    \end{align}
Since the Moyal product is associative \cite{curtrightQuantumMechanicsPhase2012,curtrightConciseTreatiseQuantum2014}, the Wigner function for the operator in \erf{noisyop} can be written,
    \begin{align}
        W_{ \op{\bar{O}} } ( \vec{x} ) 
        &= W_{ \op{\xi}(\mat{\Xi}) \op{O}  \op{\xi}(\mat{\Xi}) } (  \vec{x} ) \\
        &= W_{ \op{\xi}(\mat{\Xi}_0) \op{R} \op{O} \op{R}^\dagger \op{\xi}(\mat{\Xi}_0) } ( \mat{R}_\varphi \vec{x} ) \\
        &= \big[ W_{\op{\xi}(\mat{\Xi}_0) } \star W_{ \op{R} \op{O} \op{R}^\dagger } \star W_{\op{\xi}(\mat{\Xi}_0) } \big](\mat{R}_\varphi \vec{x})  \, ,\label{WigGeneralNoisyop}
    \end{align}
where the subscript operators $\op{R}$ are shorthand for $\op{R}(\varphi)$, and the noise matrix $\mat{\Xi}$ is diagonalized by $\mat{R}_\varphi$, \erf{diagnoisematrix}.

\subsubsection{Low noise limit} \label{subsubsection-lowerrorlimit}

In the limit of low noise, $\mat \Xi \ll \mat \id$ (all eigenvalues $\ll 1$), 
the embedded-error operator factorizes into approximately commuting exponential operators; see \erf{highsqapprox}. Its Wigner function factorizes accordingly with respect to the Moyal product \cite{curtrightQuantumMechanicsPhase2012,curtrightConciseTreatiseQuantum2014}:
    \begin{subequations} \label{NoiseMoyalsmall}
    \begin{align}
        W_{\op{\xi}( \mat{\Xi}_0 )}(\vec{x}) & \approx  \big[ W_{e^{-\frac{\Delta^2}{2}\op{p}^2}} \star W_{e^{-\frac{\kappa^2}{2}\op{q}^2}} \big] (\vec{x}) \\
        & \approx  \big[ W_{e^{-\frac{\kappa^2}{2}\op{q}^2}}\star         W_{e^{-\frac{\Delta^2}{2}\op{p}^2}} \big] (\vec{x}) \, .
    \end{align}
    \end{subequations}
In this limit, the exponential operators can be applied successively in either order by substituting either of the forms above into \erf{WigGeneralNoisyop}. 

On wavefunctions, each exponential operator applies an envelope or a convolution (depending on the basis), \erf{approxGKPlimit}. In phase space, however, the enveloping and convolution operations are not independent of one another. For example, a physical transformation that applies an envelope in one quadrature is always accompanied by blurring of the conjugate quadrature---necessary in order to ensure the state remains physical. The converse is not true, however: blurring alone is a physical operation, and as such, it does not necessitate a conjugate envelope. Such envelope-free blurring does not preserve purity, and an example is blurring from a Gaussian random displacement channel, \erf{Gaussiandispsha}. 

The resulting Moyal products can be evaluated simply using a result derived in Appendix~\ref{appendix:moyalunivariate} and summarized here. Specifically, the Moyal products for an arbitrary operator diagonal in $\op{q}$ or in $\op{p}$---$\chi(\op{q})$ and $\varphi(\op{p})$, respectively---acting on some operator $\op{O}$ can be expressed as a simple convolution (Eqs.~\eqref{qrakemap} and~\eqref{prakemap}, respectively):
    \begin{subequations} \label{Moyaldiagonal}
    \begin{align}
        &W_{\chi(\op{q})\op{O} \chi^*(\op{q})}  (q,p) 
        = \big[W_{\chi (\op{q})}\star {W}_{\op{O}} \star W_{ \chi^*(\op{q}) }\big] (q,p) \nonumber \\
        & \quad \quad =	\int dw \,  W_{\outprod{\chi}{\chi}} (q,w) W_{\op{O}} (q,p-w), \label{1qerrordet}\\
        &W_{\varphi(\op{p}) \op{O} \varphi^*(\op{p})}  (q,p) 
        = \big[W_{\varphi (\op{p})}\star {W}_{\op{O}} \star W_{\varphi^*(\op{p}) }\big] (q,p) \nonumber \\
        &\quad \quad  =  \int dw \,  W_{\outprod{\varphi}{\varphi}} (w,p) {W}_{\op{O}} (q+w,p).
    \end{align}
    \end{subequations}
Note the appearance of $W_{\outprod{\chi}{\chi}} (q,p)$ and $W_{\outprod{\varphi}{\varphi}} (q,p)$, which are Wigner functions for the (potentially unnormalized or even unnormalizable) states,
    \begin{subequations}
    \begin{align} \label{wavefunctionExpansions}
        \ket{\chi} &= \int ds \, \chi(s) \qket{s} = \sqrt{2\pi} \chi(\op{q}) \pket{0} \\
        \ket{\varphi} &= \int dt \, \varphi(t) \pket{t} = \sqrt{2\pi} \varphi(\op{p}) \qket{0} \, ,
    \end{align}
    \end{subequations}
generated by acting the diagonal operators, $\chi(\op{q})$ and $\varphi(\op{p})$, on an unbiased state in the conjugate basis. 
Note that $\ket{\varphi}$ is a state whose \emph{momentum} wavefunction is $\varphi(t)$. The relations in Eqs.~\eqref{Moyaldiagonal} have an operational interpretation in a teleportation setting: they describe the Kraus map that arises for an operator $\op{O}$ (typically taken to be a state $\op{\rho}$) undergoing single-mode teleportation with ancilla state $\ket{\chi}$ or $\op{F} \ket{\varphi}$, respectively~\cite{walshe2020continuous}.

For use with GKP encodings, we consider the embedded-error operator acting on operators, $\op{O}_\Sha$, whose Wigner functions are two-dimensional $\Sha$-functions, \erf{Osha}.
This includes but is not limited to ideal square-lattice  GKP states.
Using the decomposition of the embedded-error operator, \erf{NoiseMoyalsmall}, and the Moyal product relations, \erf{Moyaldiagonal}, we show in Appendix~\ref{Appendix:embeddederrorop} that
the Wigner function for the noisy operator $\op{\bar{O}}_\Sha = \op{\xi}(\mat{\Xi}) \op{O}_\Sha \op{\xi}(\mat{\Xi})$, is given in the low-noise limit by 
    \begin{align} 
        &W_{ \op{\bar{O}}_\Sha  }(\vec{x})  
        = \Big[ W_{\op{\xi}(\mat{\Xi}_0) } \star \Sha_{\mat{S}_R\mat{A}} [\begin{smallmatrix}
        \vec{v}_1 \\
        \vec{v}_2
        \end{smallmatrix}] \star         
            W_{\op{\xi}(\mat{\Xi}_0) } \Big](\mat{R}_\varphi \vec{x}) \\
        & \approx \frac{\pi}{\Delta \kappa} G_{\frac{1}{2} \mat{\Xi}_0 }(\mat{R}_\varphi \vec{x}) 
            \Big[  \Sha_{\mat{S}_R\mat{A}} [\begin{smallmatrix}
        \vec{v}_1 \\
        \vec{v}_2
        \end{smallmatrix}]  * G_{\frac{1}{2} \mat{\Omega} \mat{\Xi}^{-1}_0 \mat{\Omega}^\tp } \Big](\mat{R}_\varphi \vec{x}) \\
        & = \frac{\pi}{\Delta \kappa} G_{ \mat{\Sigma}_\text{env} }(\vec{x}) 
            \Rthetafxn[\mat{A}]{\vec{v}_1}{\vec{v}_2} \left( \vec{x}, 2 \pi i \mat{\Sigma}_\text{spike} \right)\, .\label{OShatransformedNoisy}
    \end{align} 
The factor $\pi/(\Delta \kappa)$ arises from the exponential operators in \erf{highqualitynoise} (see Appendix~\ref{Appendix:embeddederrorop}) and plays no physical role for approximate GKP states, whose Wigner functions must be normalized after application of the noise operator anyway. 
Equation~\eqref{OShatransformedNoisy} describes a broad two-dimensional Gaussian envelope that damps a set of Gaussian spikes, with respective covariance matrices given by\footnote{The parameters $\kappa^2$ and $\Delta^2$ are variances of Gaussian terms in the approximate GKP \emph{wavefunctions}. The elements of the covariance matrices here correspond to measured quadrature variances (proportional the square-modulus of the wavefunctions, \erf{sqrgauss}), which is the origin of the $\frac{1}{2}$ factor.} 
    \begin{subequations} \label{EnvSpikeCovariances}
    \begin{align}
        \mat{\Sigma}_\text{env} &= \frac{1}{2} \mat{\Xi}, \\
        \mat{\Sigma}_\text{spike} &= \frac{1}{2} \mat{\Omega}^\tp \mat{\Xi}^{-1} \mat{\Omega} 
        = \frac{1}{2} \frac{\mat{\Xi}}{\det \mat{\Xi}}
        \,.
    \end{align}
    \end{subequations}
The last equality holds because the $2 \times 2$ noise matrix is symmetric, $\mat{\Xi} = \mat{\Xi}^\tp $. The covariance matrices for the spikeand envelope are thus directly proportional to each other:
    \begin{align} \label{minenvelope}
        \mat{\Sigma}_\text{env} 
        =  \frac{ \mat{\Sigma}_\text{spike} }{ \det (2 \mat{\Sigma}_\text{spike}) }  \longleftrightarrow \mat{\Sigma}_\text{spike} 
        =  \frac{ \mat{\Sigma}_\text{env} }{ \det (2 \mat{\Sigma}_\text{env}) }\blk \, 
        .
    \end{align}
This critical relationship is the \emph{minimum-envelope condition}. As shown below in Sec.~\ref{subsection:physenvcondition}, in the context of approximate GKP states with a fixed $\mat{\Sigma}_\text{spike}$, smaller envelopes than allowed by \erf{minenvelope} are unphysical, and larger envelopes correspond to mixtures over minimum envelopes---such states are physical but can never be pure.

\subsection{Minimum-envelope approximate GKP states}

\begin{figure} [t]
	\includegraphics[width=0.95\linewidth]{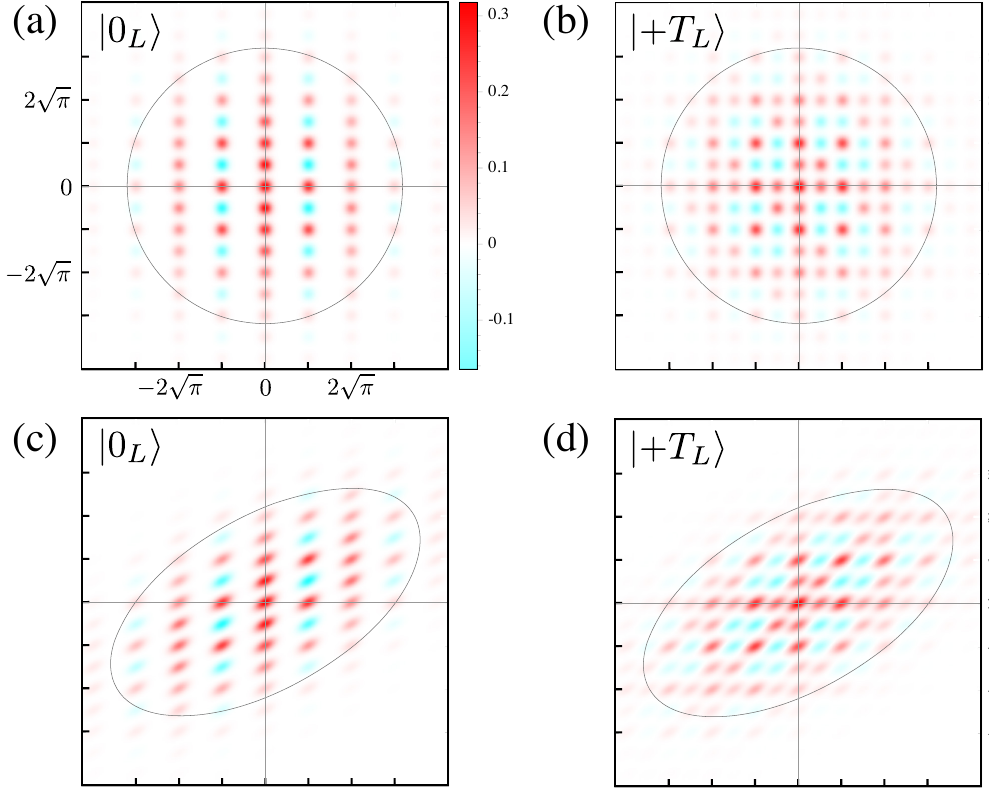} 
	\caption{\label{Fig:ApproxStates} Wigner functions for minimum-envelope approximate GKP states, Eq.~\eqref{wignerapproxGKPlownoise}. Panel (a) shows the approximate Pauli eigenstate $\ket{\bar{0}_L}$ with Bloch 3-vector $\arrow{r} = [1,0,0]^{\tp}$, and panel (b) shows the approximate $T$-type magic state $\ket{+\bar{T}_L}$ with $\arrow{r} = [\frac{1}{\sqrt{3}},\frac{1}{\sqrt{3}},\frac{1}{\sqrt{3}}]^{\tp}$. Both states have spike covariance matrix $\mat{\Sigma}_\text{spike} =\tfrac{1}{2}\Big[\begin{smallmatrix}
	(0.2)^{2} & 0 \\ 
	0 & (0.2)^{2}
	\end{smallmatrix}\Big]$
	and associated minimum envelope given by Eq.~\eqref{minenvelope}. Panels (c) and (d) show the same logical states, respectively, but with skewed spike covariance matrix $\mat{\Sigma}_\text{spike} = \tfrac{1}{2}\mat{R}^{\tp}_{\pi/3}\Big[\begin{smallmatrix}
	(0.2)^{2} & 0 \\ 
	0 & (0.4)^{2}
	\end{smallmatrix}\Big]\mat{R}_{\pi/3} $
	and associated envelope, where $\mat{R}_{\pi/3}$ is a rotation matrix, \erf{rotationmatrix}. Ovals are 2-$\sigma$ contours for the Gaussian envelopes, and ticks indicate integer multiples of $\sqrt{\pi}$.
	}
\end{figure}

\label{subsection:minenvapproxgkp}

 Minimum-envelope approximate GKP states, $\op{\bar{\rho}}_L$, are those generated by applying the embedded-error operator to an ideal GKP state $\op{\rho}_L$, \erf{GKPidealgeneralstate}:
  \begin{align} \label{Wignerapproxstate}
        \op{\bar{\rho}}_L = \frac{1}{\mathcal{N}} \op{\xi}(\mat{\Xi}) \op{\rho}_L \op{\xi}(\mat{\Xi})\, ,
    \end{align}
with normalization $\mathcal{N} = \Tr\{\op{\rho}_L [\op{\xi}(\mat{\Xi})]^2 \}$ required because the embedded-error operator is not unitary.
If the ideal GKP state is a pure state, $\op{\rho}_L = \outprod{\psi_L}{\psi_L}$, then the approximate GKP state is also a pure state, $\op{\bar{\rho}}_L = \outprod{\bar{\psi}_L}{\bar{\psi}_L}$ (as established above in the context of wavefunctions). The Wigner function for the approximate GKP state in \erf{Wignerapproxstate} is
    \begin{align} \label{wignerapproxGKP}
        \bar{W}_{\vec{r}} (\vec{x}) & \coloneqq W_{\op{\bar{\rho}}_L}( \vec{x} ) = \frac{1}{\mathcal{N} }[W_{\op{\xi}(\mat{\Xi})} \star W^{L}_{\vec{r}} \star W_{\op{\xi}(\mat{\Xi})}] (\vec{x}) \\
        &= \frac{1}{\mathcal{N} } \sum_{\mu = 0}^3 r_\mu [W_{\op{\xi}(\mat{\Xi})} \star W^{L}_{\mu} \star W_{\op{\xi}(\mat{\Xi})}] (\vec{x}) \, ,
    \end{align}
where $\mathbf{r}$ is the Bloch 4-vector parameterizing the qubit state, $W^{L}_{\vec{r}}(\vec{x})$ is the ideal-GKP Wigner function [\erf{wigneridealGKP}], $W_{\op{\xi}(\mat{\Xi})}(\vec{x})$ is the Wigner function for the embedded-error operator, and the normalization factor $\mathcal{N}$ is inherited from \erf{Wignerapproxstate}. %
The second line is written in terms of the embedded-error operator acting directly on the Wigner functions for the GKP Paulis, \erf{2DSha}, which gives rise to ``noise-broadened'' GKP Pauli Wigner functions.

The limit of low embedded error, $\mat \Xi \ll \mat \id$, is examined in general in Sec.~\ref{subsubsection-lowerrorlimit}. From \erf{OShatransformedNoisy}, the Wigner function can be written 
  \begin{align} \label{wignerapproxGKPlownoise}
    \bar{W}_{\vec{r}} (\vec{x}) 
        \approx \frac{1}{\mathcal{N} } G_{ \mat{\Sigma}_\text{env} }(\vec{x}) 
            \sum_{\mu = 0}^3 r_\mu \Rthetafxn[\sqrt{\pi}\mat{I}]{\nicefrac{\mat{\Omega}\vec{\ell}_{\mu}}{2}}{\nicefrac{\vec{\ell}_{\mu}}{2}} \left( \vec{x}, 2 \pi i \mat{\Sigma}_\text{spike} \right)\, ,
    \end{align}
with each term in the sum resulting from the embedded-error operator acting on the GKP Pauli operators. 
The features of approximate GKP states in phase space are similar to those for wavefunctions [\erf{approxGKPlimit}]: the lattice of points (a $\Sha$-function) is convolved with and enveloped by Gaussian functions. The Gaussian envelopes damp distant regions of quasiprobability and, along with the Gaussian convolutions, ensure that the Wigner function and its marginals are normalizable. For these minimum-uncertainty states, the spike and envelope covariance matrices are related by the minimum-envelope condition in Eq.~\eqref{minenvelope}. Examples are shown in Fig.~\ref{Fig:ApproxStates}.

The normalization factor in \erf{wignerapproxGKPlownoise} is given by a sum of four integrals, one for each of the noise-broadened GKP Pauli operators including the GKP identity operator:
  \begin{align} \label{normalizationapproxGKP}
        \mathcal{N} &= \int d^{2}\vec{x}\,  G_{ \mat{\Sigma}_\text{env} }(\vec{x}) 
            \sum_{\mu = 0}^3 r_\mu \Rthetafxn[\sqrt{\pi}\mat{I}]{\nicefrac{\mat{\Omega}\vec{\ell}_{\mu}}{2}}{\nicefrac{\vec{\ell}_{\mu}}{2}} \left( \vec{x}, 2 \pi i \mat{\Sigma}_\text{spike} \right) \, .
    \end{align}  
The Wigner functions for ideal GKP Pauli operators ($\mu =1,2,3$), \erf{2DSha}, are traceless (their integral over phase space is zero). This fact is visually evident in Fig.~\ref{Fig:idealGKPPauli}: each has two positive and two negative spikes within a unit cell. In \erf{normalizationapproxGKP}, the envelope arising from the embedded-error operator disturbs this property. However, when the noise is low enough, the associated integrals rapidly approach zero (see Fig.~\ref{Fig:Paulitrace} in Appendix~\ref{appendix:exactnormalization}). In this limit, the envelope can be ignored and the integrals performed, with each becoming a $\theta$-function evaluated at the origin (referred to as a $\theta$\emph{-constant} \cite{igusaThetaFunctions1972b}). The non-identity terms vanish, and the normalization is 
    \begin{align} \label{minenvconditions}
        &\mathcal{N} 
        \approx \Rthetafxn[\sqrpi\mat{I}]{\mat{0}}{\mat{0}} (\vec{0}, 2\pi     i\mat{\Sigma}_\text{env}) 
        \approx \frac{1}{\sqrt{\pi}} &\Delta^2,\kappa^2 \ll 1.
    \end{align}    \blk
Further details of the calculation are given in Appendix~\ref{appendix:exactnormalization}, and a complementary description can be found in Matsuura \emph{et al.}~\cite{matsuura2019equivalence}.

\begin{figure*}[t]
	\includegraphics[width=0.85\linewidth]{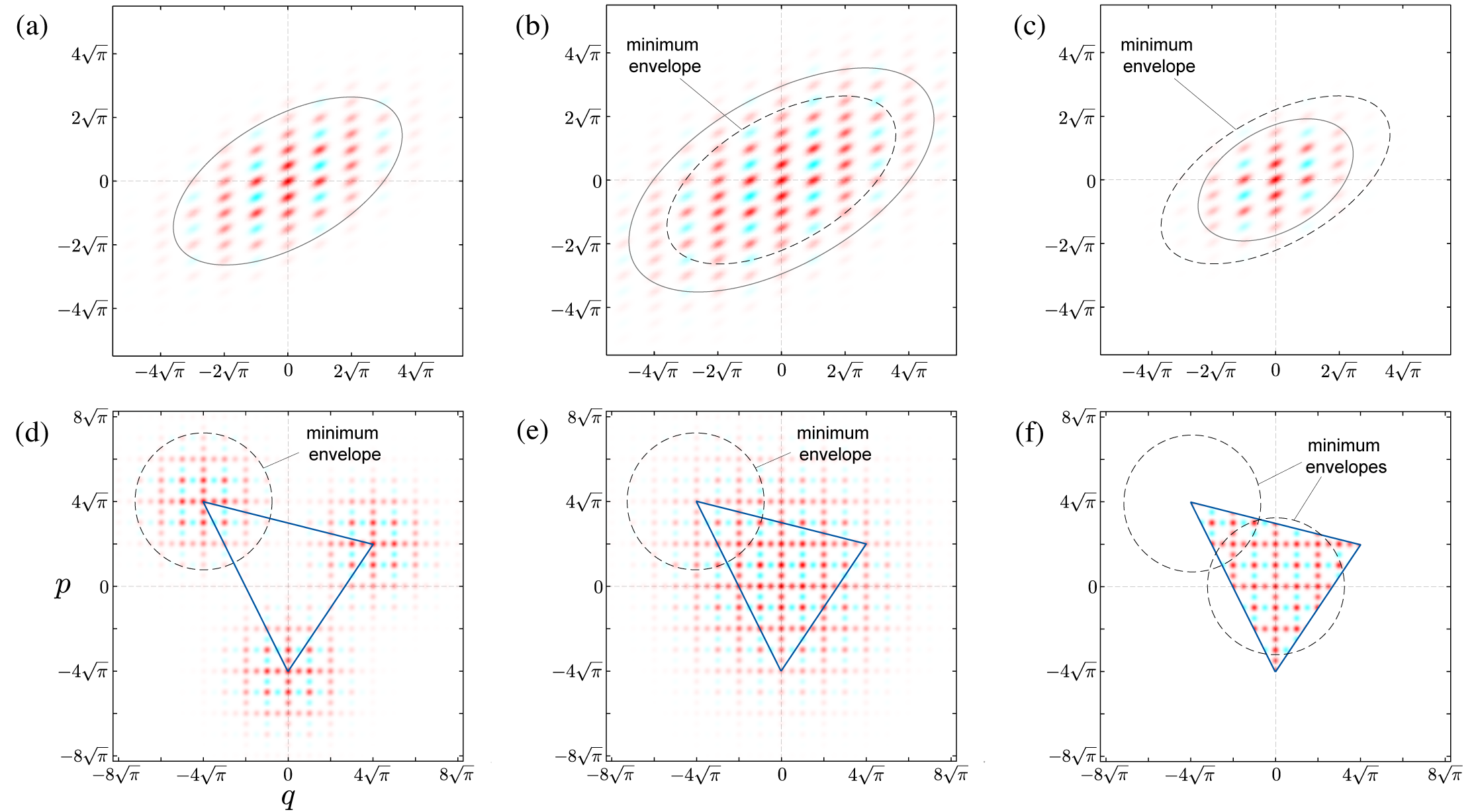} 
	\caption{\label{Envelopestudy} Envelope effects for approximate GKP states. In the first row, we plot Wigner functions of the form in Eq.~\eqref{envelopedGKP}, for fixed spike covariance 
	$\mat{\Sigma}_\text{spike} =\tfrac{1}{2} \mat{R}^{\tp}_{\pi/3}\Big[\begin{smallmatrix}
	(0.2)^{2} & 0 \\ 
	0 & (0.4)^{2}
	\end{smallmatrix}\Big] \mat{R}_{\pi/3}$ ($\mat{R}_{\pi/3}$ is a rotation matrix, \erf{rotationmatrix}),
all of which encode a logical $\ket{0_L}$. The solid ovals are 2-$\sigma$ error ellipses for the Gaussian envelopes, red is positive, blue is negative, and the colorscale is normalized for each plot separately. (a) A minimum-envelope state, Eq.~\eqref{wignerapproxGKPlownoise}, with $\mat{\Sigma}_\text{env}$ given by Eq.~\eqref{minenvelope}. (b) A larger-than-minimum envelope, which indicates mixing over minimum envelopes, and (c) an envelope that is too small, meaning that this Wigner function does not represent a physical quantum state.
	In the second row, we plot Wigner functions for a logical $\ket{+H}$ state with Bloch 3-vector $\arrow{r} = [\frac{1}{\sqrt{2}},0,\frac{1}{\sqrt{2}}]^{\tp}$ with fixed spike covariance 
	$\mat{\Sigma}_\text{spike}
	= \tfrac{1}{2}\Big[\begin{smallmatrix}
	(0.25)^{2} & 0 \\ 
	0 & (0.25)^{2}
	\end{smallmatrix}\Big]$
whose envelopes are related to a triangle, indicated in blue.(d) A "spotlight" envelope arises from a convolution of the minimum envelope with a $\delta$-function at each point using Eq.~\eqref{envelopemixing}. (e) A convolution of the minimum envelope with the enclosed triangular region. The minimum envelope is shown on the plot for reference.
	(f) A hard-boundary exclusion envelope determined by the triangle. In this case, the minimum envelope is incompatible with this Wigner function (deconvolving with the minimum envelope violates Eq.~\eqref{positiveenvelopecondition}), and such a Wigner function does not represent a physical state. This is apparent visually in the fact that the triangle contains features sharper that those of the minimum envelope, so the minimum envelope cannot ``fit into'' the triangle.
	}
\end{figure*}

\subsection{Envelope condition for physical states} 
\label{subsection:physenvcondition}

The minimum-envelope condition given in \erf{minenvelope} suggests that some enveloped $\theta$-functions correspond to approximate GKP states, pure or mixed, but when the envelopes are too small, the Wigner functions no longer represent physical states. Here, we make this statement concrete and emphasize that with it one can construct physical descriptions of GKP-encoded information directly in phase space. 

We begin by generalizing the minimum-envelope GKP Wigner function in \erf{wignerapproxGKPlownoise}. By first displacing the Wigner function of an ideal GKP state by $\vec{x}'$, applying the embedded-error operator, and then displacing back, we produce a GKP Wigner function with the spikes in the appropriate places and the same spike covariance $\mat{\Sigma}_\text{spike}$, but with a shifted envelope:
    \begin{align} 
	    \bar{W}_{\vec{r}} (\vec{x}) = \frac{1}{\mathcal{N}} G_{\mat{\Sigma}_\text{env}} (\vec{x} - \vec{x}')  \sum^{3}_{\mu=0} r_{\mu} \Rthetafxn[\sqrpi]{\nicefrac{\mat{\Omega}\vec{\ell}_{\mu}}{2} }{\nicefrac{\vec{\ell}_{\mu}}{2}} (\vec{x},2\pi i \mat{\Sigma}_{\text{spike}}) \, .
    \end{align}
This state is also a minimum-envelope state---$\mat{\Sigma}_\text{env}$ satisfies \erf{minenvelope}---meaning that if the original ideal GKP state were pure, this state will also be pure. Taking mixtures of these displacements (prior to applying the embedded-error operator) with positive distribution $h(\vec{x}) \geq 0$ produces a GKP state with Wigner function
    \begin{align} \label{envelopedGKP}
	    \bar{W}_{\vec{r}} (\vec{x}) = \frac{1}{\mathcal{N}}g(\vec{x})  \sum^{3}_{\mu=0} r_{\mu} \Rthetafxn[\sqrpi]{\nicefrac{\mat{\Omega}\vec{\ell}_{\mu}}{2} }{\nicefrac{\vec{\ell}_{\mu}}{2}} (\vec{x},2\pi i \mat{\Sigma}_{\text{spike}}) \, ,
    \end{align}
with the same, fixed spike covariance $\mat{\Sigma}_\text{spike}$, and with the envelope $g(\vec{x})$ given by a convolution of $h(\vec{x})$ with the minimum envelope $G_{\mat{\Sigma}_\text{env}} (\vec{x})$:
    \begin{align} \label{envelopemixing}
        g(\vec{x}) = [h * G_{\mat{\Sigma}_\text{env}}] (\vec x) = \int d^2 \vec{x}'\, h(\vec{x}') G_{\mat{\Sigma}_\text{env}} (\vec{x} - \vec{x}')\, .
    \end{align}
Inverting this relation using a deconvolution gives a sufficient condition for physically allowed envelopes when the spike covariance $\mat{\Sigma}_\text{spike}$ is fixed. In terms of the deblurring operator, \erf{blurringops}, this physical-envelope condition is
    \begin{align} \label{positiveenvelopecondition}
	    &\mathcal{D}^{-1}_{\mat{\Sigma}_\text{env} } g (\vec{x}) 
	    \geq 0 \, . %
    \end{align}
In short, admissible envelopes for physical GKP Wigner functions are positive semidefinite under the action of the deblurring operator associated with the minimum envelope. Minimum-envelope states have singular distributions, $h(\vec{x}) = \delta(\vec{x} - \vec{x}')$ that saturate \erf{positiveenvelopecondition} and can describe pure states. Other physical envelopes describe mixed states.

A key benefit of the physical envelope condition in Eq.~\eqref{positiveenvelopecondition} is that representations of GKP-encoded information can be constructed directly in phase space. One begins by laying out a periodic array of Gaussian spikes, described by four two-dimensional $\theta$-functions of covariance matrix $\mat{\Sigma}_\text{spike}$ (corresponding to noise-broadened Pauli Wigner functions), and a Bloch vector $\vec{r}$ determined by the state. Then, an envelope is applied, and the physical-envelope condition in \erf{positiveenvelopecondition} determines whether this envelope (a) is physically valid, (b) corresponds to a minimum-envelope state, or (c) corresponds to a mixture (either discrete or continuous) over minimum envelopes. Examples of these scenarios are given in Fig.~\ref{Envelopestudy}.

Our discussion here has focused on varying the envelope for a fixed spike covariance $\mat{\Sigma}_\text{spike}$ in order to introduce the concept of mixing over minimum envelopes. For these states, the minimum-envelope condition, Eq.~\eqref{minenvelope}, implies that larger spikes have a smaller minimum envelopes. This is contrary to the case of a minimum-uncertainty approximate GKP state fed through a displacement channel, Eq.~\eqref{displacementchannel}. In this case, the spike size is increased by the channel, but so is the envelope. This is another way to see that such a state must be a mixed state (since its envelope is no longer minimal).
\subsection{Connection between low amounts of coherent and incoherent noise}
\label{subsection:twirling}
A Gaussian mixture of ideal GKP states, \erf{mixedideal}, arises from a convolution of $\Sha$-functions that yields $\theta$-functions, which remain unnormalizable and are thus unphysical. To map these states to normalizable, approximate GKP states, \erf{wignerapproxGKPlownoise}, one applies an appropriate envelope at least as large as that specified by the minimum-envelope condition in \erf{minenvconditions}.

How is this performed in the other direction---\emph{i.e.}, what is the map that takes approximate GKP states and produces blurred ideal GKP states? The answer, given by Noh and Chamberlain~\cite{nohFaulttolerantBosonicQuantum2020}, is to take the approximate GKP state $\op{\bar{\rho}}_L$ and apply random multiples of $2\sqrt{\pi}$ shifts in position and momentum, via powers of the primitive stabilizers $\op{S}_X$ and $\op{S}_Z$ in \erf{stabilizers}:
 \begin{align}
     \op{\bar{\rho}}_L' = \sum_{\vec{n} \in \mathbb{Z}^2} [\op{S}_Z]^{n_2} [\op{S}_X]^{n_1} \op{\bar{\rho}}_L [\op{S}_X]^{n_1} [\op{S}_Z]^{n_2} \, .
 \end{align}
In phase space, the shifts are translations of the argument of the approximate GKP Wigner function,
    \begin{align}
	    W_{ \op{\bar{\rho}}'}(\vec{x}) &= \sum_{\vec{n} \in \mathbb{Z}^{2}}   W_{ \op{\bar{\rho}}_L}  (\vec{x} + 2\vec{n}\sqrpi) \\ 
	    &\propto \sum_{\vec{n} \in \mathbb{Z}^{2}}  G_{\mat{\Sigma}_{\text{env}}}  (\vec{x} + 2\vec{n}\sqrpi)  \nonumber \\
	    & \quad \times \sum^{3}_{\mu =0} r_{\mu} 
	    \Rthetafxn[\sqrpi\mat{I}]{\nicefrac{\mat{\Omega}\vec{\ell}_{\mu}}{2}}{\nicefrac{\vec{\ell}_{\mu}}{2}}  (\vec{x}, 2\pi i \mat{\Sigma}_{\text{spike}} )\, , \label{Wignerinthemiddle}
    \end{align}
where we used the quasiperiodicity of the $\theta$-function, \erf{Thetaquasiperiod}, to eliminate the $\vec{n}$ dependence. This is unsurprising, since the stabilizers are designed to act trivially on properly periodic codewords. The pulse train of Gaussians envelopes can also be written as a $\theta$-function,
    \begin{align}
        \sum_{\vec{n} \in \mathbb{Z}^{2}} G_{\mat{\Sigma}_{\text{env}}}  (\vec{x} + 2\vec{n}\sqrpi) \propto 
        \theta_{2\sqrpi\mat{I}}( \vec{x} , 2\pi i \mat{\Sigma}_{\text{env}} ).
    \end{align}
with spike covariance $\mat{\Sigma}_\text{env}$. In the high-quality limit, the eigenvalues of $\mat{\Sigma}_\text{env}$ are much larger than the square of the period~$2\sqrt{\pi}$, and $\theta_{2\sqrpi\mat{I}}( \vec{x} , 2\pi i \mat{\Sigma}_{\text{env}} ) \approx 1$. Equation~(\ref{Wignerinthemiddle}) then becomes
 \begin{align}
	    W_{ \op{\bar{\rho}}'}(\vec{x}) & \approx 
	      \sum^{3}_{\mu =0} r_{\mu} 
	    \Rthetafxn[\sqrpi\mat{I}]{\nicefrac{\mat{\Omega}\vec{\ell}_{\mu}}{2}}{\nicefrac{\vec{\ell}_{\mu}}{2}}  (\vec{x}, 2\pi i \mat{\Sigma}_{\text{spike}} )
	    \, ,
    \end{align}
which is the Wigner function for Gaussian mixture of ideal GKP states, \erf{mixedideal}. This shows the connection between coherent and incoherent noise in the high-quality GKP-state limit.

\section{GKP Error Correction in Phase Space}
\label{section:gkpec}

GKP error correction is a procedure designed to detect and correct CV-level errors on GKP-encoded quantum information. GKP error correction proceeds in two steps. First, the state to be corrected, $\op{\rho}_\text{in}$, is coupled to an ancilla mode prepared in a GKP state, $\ket{0_L}$, which is then measured in the position basis, yielding a measurement outcome $m_q$. Then, this process is repeated using another GKP ancilla, which is measured in momentum without outcome $m_p$. We refer to this part of the procedure as \emph{syndrome extraction} as it involves acquiring the measurement-outcome syndrome $\{m_q, m_p\}$. In the second part of the procedure, the syndrome is used in a decoder to determine the likelihood of a logical error having occurred. As determined by the decoder, a \emph{recovery} operation consisting of shifts in phase space is applied.

This entire procedure projects the noisy input state $\op{\rho}_\text{in}$ into the GKP code space, with CV-level noise being projected into potential qubit-level errors. When non-ideal GKP ancillae are used, the input state is projected into a subspace commensurate with the quality of these ancillae~\cite{walshe2020continuous}. That is, the error correction procedure cannot reduce the CV-level noise on the input state to below that of the ancillae.

\subsection{Kraus operator for GKP error correction}
The simplest decoder for the square-GKP code takes each measurement outcome and divides it into two pieces: the nearest integer multiple of $\sqrt{\pi}$ and a remainder. For example, the first half of GKP syndrome extraction gives outcome $m_q \in \mathbb{R}$ that splits as: 
    \begin{align}
        m_q = \lfloor m_q \rceil_{\sqrt{\pi}} + \{m_q\}_{\sqrpi} ,
    \end{align}
where the nearest-integer and remainder of real number $r$ (with respect to some $\alpha > 0$) are given by the functions
    \begin{align}
        \lfloor r \rceil_{\alpha} & \coloneqq \alpha \Big\lfloor \frac{r}{\alpha} + \frac{1}{2} \Big\rfloor \\
        \{r\}_{\alpha} & \coloneqq r - \lfloor r \rceil_{\alpha}
        \, ,
    \end{align}
and $\lfloor \cdot \rfloor$ is the floor function. Note that $\lfloor r \rceil_{\alpha}$ is an integer multiple of $\alpha$, and $\{r\}_{\alpha} \in [-\frac{\alpha}{2}, \frac{\alpha}{2} )$. Then, a recovery is given by a shift in position by the remainder, $\op{X}_\text{rec} \coloneqq \op{X}(-\{ m_q \}_{\sqrpi})$. An analogous procedure applies for the other half of GKP error correction: the outcome $m_p$ is divided as $m_p = \lfloor m_p \rceil_{\sqrt{\pi}} + \{m_p\}_{\sqrpi}$, and a momentum shift $\op{Z}_\text{rec} \coloneqq \op{Z}(-\{ m_p \}_{\sqrpi})$ is applied. 

The circuit for GKP error correction using this decoder is
    \begin{align} \nonumber 
        \begin{split}
        \Qcircuit @C=0.45cm @R=0.5cm {
    &\lstick{{ (\mathrm{out})}} &\gate{Z_\text{rec}}\cwx[2]&\qw &\qw&\qw &\targ &\gate{X_\text{rec}}\cwx[1]       &\qw &\qw &\qw                      &\ctrl{1}  &\rstick{(in)} \qw \\
    &&&&&&  & \control   &  \cw  &    &\lstick{\brasub{m_q}{p}\!\!} &\ctrl{-1} &\rstick{\ket{0_L}} \qw \\
    && \control &\cw&&\lstick{\brasub{m_p}{p}\!\!}&\ctrl{-2}&\rstick{\ket{0_L}} \qw &    &    &                         &         & \\
    }
        \end{split}\\\label{ECcircuit}
    \end{align}
which is read in a right-to-left fashion. We represent a projective measurement of observable $\op B$ with outcome $m$ using the bra $\brasub{m}{B}$.
In the circuit, a two-mode position-controlled momentum-shift gate is 
\begin{align} \label{CZgate}
    \begin{split}
        \Qcircuit @C=0.5cm @R=0.6cm {
        &                                                                             &&&&\ctrl{0}  &\qw \\
        &\ustick{\raisebox{.2em}{$\op{C}_Z \coloneqq e^{i \op{q} \otimes \op{q}} =$}} &&&&\ctrl{-1} &\qw
}
\, 
    \end{split}
\end{align}
and a two-mode position-controlled position-shift gate is
\begin{equation} \label{CXgate}
       \begin{split}
        \Qcircuit @C=0.3cm @R=0.5cm {
         &&&&&& \targ & \qw    &&&& \gate{F^\dagger} & \ctrl{1}    & \gate{F }  \qw   & \qw     \\
         &\ustick{\raisebox{.2em}{$\op{C}^{(2,1)}_X \coloneqq e^{-i \op{p} \otimes \op{q}} =\qquad$}}&&&&& \ctrl{-1} & \qw  &\ustick{\raisebox{.2em}{$=$}} &&& \qw & \ctrl{0} & \qw & \qw
    } 
    \,
       \end{split}
\end{equation}
where the second wire is the control and the first wire is the target, indicated by the superscripts on $\op{C}^{(2,1)}_X$.
We also note that $e^{-i \op{p} \otimes \op{q}} = \op{F}^\dagger_1 \op{C}_Z \op{F}_1$, where $\op{F}_1$ is the Fourier transform operator, Eq.~\eqref{GKPCliffords}, on the first mode.

The Kraus operator for GKP error correction, $\op{K}_\text{EC}(m_q, m_p)$, describes the map from the input state to the output state through the error correction procedure,
    \begin{align} \label{KrausGKPEC}
        \op{\rho}_\text{out} = \frac{1}{\Pr(m_q, m_p)} \op{K}_\text{EC}(m_q, m_p) \op{\rho}_\text{in} \op{K}^\dagger_\text{EC}(m_q, m_p).
    \end{align}
The output state is normalized by the probability to obtain the outcomes,
\begin{equation}
    \Pr(m_q,m_p) = \Tr[\op{K}^\dagger_\text{EC}(m_q, m_p) \op{K}_\text{EC}(m_q, m_p) \op{\rho}_\text{in}]\,.
\end{equation}
Since the two pieces of syndrome extraction and subsequent correction occur independently, we express the Kraus operator as
    \begin{align} \label{KrausECdecomp}
        \op{K}_\text{EC}(m_q, m_p) = \op{K}_p (m_p) \op{K}_q (m_q)
        \, .
    \end{align}
Note that a benefit of the right-to-left circuit convention we employ is that we can read off unitaries and Kraus operator (as below) from the circuit in the same order that they appear in equations.

To allow for both ideal and approximate GKP states, we consider the Kraus operator for each of the two halves of GKP syndrome extraction (one for position and one for momentum) using arbitrary ancillae, both in the pure state $\ket{\chi} = \int ds \, \chi(s) \ket{s}_q = \int dt \, \tilde{\chi}(t) \ket{t}_p $. The circuit for the first syndrome extraction is
\begin{align}
\begin{split}
\Qcircuit @C=1.5em @R=2em {
& \lstick{{ (\mathrm{out})}}          &\ctrl{1}  &\rstick{(in)} \qw \\
&\lstick{\brasub{m_q}{p}} &\ctrl{-1} &\rstick{\ket{\chi}} \qw
}
\, 
\end{split}
\end{align}
and the second is nearly identical, with the measurement replaced by $\brasub{m_p}{p}$ and the Fourier transforms in Eq.~\eqref{qGKPerrordetection} included.
The Kraus operators for syndrome extraction (before the corrective shift) are
    \begin{align} \label{qGKPerrordetection}
	\op{K}_{q, \text{syn}}(m_q) 
	    &\coloneqq \brasub{m_q}{p} \op{C}_Z \ket{\chi} 
	     = \tilde{\chi}  (m_q - \op{q}) \, ,
	     \\
	 \op{K}_{p, \text{syn}}(m_p) 
	    &\coloneqq \brasub{m_p}{p} \op{F}^\dagger_1 \blk \op{C}_Z \op{F}_1  \ket{\chi} 
	     = \tilde{\chi}  ( m_p - \op{p} ) \, .
    \end{align}

For GKP error correction, we use pure, approximate GKP states $\ket{\bar{0}_L}$, Eq.~\eqref{approximateGKPstate}, with symmetric noise, $\mat{\Xi} \propto \mat{I}$, as ancillae. Such states have the property that $\op{F} \ket{\bar{0}_L} = \op{F}^\dagger \ket{\bar{0}_L} = \ket{\bar{+}_L}$, where the noise in $\ket{\bar{+}_L}$, $\mat{\Sigma}_\text{spike}$ and $\mat{\Sigma}_\text{env}$, is the same as for $\ket{\bar{0}_L}$. 
Ideal GKP states have this property trivially.
These GKP ancillae, $\ket{\bar{0}_L} = \int ds \, \bar{\psi}_{0}(s) \qket{s}$, with position wavefunction $\bar{\psi}_{0}(s)$ that represents either ideal GKP states [Eq.~\eqref{GKPwavefunctions}] or approximate GKP states [Eq.~\eqref{approxGKPlimit}], yield Kraus operators
    \begin{subequations} \label{GKPerrorcorrection}
    \begin{align} 
	    \op{K}_{q}(m_q) &= \op{X}(\{m_q\}_{\sqrt{\pi}}) \bar{\psi}_{+} (m_q - \op{q}) \, ,\\
	    \op{K}_{p}(m_p) &= \op{Z}(\{m_p\}_{\sqrt{\pi}}) \bar{\psi}_{+} (m_p - \op{p}) \, ,
    \end{align}
    \end{subequations}
where $\bar{\psi}_{+}(s)$ is the position wavefunction for $\ket{\bar{+}_L}$, since $\tilde{\bar{\psi}}_{0}(s) = \bar{\psi}_{+}(s)$ for these states. Note that we have dropped the $L$ subscript on wavefunctions for brevity. Together, these operators give the Kraus operator for GKP error correction using the simple decoder in Eq.~\eqref{KrausGKPEC}.

We can transform the Kraus operators into a useful form by extracting the outcome-dependent shifts in Eqs.~\eqref{GKPerrorcorrection}, \emph{e.g.},~$\bar{\psi}_+(m_q - \op{q}) = \op{X}^\dagger(m_q) \bar{\psi}_+(\op{q}) \op{X}(m_q)$, where we also use the fact that $\bar{\psi}_+(x) = \bar{\psi}_+(-x)$. Performing the analogous transformation on $\bar{\psi}_+(m_p - \op{p})$, we get
    \begin{subequations} 
    \begin{align} 
	    \op{K}_{q}(m_q) %
	    &= \op{X}(-\lfloor m_q \rceil_{\sqrt{\pi}})  \bar{\psi}_+(\op{q}) \op{X}(m_q) \\
	    \op{K}_{p}(m_p) &= \op{Z}(-\lfloor m_p \rceil_{\sqrt{\pi}})  \bar{\psi}_+(\op{p}) \op{Z}(m_p) \, .
    \end{align}
    \end{subequations}
Each describes a shift (position or momentum) by the respective measurement outcome on the input state, followed by application of  $\bar{\psi}_+(\op{q})$ or $\bar{\psi}_+(\op{p})$, and then finally by another shift by an integer multiple of $\sqrt{\pi}$. When combined together, the full Kraus operator is
    \begin{align}  \label{KEC_rewrite}
	    \op{K}_\text{EC}(m_q, m_p) = 
	     & e^{i \phi} \op{X}(-\lfloor m_q \rceil_{\sqrt{\pi}}) \op{Z}(-\lfloor m_p \rceil_{\sqrt{\pi}})  \nonumber \\*
	    & \quad \times \bar{\psi}_+(\op{p})  \bar{\psi}_+(\op{q}) \op{X}(m_q) \op{Z}(m_p) \, ,
    \end{align}
with phase $\phi = - \{ m_q \}_{\sqrt{\pi}} \lfloor m_p \rceil_{\sqrt{\pi}}$. The important term in this expression is $\bar{\psi}_+(\op{p})  \bar{\psi}_+(\op{q})$, which becomes the projector onto the square-lattice GKP subspace when the ancillae are ideal GKP states~\cite{allGaussianPRL,walshe2020continuous}, as we show in Sec.~\ref{Sec:ECwithidealstates}.\\

\subsection{GKP error correction in phase space}
\label{subsection:gkpecphasespace}

GKP error correction can be described directly in phase space using the Moyal product, Eq.~\eqref{MoyalProductdef}, 
\begin{align} \label{ECMoyal}
 	&\op{K}_\text{EC}(m_q, m_p) \op{\rho}_\text{in} \op{K}^\dagger_\text{EC}(m_q, m_p) 
  	 \nonumber \\ 
  	 &\quad \quad \quad \to \big[W_{\op{K}_\text{EC}} \star W_{\op{\rho}_\text{in}} \star W_{\op{K}^\dagger_\text{EC}} \big] (q,p)
	\, ,
\end{align}
where $W_{\op{K}_\text{EC}}(q,p)$ is the Wigner function for the GKP error correction Kraus operator. Since $\op{K}_\text{EC}(m_q, m_p)$ is the product of two operators, Eq.~\eqref{KrausECdecomp}, we can also separate its Wigner function using the Moyal product:
\begin{align}
	W_{\op{K}_\text{EC}}(q,p) = \big[ W_{\op{K}_p} \star W_{\op{K}_q} \big] (q,p)
	\, ,
\end{align}
where $W_{\op{K}_q}(q,p)$ and $W_{\op{K}_p}(q,p)$ are the Wigner functions for the two halves of GKP error correction.
This allows us to apply each half of GKP error correction separately in phase space.

To isolate the key features of GKP error correction, we consider the specific case of outcomes $\{m_q, m_p\} = \{0,0\}$, noting that the shifts associated with other outcomes in the general expression in Eq.~\eqref{KEC_rewrite}, can be applied in phase space using Eq.~\eqref{GaussianWignertransformation}.
In this case, $\op{K}_\text{EC}(0, 0) = \bar{\psi}_+(\op{p})  \bar{\psi}_+(\op{q})$, with each half of GKP error correction given by a Kraus operator diagonal in either $\op{q}$ or $\op{p}$. Then, we make use of Eqs.~\eqref{Moyaldiagonal}, which show that Moyal products with such diagonal operators describe phase-space convolutions.  
Each half of GKP error correction
gives a map in phase space:
    \begin{subequations} \label{rakes}
    \begin{align}
        W_{\bar{\psi}_+(\op{q}) \op{\rho} \bar{\psi}^*_+(\op{q})}  (q,p) & =
        \int dw \; W_+ (q,w) W_{\op{\rho}} (q,p-w)  \, , \label{qrake} 
        \\
        W_{\bar{\psi}_+(\op{p}) \op{\rho} \bar{\psi}^*_+(\op{p })} (q,p) & =
        \int dw \;  W_0 (w,p) W_{\op{\rho}} (q+w,p) \, , \label{prake}
    \end{align}
    \end{subequations}
where $W_+(q,p)$ is the Wigner function for GKP state~$\ket{\bar{+}_L}$, and $W_0(q,p)$ is the Wigner function for GKP state~$\ket{\bar{0}_L}$. 
This Wigner function is described either by a two-dimensional $\Sha$-function for an ideal GKP ancilla, Eq.~\eqref{wigneridealGKP}, or by a two-dimensional $\theta$-function with a Gaussian envelope for an approximate GKP ancilla, Eq.~\eqref{wignerapproxGKPlownoise}.
The phase-space map for the error correction circuit in Eq.~\eqref{ECcircuit} with $\{m_q, m_p\} = \{0, 0\}$ is found by applying Eq.~\eqref{qrake} followed by Eq.\eqref{prake}, 
\begin{widetext}
    \begin{equation} \label{FullGKPEC_Wigner}
        W_{ \bar{\psi}_+(\op{p}) \bar{\psi}_+(\op{q}) \op{\rho}_\text{in} \bar{\psi}_+(\op{q}) \bar{\psi}_+(\op{p}) }  (q,p)
        \propto  \iint dw'dw'' \,  W_{0} (w',p) W_{+} (q+w',w'') W_{\op{\rho}_\text{in}} (q+w',p-w'')  \, .
    \end{equation}
\end{widetext}
Noting that the Wigner functions for the symmetric $\ket{0_L}$ and $\ket{+_L}$ states considered here are related by swapping their arguments, $W_0(q,p) = W_+(p,q)$,
we can interpret the two convolutions together as a map that performs the same operation in position as in momentum, which periodically filters support in a one quadrature while replicating periodic features of the ancilla in the conjugate quadrature. We discuss this process more thoroughly in the following section. 

The phase-space error-correction formula in Eq.~\eqref{FullGKPEC_Wigner} can be used for other measurement outcomes by simply using $\op{\rho}_\text{in} \rightarrow \op{X}(m_q) \op{Z}(m_p) \op{\rho}_\text{in} \op{Z}^\dagger (m_p) \op{X}^\dagger (m_q)$ and then applying the corrections (displacements) specified in \erf{KEC_rewrite} to get the final output state. 
Note that if one were to perform the two halves of error correction in the other order---$p$ correction before $q$ correction---then the phase-space map is different. A more general form of Eq.~\eqref{FullGKPEC_Wigner} that allows for arbitrary pure (and different) ancillae is given in Eq.~\eqref{WignerMapqp}.
Finally, we note that for mixed states (ancillae and/or the input state), the linearity of Wigner functions enables allows us to consider syndrome extraction as a weighted sum of error-corrected pure states.

\subsubsection{Phase-space effects of GKP error correction} \label{subsubsection:phasespaceffectsEC}

The effects of GKP error correction can be understood by inspecting each half of GKP error correction separately. Consider the first half, described by Eq.~\eqref{qrake}, that corrects CV errors in position.
The input state $W_{\op{\rho}_\text{in}}(q,p)$ experiences a convolution in momentum with $W_+(q,p)$, weighted by the point-wise product in position. 
We refer to this phase-space operation as ``raking'' in a single quadrature, since the periodic spikes of the GKP ancilla act to clear away any parts of $W_{\op{\rho}_\text{in}}(q,p)$ that do not align with the $\sqrt{\pi}$-periodic GKP grid in position. 
(The other half of GKP error correction performs raking in momentum). 
Second, the momentum convolution periodically replicates features of $W_{\op{\rho}_\text{in}}(q,p)$ along the $p$-direction in phase space (up to the envelope associated with the ancilla state). This feature arises from 
the convolution of a function $f(x)$ with a univariate $\Sha$-function, 
\begin{align}
	\big[f  * \Sha_{T} 
	[\begin{smallmatrix}
	v_{1}\\
	v_{2}
	\end{smallmatrix}]
	\big] (x) = \sqrt{\abss{T}} \sum_{n \in \mathbb{Z}} e^{-2 \pi i  v_{1} n} f \big(x + nT + v_{2}T \big)
	\, ,
\end{align}
that periodically replicates the function $f$ over the comb while respecting the characteristics of the $\Sha$-function that give half-period displacements and periodic negativity.

\begin{figure} [t]
	\includegraphics[width=0.9\linewidth]{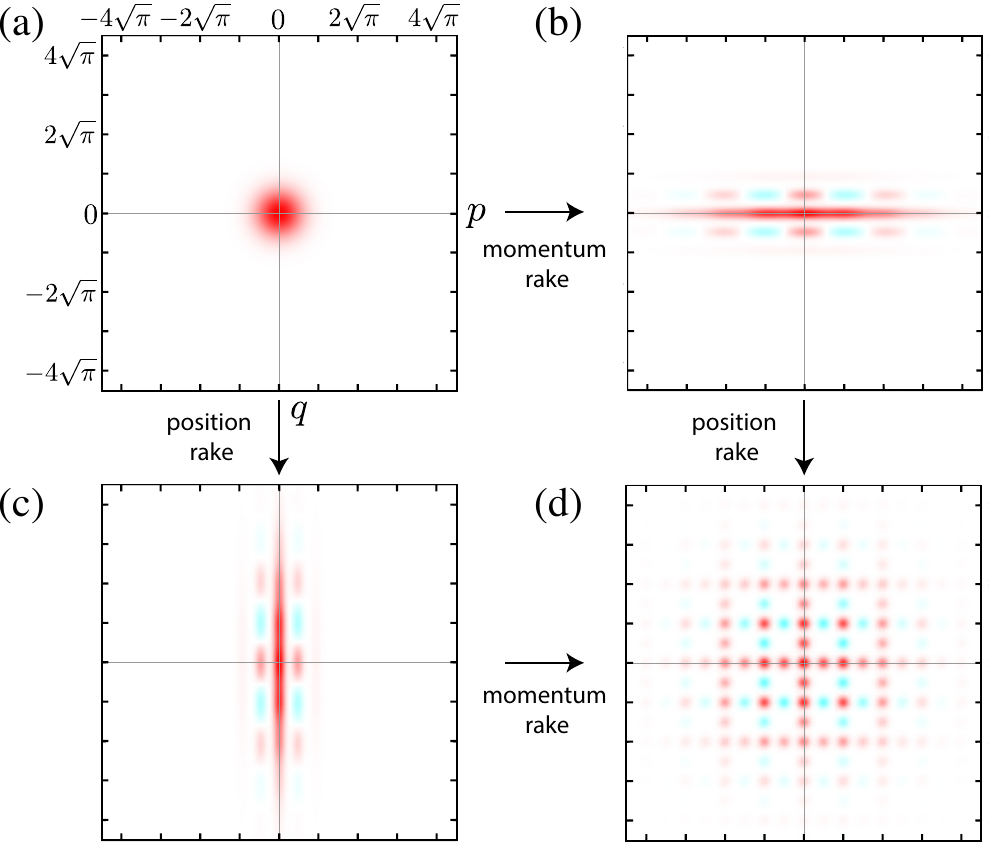}
	\caption{\label{Fig:rake} Applying GKP error correction, step-by-step, to an input vacuum state for syndrome $\{m_q, m_p\} = \{0,0\}$. (a) Vacuum state. (b) Momentum-raked vacuum. (c) Position-raked vacuum. (d) Vacuum raked in both quadratures gives a GKP $H$-type magic state~\cite{allGaussianPRL}. As discussed in Sec.~\ref{section:gkpec}, when high-quality GKP ancilla states are used for error correction, the two raking operations approximately commute, and we plot~(d) under this assumption.
	}
\end{figure}

For high-quality approximate GKP states, raking from Eq.~\eqref{qrake}
multiplies the Gaussian spikes of the input state by the Gaussian spikes of the ancilla. A product of Gaussians yields another Gaussian, 
    \begin{align}
    	&G_{\mat{\Sigma}_{1}}(\vec{x}) G_{\mat{\Sigma}_{2}}(\vec{x}) \propto G_{\mat{\Sigma}_{12}}(\vec{x}),
    \end{align}
with the key feature that the \emph{precision} (inverse covariance matrix) is additive:
    \begin{align}
	    \mat{\Sigma}_{12}^{-1} = \mat{\Sigma}_{1}^{-1} + \mat{\Sigma}_{2}^{-1} .
    \end{align}
This means that the precision of the output Wigner function's spikes is bounded from below by the precision of the ancilla spikes, regardless of whether the spikes have the same mean. Since the precision of a Gaussian can only increase under multiplication with another Gaussian, the quality of the ancilla states ($\mat{\Sigma}_\text{spike}$) sets the lower bound for the quality of the output state after error correction.

In summary, each half of GKP error correction rakes an input Wigner function, which periodically damps support in each quadrature and replicates in the other, with the noise in the GKP ancillae determining the strength of the effect. 
When performed consecutively, the two halves produce a projection into an approximate GKP subspace defined by the ancillae~\cite{walshe2020continuous}.
This projection can be applied to any state (not just noisy GKP states), as illustrated in Fig.~\ref{Fig:rake}, where an input vacuum state is consecutively raked in both quadratures for syndrome $\{0, 0\}$. 
Additionally, this projection applies even for $\{m_q, m_p\} \neq \{0,0\}$, since the error-correction procedure (including the corrective shifts) realigns the spikes with the phase-space grid that defines square-lattice GKP encodings. Although the quality of a damaged GKP state is restored to that of the GKP ancillae and the state is realigned with the proper grid, in general there may still be qubit-level errors incurred during this process. In a fault-tolerant setting, these errors are dealt with by concatenation~\cite{terhalScalableBosonicQuantum2020, nohFaulttolerantBosonicQuantum2020, hangglienchancednoise2020, bourassaBlueprintScalablePhotonic2020}.
An equivalent perspective on using GKP error correction for error mitigation and for GKP magic-state preparation was given by Fabre \emph{et al.}~\cite{fabre2020} using a modular decomposition of the phase plane.
\blk

\subsubsection{Error correction with ideal GKP ancillae} \label{Sec:ECwithidealstates}

For ideal GKP ancillae in the GKP error-correction circuit, Eq.~\eqref{ECcircuit}, analysis of an error-corrected state is simpler due to the properties of $\Sha$-functions under integration. 
In this setting, error syndromes arise only from the input state.
Just as above, we focus on $\{m_q, m_p\} = \{0,0\}$, noting that outcome-dependent shifts can be applied first to the input state and then after the error-correction projection as described by Eq.~\eqref{KEC_rewrite}. In this case, the Kraus operator for ideal GKP ancillae is
    \begin{align} 
        \op{K}_\text{EC}(0, 0) = \psi_+(\op{p}) \psi_+(\op{q}) = \psi_+(\op{q}) \psi_+(\op{p})
        = \op{\Pi}_L \label{idealEC}
        \, ,
    \end{align}
where $\op{\Pi}_L$ is the projector onto the square-lattice GKP subspace~\cite{allGaussianPRL, walshe2020continuous},
    \begin{align}
        \op{\Pi}_L \coloneqq \outprod{0_L}{0_L} + \outprod{1_L}{1_L}
        \, .
    \end{align}    
Thus the error-correction map, Eq.~\eqref{KrausGKPEC}, 
    \begin{align}
         \op{K}_\text{EC}(0, 0) \op{\rho}_\text{in} \op{K}^\dagger_\text{EC}(0, 0) \propto \op{\Pi}_L \op{\rho}_\text{in} \op{\Pi}_L
    \end{align}
simply projects the input state into the ideal GKP subspace.
In phase space, the convolutions in Eq.~\eqref{FullGKPEC_Wigner}, can be evaluated using the Wigner functions for the ideal GKP ancillae, Eq.~\eqref{GKPwavefunctions}, giving the error-corrected Wigner function in terms of two-dimensional $\Sha$-functions,
\begin{align} 
	&W_{\op{K}_\text{EC}\op{\rho}_\text{in} \op{K}^{\dagger}_\text{EC}} (\vec{x}) \nonumber \\
	& \qquad \propto\Sha_{\qql\mat{I}}(\vec{x}) \sum_{\vec{m} \in \mathbb{Z}^{2}} e^{ -2 \pi i (\frac{1}{\sqrpi} \vec{x}^\tp \mat{\Omega}\vec{m})} W_{\op{\rho}_\text{in}} (\vec{x} +\vec{m} \sqrpi).   \label{2ecout} 
\end{align}
Other measurement outcomes are found by using a modified input state that has been displaced by $\op{X}(m_q) \op{Z}(m_p)$, as is evident in Eq.~\eqref{KEC_rewrite}. 
This $\Sha$-function restricts the support of the distribution to multiples of $\qql$ in both $q$ and $p$. In the extreme limit of perfect ancillas, the only non-zero quasiprobability that can exist in the output is in some way associated with the encoding. In fact, one can consider the identity: 
    \begin{align}
        \Sha_{\qql\mat{I}} (\vec{x}) =  \frac{1}{2}  \sum_{\mu=0}^{3} \Shafxn[\sqrpi\mat{I}]{\vec{0}}{\nicefrac{\vec{\ell}_{\mu}}{2}} (\vec{x}).
    \end{align}
Simply, the only non-zero coordinates in the output distribution are exclusively within the set of encoded coordinates, with distinct subsets of non-zero coordinates for each component of the encoded Bloch 4-vector. We can expand in this basis and use the properties of each $\Sha$-function
\begin{align}
    &W_{\op{K}_\text{EC}\op{\rho}_\text{in} \op{K}^{\dagger}_\text{EC}} (\vec{x})
    \propto \frac{1}{2} \sum^{3}_{\mu=0} \Shafxn[\sqrpi\mat{I}]{\vec{0}}{\nicefrac{\vec{\ell}_{\mu}}{2}}(\vec{x}) \nonumber \\
    &\quad \times \sum_{\vec{m} \in \mathbb{Z}^{2}} e^{- 2 \pi i (\frac{1}{\sqrpi} \vec{x}^\tp \mat{\Omega}\vec{m})} W_{\op{\rho}_\text{in}} (\vec{x} +\vec{m} \sqrpi) \\
    &=\frac{1}{2} \sum^{3}_{\mu=0} \sum_{\vec{n},\vec{m} \in \mathbb{Z}^{2}} \delta \left(\vec{x} + \vec{n}\sqrpi + \vec{\ell}_{\mu}\frac{\sqrpi}{2}\right) e^{ 2 \pi i ((\frac{1}{2}\vec{\ell}_{\mu})^{\tp} \mat{\Omega}\vec{m})}\nonumber \\
    &\quad \times W_{\op{\rho}_\text{in}} \left(-\vec{n}\sqrpi - \vec{\ell}_{\mu}\frac{\sqrpi}{2} +\vec{m} \sqrpi\right),
\end{align}
which with the new index $\vec{j} \coloneqq \vec{n} - \vec{m}$ has the form
\begin{align}
    &=\frac{1}{2} \sum^{3}_{\mu=0} \sum_{\vec{j},\vec{n} \in \mathbb{Z}^{2}} \delta \bigg(\vec{x} + \vec{n}\sqrpi + \vec{\ell}_{\mu}\frac{\sqrpi}{2}\bigg) e^{- 2 \pi i (\frac{1}{2}(\mat{\Omega}\vec{\ell}_{\mu})^{\tp} \vec{n})} \nonumber \\
    &\quad \times e^{ 2 \pi i (\frac{1}{2}(\mat{\Omega}\vec{\ell}_{\mu})^{\tp} \vec{j})} W_{\op{\rho}_\text{in}} \bigg(-\vec{j}\sqrpi - \vec{\ell}_{\mu}\frac{\sqrpi}{2}\bigg)\\
    &=\frac{1}{2} \sum^{3}_{\mu=0} W^{L}_{\mu} (\vec{x})
    \underbrace{\sum_{\vec{j} \in \mathbb{Z}^{2}} e^{ 2 \pi i (\frac{1}{2}(\mat{\Omega}\vec{\ell}_{\mu})^{\tp} \vec{j})} W_{\op{\rho}_\text{in}} \bigg(-\vec{j}\sqrpi - \vec{\ell}_{\mu}\frac{\sqrpi}{2}\bigg)
    }_{r_\mu}
\end{align}
Notice that this is just an expansion in GKP Pauli Wigner functions, $W^{L}_{\mu}$, so the remaining sum must represent the Bloch-vector coefficients~$r_{\mu}$ of the output state, as indicated. To make the connection to what we know already about Bloch-vector components, we rewrite the sum as an integral and obtain
\begin{align}
  r_\mu &=
  \int d^2 \vec{x}\, \Shafxn[\sqrpi\mat{I}]{\nicefrac{\mat{\Omega}\vec{\ell}_{\mu}}{2}}{\nicefrac{\vec{\ell}_{\mu}}{2}}(\vec{x})
  W_{\op{\rho}_\text{in}} (\vec{x})
\\
    &=
    \int d^2 x\, W^{L}_{\mu} (\vec{x}) W_{\op{\rho}_\text{in}} (\vec{x})
\\
    &= \Tr [\op{\sigma}^L_\mu \op{\rho}_\text{in}].
\end{align}
One might wonder where the projection onto the GKP subspace, $\op \Pi_L$, has gone. In fact, it gets absorbed into the GKP Pauli operators when tracing the error-corrected state~$\op{\Pi}_L \op{\rho}_\text{in} \op{\Pi}_L$ with the GKP Pauli operators~\cite{allGaussianPRL}:
    \begin{align}
         r_\mu = \Tr [\op{\sigma}^L_\mu \op{\Pi}_L \op{\rho}_\text{in} \op{\Pi}_L] = \Tr [\op{\sigma}^L_\mu \op{\rho}_\text{in}],
    \end{align}
since $\op{\Pi}_L \op \sigma^L_\mu \op{\Pi}_L = \op \sigma^L_\mu$, which justifies the final form of Eq.~\eqref{2ecout}:
\begin{align}
\label{eq:2ecoutBloch}
    	W_{\op{K}_\text{EC}\op{\rho}_\text{in} \op{K}^{\dagger}_\text{EC}} (\vec{x}) %
    	 &=  \frac{1}{2} \sum_{\mu=0}^{3} W^L_{\mu}(\vec{x})
    	 \int d^2 \vec{x}'\, W^{L}_{\mu} (\vec{x}') W_{\op{\rho}_\text{in}} (\vec{x}')
    	 ,
    \end{align}
up to normalization.

\section{Teleportation-based GKP error correction}
\label{section:magicstate}

A key result in Ref.~\cite{allGaussianPRL} is that applying GKP error correction to unencoded states---particularly those that are simple to prepare, like the vacuum state---probabilistically produces GKP magic states, which are useful for universal, fault-tolerant quantum computing. One can use the complementary phase-space description of the GKP error correction circuit in Eq.~\eqref{ECcircuit}, presented in the previous section, to find the same result.

An alternate method of implementing GKP error correction involves teleporting the input state through a GKP Bell pair~\cite{walshe2020continuous}, which can also be interpreted as performing heterodyne detection on one half of the GKP Bell pair~\cite{allGaussianPRL}. We review this protocol in phase space.
The teleportation circuit for GKP error correction is,
\begin{align} 
\begin{split} \label{ECcircuitteleport}
	\Qcircuit @C=2em @R=2.2em {
	\lstick{\brasub{m_q}{q}} &\ustick{\qquad W_{c}}\qw[-1] &\bsbal{1} &\ustick{\qquad W_{b}}\qw[-1]&\qw[-1]&\qw[-1]&\ustick{\mkern-34mu W_{a}}\qw[-1] & \rstick{\mkern-34mu \op{\rho}_\text{in}} \\
	\lstick{\brasub{m_p}{p}} &\qw[-1]                      &\qw       &\qw[-1]&\ctrl{1}&\qw[-1]&\qw & \rstick{\mkern-34mu\ket{+_{L}}} \\
	\lstick{{ (\mathrm{out})}}         &\qw[-1]                      &\qw[-1]   &\qw[-1]&\targ&\qw[-1]&\qw& \rstick{\mkern-34mu\ket{0_{L}}}   \POS"1,6"."3,6"!C*+<0em,2em>\frm{--} \POS"1,4"."3,4"!C*+<0em,2em>\frm{--} \POS"1,2"."3,2"!C*+<0em,2em>\frm{--}
}
\end{split}
\end{align}
where the circuit is read right-to-left, and the homodyne measurements in bases $q$ and $p$ with outcomes $m_q$ and $m_p$ are depicted in using bras. The arrow between modes 1 and 2 indicates the 50:50 beamsplitter~\cite{walshe2020continuous},
\begin{equation}\label{BScircuit}
\begin{split}
    \Qcircuit @C=1.25em @R=2.6em {
	                                                                                                                           &&&&&& \bsbal{1} & \rstick{j} \qw  \\
	  \raisebox{2.9em}{$\op{B}_{jk} \coloneqq e^{-i \frac{\pi}{4}(\op{q}_j \otimes \op{p}_k - \op{p}_j \otimes \op{q}_k )} = $}&&&&&& \qw       & \rstick{k} \qw \\
		}
\end{split}		
\end{equation}
and the GKP Bell pair on modes 2 and 3 is generated by coupling two $\ket{+_L}$ GKP states with a $\op{C}^{(1,2)}_X$ gate, Eq.~\eqref{CXgate}. The vertical dashed lines indicate places along the circuit where we calculate the joint Wigner function below.

This circuit can be analyzed easily in phase space by virtue of the fact that the measurements and unitaries are all Gaussian. Gaussian unitary action is given by a transformation of the Wigner function's arguments, Eq.~\eqref{GaussianWignertransformation}, using a symplectic matrix $\mat{S}_{\op{U}}$ associated with the unitary. The symplectic matrices for the beam splitter in Eq.~\eqref{BScircuit} and the $\op{C}^{(1,2)}_X$ gate are
    \begin{align}
        \mat{S}_{\op{B}_{jk}} &= \frac{1}{\sqrt{2}}
    \begin{bmatrix}
        1 & -1 & 0 & 0 \\
        1 & 1 & 0 & 0 \\
        0 & 0 & 1 & -1 \\
        0 & 0 & 1 & 1
    \end{bmatrix}
    ,
    \\
    \mat{S}_{\op{C}^{(1,2)}_X} &= 
    \begin{bmatrix}
        1 & 0 & 0 & 0 \\
        1 & 1 & 0 & 0 \\
        0 & 0 & 1 & -1 \\
        0 & 0 & 0 & 1
    \end{bmatrix}
    .
    \end{align}
We will proceed step-by-step through the teleportation circuit beginning from the three-mode input state
    \begin{align}
	    W_{a} (\vec{x}) \coloneqq W_{\op{\rho}_\text{in}} (\vec{x}_{1}) W^{L}_{+} (\vec{x}_{2}) W^{L}_{0}(\vec{x}_{3}),
    \end{align}
indicated at the top of the circuit, Eq.~\eqref{ECcircuitteleport}. The arguments of the Wigner functions, $\vec{x} = [q_1, q_2, q_3, p_1, p_2, p_3]^\tp$, are the phase space coordinates of the joint system and $\vec{x}_{j} = [q_j, p_j]^\tp$ for each mode $j$, respectively. Note that since unitaries above are two-mode; evolution along the circuit at each step requires padding the symplectic matrix with a $2 \times 2$ identity matrix $\mat{I}_{j}$ for the unaffected mode. First, applying the $\op{C}^{(2,3)}_X$ gate creates the GKP Bell pair, manifesting as a transformation on the Wigner-function coordinates of the second and third mode,
    \begin{align}
    	W_{b} (\vec{x}) &= W_{a} \big((\mat{I}_{1} \oplus \mat{S}^{-1}_{\op{C}^{(2,3)}_X})\vec{x}\big) \\
    	 &= W_{\op{\rho}_\text{in}} (\vec{x}_{1}) W^{L}_{+} (\vec{x}_{2} + \vec p_3 ) W^{L}_{0}(\vec{x}_{3} - \vec q_2)  ,
    \end{align}
where, in a slight abuse of notation, $\vec p_3 = [0,p_3]^\tp$, and $\vec q_2 = [q_2, 0]^\tp$.\footnote{A GKP Bell pair can also be created by sending two GKP qunaught states, also called sensor states~\cite{duivenvoorden2017single}, through a beamsplitter~\cite{walshe2020continuous}.} 
Next, the beamsplitter $\op{B}_{12}$ is applied, 
    \begin{align}
    	W_{c}(\vec{x}) &= W_{b} \big((\mat{S}^{-1}_{\op{B}_{12}}\oplus\mat{I}_{3})\vec{x}\big)\\
    	 &=W_{\op{\rho}_\text{in}} (\vec{x}_{+}) W^{L}_{+} (\vec{x}_{-} + \vec{p}_{3}) W^{L}_{0} (\vec{x}_{3} - \vec{q}_{-}) 
    	 \, , 
    \end{align}
which transforms the phase-space coordinates on the first and second modes, 
    \begin{align} \label{coordinate}
	&\vec{x}_{\pm} \coloneqq \frac{1}{\sqrt{2}}
	    \begin{bmatrix}
	        q_2 \pm q_1 \\
	        p_2 \pm p_1
	    \end{bmatrix} \, .
    \end{align}

Finally, modes 1 and 2 are measured via homodyne detection, giving outcomes $m_q$ and $m_p$. Homodyne measurement is also a Gaussian operation, and its effect in phase space is simple. For destructive homodyne measurement, set the argument of the measured quadrature equal to the measurement outcome, and then integrate over the conjugate argument.\footnote{Alternatively, the Wigner-function maps in Eqs.~\eqref{rakes} can be used to describe non-destructive position and momentum homodyne measurements using as $\psi$ the wavefunctions for the appropriate quadrature eigenstates describing the projective measurements.} Doing this on modes 1 and 2 gives the post-measurement state on mode 3,
    \begin{align}
        &W_{{ (\mathrm{out})}}(\vec{x}_{3}) \nonumber \\
        &= \int dp_{1} dq_{2} \; W_{\op{\rho}_\text{in}} (\vec{x}'_{+}) W^{L}_{+} (\vec{x}'_{-} + \vec{p}_{3}) W^{L}_{0} (\vec{x}_{3} - \vec{q}_{-}') 
        \, ,
    \end{align}
where the measurement outcomes have been substituted into Eq.~\eqref{coordinate}, 
    \begin{align}
    	&\vec{x}'_{\pm} = \frac{1}{\sqrt{2}}
    	\begin{bmatrix}
    	    q_2 \pm m_q \\
    	    m_p \pm p_1
    	\end{bmatrix} \,,
    \end{align}
and analogously, $\vec q_-' = \tfrac {1} {\sqrt 2}[q_2 - m_q, 0]^\tp$. Performing the integration gives
\begin{align}
&W_{{ (\mathrm{out})}}(\vec{x}) \nonumber\\
&\quad\propto \Sha_{\qql\mat{I}}(\vec{x}) \sum_{\vec{j} \in \mathbb{Z}^{2}} e^{2\pi i (\frac{1}{\sqrpi}\vec{x}^\tp \mat{\Omega} \vec{j})} W_{\op{\rho}_\text{in}} (\vec{x} + \sqrt 2 \vec m + \vec{j}\sqrpi),
\end{align}
where $\vec m = [m_q, m_p]^\tp$.
This expression is of the same form as Eq.~\eqref{2ecout}. Thus, it can also be rewritten---using the same techniques used to derive Eq.~\eqref{eq:2ecoutBloch}---to give the output state (up to normalization),
    \begin{align}
    \label{eq:vacECBloch}
    &	W_{{ (\mathrm{out})}} (\vec{x}) 
        = \frac{1}{2} \sum_{\mu=0}^{3} r_\mu W^L_{\mu}(\vec{x})
    	 \, ,
    \end{align}
with GKP Bloch vector components
    \begin{equation} \label{GKPBVcomps}
        r_\mu = \int d^2 \vec{x}'\, W^{L}_{\mu} (\vec{x}') W_{\op{\rho}_\text{in}} (\vec{x}' + \sqrt 2 \vec m)
        \, .
    \end{equation}
This agrees with the result reported in Eq.~(5) of Ref.~\cite{allGaussianPRL} for a displacement of $\sqrt 2 \vec m$. \blk
The $\sqrt 2$ comes about due to the definition of $\op V$ [Eq.~\eqref{shiftopV}] versus the Glauber displacement operator $\op D$, which appears when deriving this same result using circuit identities~\cite{walshe2020continuous}. 
The GKP Bloch vector components, Eq.~\eqref{GKPBVcomps}, depend on both the input state and the measurement outcomes, with the result that the Bloch vector does not, in general, correspond to a GKP Pauli eigenstate.
Consider the case where $\op{\rho}_\text{in}$ is the vacuum state. In this setting, the circuit in Eq.~\eqref{ECcircuitteleport} has two equivalent interpretations. In one interpretation, it describes teleportation of the vacuum state through a GKP Bell pair, and, in the other, it describes heterodyne detection on one half of a GKP Bell pair.
For both, the error-corrected state for measurement outcomes $\vec{m} = \{0,0\}$ is an approximate Hadamard eigenstate~\cite{allGaussianPRL}, which is a GKP magic state---\emph{i.e.}, a state that can be used to teleport an encoded non-Clifford gate to achieve universal quantum computation. 
The creation of the magic state is visualized in phase space by raking the vacuum state in both quadratures, as illustrated in Fig.~\ref{Fig:rake}.
While other outcomes~$\vec m$ produce other states, almost all of these are outside the Paulihedron (convex hull of Pauli eigenstates), which means they can be distilled into higher-quality magic states~\cite{campbell2010,Bravyi:2005dx}. The only exceptions, which are rare~\cite{allGaussianPRL}, are those outcomes that give states that are too close to Pauli eigenstates. Magic-state production through GKP error correction is not limited to input vacuum states, either. Even some mixed input states can be used, as was shown for thermal states with low enough thermal occupancy~\cite{allGaussianPRL}.

\blk

\section{Discussion}
\label{sec:discussion}

We have laid out a broadly applicable description of ideal and approximate GKP states in phase space using 
$\theta$-functions and their limiting case, $\Sha$-functions.
GKP states in phase space decompose into a sum of two-dimensional $\Sha$-functions,
each representing an encoded GKP Pauli operator. 
We have shown that the set of GKP-Pauli Wigner functions are related to each other via simple half-period transformations on their characteristics, which can be generated by Gaussian unitary operations. 

The fact that Gaussian operations correspond to simple, linear transformations on a Wigner function's arguments allows straightforward modeling of GKP Clifford quantum computation, as well as lattice transformations between different types of GKP encodings---\emph{e.g.},~square-lattice to hexagonal-lattice and back again. 
Furthermore the remaining operations required for GKP error correction can also be implemented using Gaussian operations, singling out the GKP code as the only known bosonic code to admit a fault-tolerant, all-Gaussian, universal gate set%
~\cite{allGaussianPRL, yamasakicostreducedaps2020}.

High-quality approximate GKP states are represented in phase space as enveloped 2-dimensional $\theta$-functions. We derived conditions that relate the covariance matrix for the spikes to that of the envelope. With this, one can immediately write down a Wigner function for a physical approximate GKP state simply by specifying the state's Bloch vector $\vec{r}$---which could represent a pure or mixed logical state---and the spike covariance matrix, thus circumventing the need to first write down an approximate GKP wave function and then calculate its Wigner function~\cite{GKP,matsuura2019equivalence,garcia2020wigner}.

A useful feature of $\theta$-functions and $\Sha$-functions is how they relate simply via Gaussian blurring and deblurring.  Gaussian blurring in phase space correponds to a purely positive-imaginary offset to the $\mat{\tau}$ parameter while preserving the characteristics of the function (which encode the discrete quantum information). In the context of quasiprobability distributions, this operation also maps between the Wigner function and other distributions, such as the Husimi $Q$-function and the Glauber-Sudarshan $P$-function. In this way, the GKP Wigner functions we have presented may be extended to the whole spectrum of $s$-ordered quasiprobability functions~\cite{cahillglauberorder1969,cahillglauberdensity1969} through transformations of $\mat{\tau}$\footnote{For well-prepared GKP states, however, some orderings ($P$-function, for instance) will result in values of~$\mat \tau$ outside the Siegel upper-half space, making the resulting functions ill defined.}. Moreover, as pure loss can be transformed into an effective Gaussian noise channel through phase-insensitive amplification~\cite{albertPerformanceStructureSinglemode2018}, we can also consider this process in terms of a valid
$\mat{\tau}$ parameter proportional to the effective additive Gaussian noise. Lastly, although we have presented, throughout this work, results for a square-lattice GKP code, we re-emphasize that any of these properties can equally be expressed in the context of other GKP lattices, including the hexagonal GKP code~\cite{GKP,harrington2001achievable,albertPerformanceStructureSinglemode2018}. 

In general, there are several technicalities in describing general GKP phase-space distributions, as the effects of finite squeezing at higher noise levels and how the higher dimensional encoded spaces are represented in phase space is more intricate \cite{matsuura2019equivalence}. By working within a high-squeezing approximation and restricting to qubits, we may simplify the phase space features of generically encoded qubits into a basis of $\theta$-functions in proportion to the encoded Bloch vector. In this way, we provide a compact representation of GKP-encoded information, allowing for both pure and mixed states to be handled with the same mathematical tools. These tools, furthermore, generate
an intuitive, visual representation of
GKP error correction in phase space as ``raking'' and exemplified by a demonstration of magic state generation from the vacuum.  \blk

Although we have presented results for a GKP ancilla in the application of GKP error correction, there are many alternative choices for ``raking'' functions. From a resource-based perspective, the Wigner negativity~\cite{veitch2012negative,garcia2020wigner} of the ancillary mode can be effectively transferred into an output mode with this procedure~\cite{yamasakicostreducedaps2020,Garcia-Alvarez2020} such that, in general, this may have other applications in quantum information theory. There also remains the open question for future research in using generalized forms of these operations for other forms of bosonic state preparation.

\begin{acknowledgments}
We thank  Giacomo Pantaleoni for helpful discussions about $\theta$-functions. This work is supported by the Australian Research Council Centre of
Excellence for Quantum Computation and Communication Technology (Project No. CE170100012).
\end{acknowledgments}

\appendix

\section{Relation between Gaussian pulse trains and $\theta$-functions}

We expand a Gaussian pulse train with period $T \in \mathbb{R}^{+}_{\neq0}$ as\footnote{We consider only positive periods here for the sake of brevity. Negative-period pulse trains are simply reflections of pulse trains of positive period; however in $\theta$-function form, the direction of the second characteristic is also reflected. Although some pulse trains may be even functions in some cases, such as for  $[1/2,1/2]^\tp$ characteristics, the pulse train is odd (see Fig.~\ref{shahthetatable}).}
\begin{align}
&\sum_{n \in \mathbb{Z}} e^{-2\pi i n v_{1}} G_{\sigma^{2}} (x + nT + v_{2}T) \nonumber \\
&= \frac{1}{T} \frac{1}{\sqrt{-i\tau_{f}}} e^{-\frac{i\pi}{\tau_{f}} (z_{s} + v_{2})^{2}} \sum_{n \in \mathbb{Z}} e^{2\pi i (\frac{1}{2} n^{2} \frac{-1}{\tau_{f}} - \frac{n}{\tau_{f}}(z_{s}+v_{2}))}e^{-2\pi i n v_{1}}, \label{pulsetrainderivation1}
\end{align}
where we have defined new parameters,
\begin{align} \label{parameterswap}
z_{s} \coloneqq \frac{x}{T}\, \quad \quad \tau_{f} \coloneqq \frac{2\pi i \sigma^{2}}{T^{2}}
\, .
\end{align}
Absorbing the translation $v_{2}$ into a new argument $z_{s}' \coloneqq z_{s} + v_{2}$ and reindexing the sum as $m=-n$, Eq.~\eqref{pulsetrainderivation1} can then be written as a $\theta$-function,
\begin{align}
&\frac{1}{T} \frac{1}{\sqrt{-i\tau_{f}}} e^{ - \frac{i \pi {z}_{s}'^{2}}{\tau_{f} }} \sum_{m \in \mathbb{Z}} e^{2\pi i (\frac{1}{2} m^{2} \frac{-1}{\tau_{f}} + m(\frac{z'_{s}}{\tau_{f}}+v_{1}))} \nonumber \\
&=\frac{1}{T} \frac{1}{\sqrt{-i\tau_{f}}} e^{ - \frac{i \pi {z}_{s}'^{2}}{\tau_{f} }} \thetafxn{0}{v_{1}} \bigg(\frac{z'_{s}}{\tau_{f}} , \frac{-1}{\tau_{f}}\bigg).
\end{align}
Using an inverse Jacobi identity \cite{igusaThetaFunctions1972b,mumfordIntroductionMotivationTheta1983,mumfordBasicResultsTheta1983}, 
\begin{align}
\theta \begin{bmatrix}
0 \\ 
v_{2}
\end{bmatrix} \bigg(\frac{z}{\tau}, \frac{i^{2}}{\tau} \bigg) = \sqrt{-i\tau} e^{\frac{i \pi z^{2}}{\tau}} \theta \begin{bmatrix}
v_{2} \\ 
0
\end{bmatrix} (z , \tau) ,
\end{align}
which describes the how $\theta$-functions are modified under the transformation $\tau \to \frac{i^{2}}{\tau}$,
gives
\begin{align}
\frac{1}{T}\frac{1}{\sqrt{-i \tau_{f}}} e^{ - \frac{i \pi {z}_{s}'^{2}}{\tau_{f}}} \thetafxn{0}{v_{1}} \bigg( \frac{z'_{s}}{\tau_{f}}, \frac{1}{\tau_{f}}\bigg) = \frac{1}{T} \thetafxn{v_{1}}{0} (z'_{s} , \tau_{f}) . 
\end{align}
Restoring $z_{s}$ from $z'_{s}$, we have the result:
\begin{align}
\sum_{n \in \mathbb{Z}} e^{-2\pi i n v_{1}} G_{\sigma^{2}} (x + nT + v_{2}T) &= \frac{1}{T} \thetafxn{v_{1}}{v_{2}} (z_{s} , \tau_{f})
\, .
\end{align}
Using Eq.~\eqref{1dthetaT}, this $\theta$-function can be written as the $\theta_T$-function in Eq.~\eqref{Gaussianpulsetrain}.

\section{GKP wave-function normalization for low-noise states} \label{appendix:approxnormalization}

In this section we explicitly derive the $L^{2}$ normalization for a damped Gaussian pulse train in terms of $\theta$-function of arbitrary characteristics in the limit of low noise. For the function
\begin{align} \label{somefunction}
    f(x) \coloneqq \frac{1}{\sqrt{\mathcal{N}}} G_{\kappa^{-2}} (x) \sum_{n \in \mathbb{Z}} e^{-2\pi i n v_{1}} G_{\Delta^{2}} (x + (n + v_{2})T),
\end{align}
the square-normalization factor $\mathcal{N} = \int dx \,  |f(x)|^2$ is 
\begin{align}
    \mathcal{N} &=  \frac{1}{2\sqrt{\pi\kappa^{-2}}} \int dx\; G_{\frac{\kappa^{-2}}{2}} (x) \sum_{n,m \in \mathbb{Z}} e^{2\pi i (m-n) v_{1}} \nonumber \\ 
    &\times G_{\Delta^{2}} (x + (n + v_{2})T) G_{\Delta^{2}} (x + (m + v_{2})T).
\end{align}
where we used
\begin{align}
	\big[G_{\kappa^{-2}} (x)\big]^{2} = \frac{1}{2\sqrt{\pi\kappa^{-2}}} G_{\frac{\kappa^{-2}}{2}} (x)
	\, . \label{sqrgauss}
\end{align}
For small $\Delta^2$ (compared to the period $T$), any product of Gaussians arising from the pulse trains is approximately zero unless the Gaussians share the same mean, which can be expressed as\footnote{The order of this approximation can be found by asymptotically expanding the product of neighboring Gaussians in the limit of small $\Delta$. This product is proportional to $\Delta^{-1} e^{-\frac{\pi}{\Delta^{2}}}$, which gives an expansion of $\mathcal{O}(\Delta^{2n-1})$ in the limit of $\Delta \to 0^{+}$, for any positive integer~$n\in\mathbb{Z}^{+}$.}
\begin{align}
	&G_{\Delta^{2}} (x + (n + v_{2})T)  G_{\Delta^{2}} (x + (m + v_{2})T) \nonumber \\
	&\quad \quad \propto \delta_{n,m} \big[G_{\Delta^{2}} (x + (n + v_{2})T)\big]^{2},
\end{align}
such that
\begin{align}
    \mathcal{N} 
    &\approx  \frac{1}{2\sqrt{\pi\kappa^{-2}}} \int dx\;  G_{\frac{\kappa^{-2}}{2}} (x) \sum_{n \in \mathbb{Z}} \big[G_{\Delta^{2}} (x + (n + v_{2})T)\big]^{2} \\
    &= \frac{1}{4 \pi } \frac{\kappa}{\Delta}  \int dx\;  G_{\frac{\kappa^{-2}}{2}} (x) \sum_{n \in \mathbb{Z}} G_{\frac{\Delta^{2}}{2}} (x + (n + v_{2})T). 
\end{align}
In the limit of small spikes, $\Delta^2 \rightarrow 0$, the Gaussian $G_{\frac{\Delta^{2}}{2}} (x + (n + v_{2})T)$ is a nascent $\delta$-function, giving
\begin{align}
    \mathcal{N} 
    &\approx  \frac{1}{4 \pi} \frac{\kappa}{\Delta}   \sum_{n \in \mathbb{Z}} G_{\frac{\kappa^{-2}}{2}} (- (n + v_{2})T) \label{RiemannSumSetup} \\ 
    &= \frac{1}{4 \pi T} \frac{\kappa}{\Delta} \thetafxn[]{0}{-v_{2}} \bigg(0 , \frac{i\pi \kappa^{-2} }{T^{2}}\bigg)
    \, .
\end{align}
 For small $\kappa$, the $\theta$-constant is approximately 1 for all characteristics $v_{2}$, such that 
 \begin{align} \label{normfacotr_appendix}
      \mathcal{N} 
      &\approx  \frac{1}{4 \pi T} \frac{\kappa}{\Delta}.
 \end{align}

 \begin{figure}[t]
	\includegraphics[width=1\linewidth]{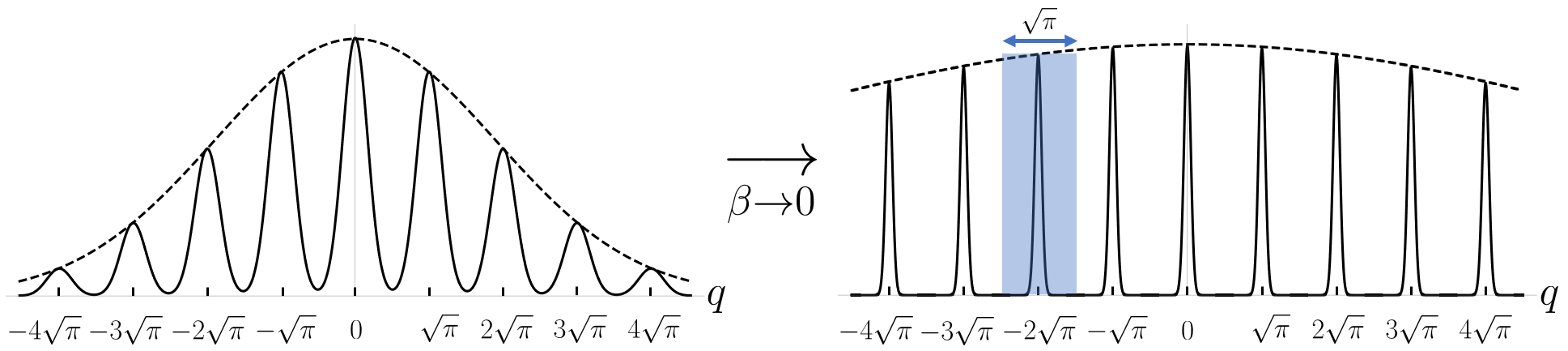}
	\caption{\label{fig:riemannsum} Illustration of the Riemann sum concept in Eq.~\eqref{Riemannfirst}. For symmetric noise, $\kappa^2 = \Delta^2 = \beta$, the envelope (dashed line) in the position wave function of a normalized approximate GKP state flattens, such that the area under the envelope can be related to a sum of rectangles, each with width $\sqrpi$ and height given by the value of the envelope at the center of every spike in the $\theta$-function.}
\end{figure}

An alternative derivation of the normalization result can be found be considering Eq.~\eqref{RiemannSumSetup} as a Riemann sum. By multiplying the periodically sampled Gaussian by the period $T$, the sum in the expression gives an approximation of the area under a normalized Gaussian,
 \begin{align} \label{Riemannfirst}
       \sum_{n \in \mathbb{Z}} T G_{\frac{\kappa^{-2}}{2}} (- (n + v_{2})T) 
     \approx  \int dx \, G_{\frac{\kappa^{-2}}{2}} (x - v_{2}T) = 1.
 \end{align}
This leads to the normalization in Eq.~\eqref{normfacotr_appendix}. An illustration of this procedure is given in Fig.~\ref{fig:riemannsum}.
 Explicitly including the normalization factor, the damped Gaussian pulse train $f(x)$ in Eq.~\eqref{somefunction} can be written as
 \begin{align}
     f(x)  &=\sqrt{4 \pi \abss{T}} \sqrt{\frac{\Delta}{\kappa}} G_{\kappa^{-2}} (x)  \frac{1}{T} \thetafxn[]{v_{1}}{v_{2}} \bigg(\frac{x}{T},\frac{2\pi i\Delta^{2}}{T^{2}}\bigg)\\
     &=\sqrt{4 \pi} \sqrt{\frac{\Delta}{\kappa}} G_{\kappa^{-2}} (x) \thetafxn[T]{v_{1}}{v_{2}} (x, 2 \pi i \Delta^{2}),
 \end{align}
where the second line uses the $\theta_T$-function relation, Eq.~\eqref{1dthetaT}.

\section{Relations between enveloped $\Sha$-functions and enveloped $\theta$-functions}
\label{appendix:multiplyingGaussians}

The product of the two normalized Gaussian distributions $G_{\sigma^{-2}}(x)$ and $G_{\sigma^{2}}(x - \mu)$ is
    \begin{align} \label{gaussianproduct}
        G_{\sigma^{-2}}(x)  G_{\sigma^{2}}(x-\mu)  = G_{\sigma^{2}+\sigma^{-2}} (\mu) G_{(\sigma^{2} + \sigma^{-2})^{-1}}(x-\mu'),
    \end{align}
where $G_{\sigma^{2}+\sigma^{-2}} (\mu) $ is the normalization factor, and the mean $\mu'$ is given by 
    \begin{align}
        \mu' = (\sigma^{2} + \sigma^{-2})^{-1} \sigma^{-2}\mu.
    \end{align}
For $\sigma^2 \ll 1$, the approximations
    $\sigma^{2} + \sigma^{-2} \approx \sigma^{-2}$, $(\sigma^{2} + \sigma^{-2})^{-1} \approx \sigma^{2}$,
and $\mu' \approx \mu$ gives for Eq.~\eqref{gaussianproduct} an approximate product,
\begin{align}
    G_{\sigma^{-2}}(x) G_{\sigma^{2}}(x-\mu)  \approx G_{\sigma^{-2}} (\mu) G_{\sigma^{2}}(x-\mu)
    \, . \label{approxspikeproduct}
\end{align}
This is an instance of a narrow Gaussian behaving as a nascent $\delta$-function.

For a Gaussian pulse train with means $\mu_n = nT$ for $n \in \mathbb{Z}$, this approximation gives
\begin{align}
    \sum_{n \in \mathbb{Z}} G_{\sigma^{-2}}(x) G_{\sigma^{2}}(x-nT)  \approx  \sum_{n \in \mathbb{Z}} G_{\sigma^{-2}} (nT) G_{\sigma^{2}}(x-nT),
\end{align}
which can be expressed in terms of convolutions as
\begin{align}
     &G_{\sigma^{-2}}(x) [\Sha_{T} * G_{\sigma^{2}}](x) \nonumber \\
     &\quad \approx  \int dy\, G_{\sigma^{-2}}(y) \Sha_{T} (y) G_{\sigma^{2}}(x-y) \\
     &\quad = [ (G_{\sigma^{-2}}\Sha_{T}) * G_{\sigma^{2}}](x)
    \, .
\end{align}
This indicates the approximate result that small spike variance and large envelope variance, $\sigma^{2} \ll \sigma^{-2}$, an enveloped $\Sha$-function under convolution is approximately an enveloped $\theta$-function.

\section{Normalization of approximate GKP Wigner functions}\label{appendix:exactnormalization}

\begin{figure} [t]
	\includegraphics[width=0.8\linewidth]{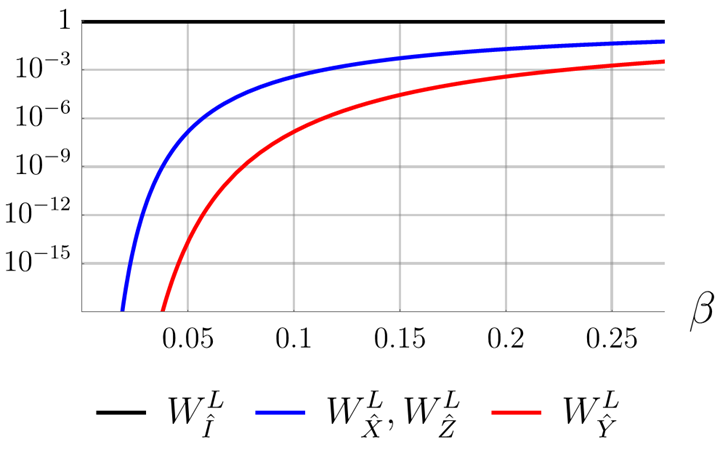}
	\caption{Trace of the Wigner functions (phase-space integral) of each noise-broadened Pauli operator (including identity) for symmetric noise, $\mat \Xi = \text{diag}(\beta, \beta)$. For small $\beta$, only the trace of the identity is nonvanishing. \label{Fig:Paulitrace}}
\end{figure}

The Wigner function of a high-quality approximate GKP state is given in Eq.~\eqref{wignerapproxGKPlownoise}.  
In this limit, $\Delta^2, \kappa^2 \ll 1$, the Gaussian for each spike, $G_{\mat{\Sigma_{\text{spike}}}}(\vec{x})$, acts as a nascent $\delta$-function, and we may approximate the total integral of the Wigner function for each noise-broadened GKP Pauli operator as
\begin{align}
    &\int d^{2} \vec{x}\;  W^{L}_{\mu} (\vec{x}) \nonumber \\
    & =\int d^{2}\vec{x} \; G_{\mat{\Sigma}_{\text{env}}} (\vec{x}) \Rthetafxn[\sqrpi\mat{I}]{\nicefrac{\mat{\Omega}\vec{\ell}_{\mu}}{2}}{\nicefrac{\vec{\ell}_{\mu}}{2}} (\vec{x}, 2\pi i \mat{\Sigma}_{\text{spike}})\\
   &=\sqrpi \int d^{2}\vec{x} \;  G_{\mat{\Sigma}_{\text{env}}} (\vec{x}) \nonumber \\
   &\quad \times \sum_{\vec{n} \in \mathbb{Z}^{2}} G_{\mat{\Sigma_{\text{spike}}}} \bigg(\vec{x} + \vec{n}\sqrpi + \vec{\ell}_{\mu}\frac{\sqrpi}{2}\bigg) e^{-i\pi \vec{n}^{\tp}\mat{\Omega} \vec{\ell}_{\mu}} \\
   & \approx  \sqrpi \sum_{\vec{n} \in \mathbb{Z}^{2}} G_{\mat{\Sigma}_{\text{env}}} \bigg(- \vec{n}\sqrpi - \vec{\ell}_{\mu}\frac{\sqrpi}{2}\bigg) e^{-i\pi \vec{n}^{\tp}\mat{\Omega} \vec{\ell}_{\mu}}.
\end{align}
Multiplying this by the area $(\sqrpi)^{2}$ gives an approximation to a Riemann sum, similar to Eq.~\eqref{Riemannfirst},
\begin{align}
    &\int d^{2}\vec{x}\; W^{L}_{\mu} (\vec{x}) \nonumber \\
    & \quad \approx  \frac{1}{\sqrpi} \sum_{\vec{n} \in \mathbb{Z}^{2}} \pi  G_{\mat{\Sigma}_{\text{env}}} \bigg(- \vec{n}\sqrpi - \vec{\ell}_{\mu}\frac{\sqrpi}{2}\bigg) e^{-i\pi \vec{n}^{\tp}\mat{\Omega} \vec{\ell}_{\mu}}\\ 
    &\quad\approx \frac{1}{\sqrpi} \int d^{2}\vec{x} \;  G_{\mat{\Sigma}_{\text{env}}} \bigg(\vec{x} - \vec{\ell}_{\mu}\frac{\sqrpi}{2}\bigg) e^{i\sqrpi \vec{x}^{\tp}\mat{\Omega} \vec{\ell}_{\mu}}\\
    &\quad =\frac{1}{\sqrpi} e^{-\frac{i\pi}{2} \vec{\ell}^{\tp}_{\mu}\mat{\Sigma}_{\text{env}}\vec{\ell}_{\mu}}. \label{Wignorm1}
\end{align}
In the limit of low noise, this factor vanishes for all Pauli Wigner functions other than the identity, and the normalization factor for any GKP state approaches $1/\sqrt{\pi}$. This is indeed as expected, since the Pauli operators are traceless. This behavior is plotted in Fig.~\ref{Fig:Paulitrace} for GKP states with symmetric noise.

Matsuura \emph{et al.}~\cite{matsuura2019equivalence} considered various types of approximate GKP states outside of the low-noise limit, where the parameter
    \begin{align}
        \gamma \coloneqq \sqrt{1 - \Delta^{2}\kappa^{2}}
    \end{align}
determines how close to ``low noise'' the state is. For $\Delta^2,\kappa^2  \ll 1$, then $\gamma \to 1$, which is the low-noise situation we consider throughout this paper. 
The normalization for their 
approximate-GKP Wigner functions, 
is given by a pair of $\theta$-constants, each arising from the total integral of a damped $\theta$-function with periodicity $2\sqrpi$ in position and $\qql$ in momentum. 
Here, we 
connect our formulation to that in Ref.~\cite{matsuura2019equivalence}. This procedure gives the normalization for low-noise approximate-GKP Wigner functions as an analytic sum of four $\theta$-constants, one arising from each GKP Pauli operator (including identity).

We begin with a 2-dimensional $\Sha$-function with period $\gamma \sqrt{\pi}$ and characteristics $\{ \vec{v}_{1},\vec{v}_{2} \}$ damped by an (unnormalized) Gaussian,
\begin{align} \label{dampedShatilde}
	\tilde{\Sha}_{\gamma\sqrpi\mat{I}, \mat{\nu}} \begin{bmatrix}
	\vec{v}_{1} \\
	\vec{v}_{2}
	\end{bmatrix} (\vec{x}) \coloneqq  e^{-\frac{1}{2} \vec{x}^{\tp} \mat{\nu}^{-1} \vec{x}} \Sha_{\gamma\sqrpi\mat{I}} \begin{bmatrix}
	\vec{v}_{1} \\
	\vec{v}_{2}
	\end{bmatrix} (\vec{x}) 
	\, ,
\end{align}
where the covariance matrix of the Gaussian is defined to be\footnote{We consider the case of $\varphi=0$, since rotations that diagonalize the covariance matrix preserve the size and shape of error ellipses.}
    \begin{align}
        \mat{\nu} 
        &\coloneqq \begin{bmatrix}
        \frac{1}{2\kappa^{2}} - \frac{\Delta^{2}}{2} & 0 \\
0 & \frac{1}{2\Delta^{2}} - \frac{\kappa^{2}}{2}
        \end{bmatrix} = \mat{\Sigma}_\text{env} - \mat{\Sigma}_\text{spike}
        \, .
    \end{align}
Integrating over the phase plane gives a $\theta$-constant
\begin{align} \label{integraldampedSha}
	&\int d^{2}\vec{x}\; \tilde{\Sha}_{\gamma\sqrpi\mat{I},\mat{\nu}} \begin{bmatrix}
	\vec{v}_{1} \\
	\vec{v}_{2}
	\end{bmatrix} (\vec{x})
= \frac{\pi}{\Delta\kappa} \Rthetafxn[\frac{\sqrpi}{\gamma}\mat{I}]{\vec{v}_{1}}{\vec{v}_{2}}(\vec{0}, \tfrac{2\pi i}{\gamma^{2}} \mat{\Sigma}_{\text{env}}) . 
\end{align} 
Applying a blurring operation (Weierstrass transform) with covariance matrix $\mat{\Sigma}_{\text{spike}}$, Eq.~\eqref{blurringops}, to the damped $\Sha$-function in Eq.~\eqref{dampedShatilde} gives 
    \begin{align} \label{convolveddampedsha}
	    &\mathcal{D}_{\mat{\Sigma}_{\text{spike}}} \tilde{\Sha}_{\gamma\sqrpi\mat{I}} 
	    \begin{bmatrix}
	    \vec{v}_{1}  \\ 
	    \vec{v}_{2}
	    \end{bmatrix} (\vec{x}) \nonumber \\
	    & \quad \quad = \frac{\pi}{\Delta \kappa} G_{\mat{\Sigma}_{\text{env}}} (\vec{x}) \Rthetafxn[\frac{\sqrpi}{\gamma}\mat{I}]{\vec{v}_{1}}{\vec{v}_{2}} (\vec{x}, \tfrac{2\pi i}{\gamma^{2}} \mat{\Sigma}_{\text{spike}}) .
    \end{align}
Note that this expression differs from the low-noise square-lattice GKP Wigner functions in Eq.~\eqref{wignerapproxGKPlownoise}. The Gaussian envelope is the same, but there are two differences. First, the period in both quadratures is $\frac{1}{\gamma^2} \sqrt{\pi} \geq \sqrt{\pi}$ meaning that the spikes are slightly off the intended grid. Second, the covariance of each spike is $ \frac{1}{\gamma^2} \mat{\Sigma}_\text{spike} \geq \mat{\Sigma}_\text{spike}$. In the low-noise limit of $\gamma \to 1$, the forms agree. However, the blurring operation does not change the total integral, meaning that the functions in Eq.~\eqref{convolveddampedsha} are normalized by the $\theta$-constant in Eq.~\eqref{integraldampedSha}.

This procedure can be performed to identify Wigner functions for each of the GKP Pauli matrices. For low noise, $\gamma \to 1$, the expression for a normalized, approximate GKP state parameterized by Bloch 4-vector $\vec{r}$ is
\begin{align}
	\tilde{W}_{\vec{r}} (\vec{x}) \approx \frac{1}{\mathcal{N}} G_{\mat{\Sigma}_{\text{env}}} (\vec{x}) \sum^{3}_{\mu=0} r_{\mu} \Rthetafxn[\sqrpi \mat{I}]{\nicefrac{\mat{\Omega}\vec{\ell}_{\mu}}{2}}{\nicefrac{\vec{\ell}_{\mu}}{2}} (\vec{x},2\pi i \mat{\Sigma}_{\text{spike}})
\end{align}
where the normalization is the sum of four $\theta$-constants,
\begin{align}
    \mathcal{N} = \sum^{3}_{\mu=0} r_{\mu} \Rthetafxn[\sqrpi \mat{I}]{\nicefrac{\mat{\Omega}\vec{\ell}_{\mu}}{2}}{\nicefrac{\vec{\ell}_{\mu}}{2}} (\vec{0}, 2 \pi i \mat{\Sigma}_{\text{env}}) \approx \frac{1}{\sqrpi},
\end{align}
which agrees with our approximation earlier, Eq.~\eqref{Wignorm1}.

\section{Moyal product for univariate Wigner functions} \label{appendix:moyalunivariate}

In this Appendix, we show that Wigner-function transformations generated by arbitrary operators that are diagonal in position or momentum are simple convolutions.
For arbitrary operator $\op{O}$ and diagonal operators 
    \begin{subequations} \label{diagonalops}
    \begin{align}
        \chi (\op{q})    &\coloneqq \int ds \; \chi (s) \outprodsubsub{s}{s}{q}{q}, \\
        \varphi (\op{p}) &\coloneqq \int dt \; \varphi (t) \outprodsubsub{t}{t}{p}{p}, \blk
    \end{align} 
    \end{subequations}
we wish to find the Wigner functions for the transformed operators:
    \begin{align}
        \chi(\op{q}) \op{O} [\chi(\op{q})]^\dagger &\rightarrow W_{\chi(\op{q}) \op{O} \chi^*(\op{q})}(q,p)\, ,  \label{diagonalqgoal}\\ 
        \varphi(\op{p}) \op{O} [\varphi(\op{p})]^\dagger &\rightarrow W_{\varphi(\op{p}) \op{O} \varphi^*(\op{p})}(q,p) \label{diagonalpgoal}
        \, ,
    \end{align}
noting that $[\chi(\op{q})]^\dagger = \chi^*(\op{q})$ and $[\varphi(\op{p})]^\dagger = \varphi^*(\op{p})$. We approach this task using an alternate form of the Moyal product \cite{curtrightQuantumMechanicsPhase2012,curtrightConciseTreatiseQuantum2014}, Eq.~\eqref{MoyalProductdef},
    \begin{align} 
        W_{\op{A}\op{B}}(q,p) 
        &= \big[W_{\op{A}} \star W_{\op{B}}\big](q,p)  \\
        &= 2 \pi\,  W_{\op{A}}(q,p) e^{\frac{i}{2} \overleftarrow{\vec{\nabla}}^\tp \mat{\Omega}\overrightarrow{\vec{\nabla}}} W_{\op{B}}(q,p)
        \, , \label{boppform}
    \end{align}
where the \emph{Bopp operators} $\overleftarrow{\vec{\nabla}}$ and $\overrightarrow{\vec{\nabla}}$ are defined by partial derivatives of position and momentum acting in the direction of the arrow \cite{curtrightQuantumMechanicsPhase2012}
    \begin{align}
        \overleftarrow{\vec{\nabla}} \coloneqq 
        \begin{bmatrix}
            \overleftarrow{\partial_{q}} \\
            \overleftarrow{\partial_{p}}
        \end{bmatrix},	\quad \quad
        \overrightarrow{\vec{\nabla}} \coloneqq 
        \begin{bmatrix}
            \overrightarrow{\partial_{q}} \\
            \overrightarrow{\partial_{p}}
        \end{bmatrix}
        \, .
    \end{align}
The Wigner functions for the operators in Eqs.~\eqref{diagonalops} are functions of only $q$ or $p$, respectively,
    \begin{align}
        W_{\chi(\op{q})} (q,p)    &= \frac{1}{2\pi} \chi (q), \\
        W_{\varphi(\op{p})} (q,p) &= \frac{1}{2\pi} \varphi (p) .
    \end{align}
Such univariate functions are useful when calculating Moyal products using the Bopp-operator form of the Moyal product, Eq.~\eqref{boppform}.
We begin by considering Eq.~\eqref{diagonalqgoal} and separating the Moyal products,
    \begin{align}
        W_{ \chi(\op{q})\op{O} \chi^*(\op{q}) } (q,p) 
        = \Big[W_{\chi (\op{q})}\star {W}_{\op{O}} \star W_{ \chi^*(\op{q}) }\Big] (q,p)  
    \end{align}
Using Eq.~\eqref{boppform}, performing the first Moyal product gives
    \begin{align}
        W_{\chi(\op{q})\op{O}}(q,p) &\coloneqq \big[W_{\chi(\op{q})} \star W_{\op{O}} \big] (q,p)\\
        &= \chi (q + \tfrac{i}{2} \overrightarrow{\partial_{p}}) W_{\op{O}} (q,p)
        \, ,\label{firstBopp}
    \end{align}
where $W_{\op{O}} (q,p)$ is the Wigner function for $\op{O}$.
Then, performing the second Moyal product gives,
    \begin{align}
        \big[W_{\chi(\op{q})\op{O}} \star W_{\chi^*(\op{q})} \big] (q,p) &= W_{\chi(\op{q})\op{O}} (q,p) \chi^{*}(q - \tfrac{i}{2}\overleftarrow{\partial_{p}}) \\
        &=\chi^* (q - \tfrac{i}{2}\overrightarrow{\partial_{p}})W_{\chi(\op{q})\op{O}} (q,p). \label{secondBopp}
    \end{align}
Plugging Eq.~\eqref{firstBopp} into Eq.~\eqref{secondBopp} gives
    \begin{align} \label{boppflippedwigner}
         W_{ \chi(\op{q})\op{O} \chi^*(\op{q}) } (q,p) 
        = \chi^{*}(q - \tfrac{i}{2}\partial_{p})\chi (q + \tfrac{i}{2} \partial_{p}) W_{\op{O}} (q,p), 
    \end{align}
where we drop the arrow in our notation, since the partial derivatives all act on $W_{\op{O}}(q,p)$.
We now evaluate Eq.~\eqref{boppflippedwigner}. First, we introduce a $\delta$-function
and then change variables, $w = p-p'$, to get\footnote{Decompositions of Dirac $\delta$-functions in this form and their applications is discussed far more rigorously in Ref.~\cite{kempfNewDiracDelta2014a}.}
    \begin{align}
        & W_{ \chi(\op{q})\op{O} \chi^*(\op{q}) } (q,p) \nonumber \\
        & = \int dp' \, \chi^* (q - \tfrac{i}{2}\partial_{p})\chi (q + \tfrac{i}{2} \partial_{p}) \delta(p-p')  W_{\op{O}} (q,p') \\
        &= \int dw \;  W_{\op{O}} (q,p-w) \chi^{*}(q - \tfrac{i}{2}\partial_{w})\chi (q + \tfrac{i}{2} \partial_{w}) \delta(w) \label{isthismyfinalform}
        \, .
\end{align}
Using the Wigner function for a pure (unnormalized) state 
    \begin{align} \label{genericstate}
        \ket{\psi} = \int ds \, \psi(s) \qket{s} = \int dt \, \tilde{\psi}(t) \pket{t}
        \, ,
    \end{align}
in terms of its position wave function $\psi(q)$,
    \begin{align} \label{Wignerfromwavefunctions}
        W_{\outprod{\psi}{\psi}} (q,p) &= \frac{1}{\pi} \int dy\, \psi^{*} (q + y) \psi(q-y) e^{2ipy} \\ 
        &= \frac{1}{\pi}  \psi^* (q - \tfrac{i}{2}\partial_{p}) \psi(q + \tfrac{i}{2}\partial_{p}) \int dy \, e^{2ipy}, \\
         &=  \psi^* (q - \tfrac{i}{2}\partial_{p}) \psi(q + \tfrac{i}{2}\partial_{p}) \delta(p), 
    \end{align}
it is straightforward to write Eq.~\eqref{isthismyfinalform} as a convolution between two Wigner functions,
    \begin{align} \label{qrakemap}
        W_{\chi(\op{q}) \op{O} \chi^*(\op{q})} (q,p) 
        = \int dw \, W_{\op{O}} (q,p-w) W_{\outprod{\chi}{\chi}} (q,w), 
    \end{align}
where we have defined
    \begin{align} \label{appendixchi}
        \ket{\chi} \coloneqq \int ds \, \chi(s) \qket{s} = \sqrt{2 \pi} \chi(\op{q}) \ket{0}_p \, ,
    \end{align}
which is not, in general, a normalized state. 

For maps given by diagonal operators of $\op{p}$, Eq.~\eqref{diagonalpgoal}, a similar procedure gives
    \begin{align}
        &W_{ \varphi(\op{p})\op{O} \varphi^*(\op{p}) } (q,p) \nonumber \\*
        &= \int dw \,  W_{\op{O}} (q+w,p) \varphi^{*}(p - \tfrac{i}{2}\partial_{w})\varphi (p + \tfrac{i}{2} \partial_{w}) \delta(-w) 
        \, . 
    \end{align}
Using the Wigner function for a pure (unnormalized) state $\ket{\psi}$, Eq.~\eqref{genericstate}, 
in terms of its momentum wavefunction $\tilde{\psi} (p)$,
    \begin{align} \label{Wignerfrommomwavefunctions}
        W_{\outprod{\psi}{\psi}}(q,p) &= \frac{1}{\pi} \int dx \, \tilde{\psi}^*(p + x) \tilde{\psi} (p -x) e^{-2ixq}\\
        &=\tilde{\psi}^{*}( p - \tfrac{i}{2} \partial_{q}) \tilde{\psi} (p + \tfrac{i}{2} \partial_{q}) \delta(-q),
    \end{align}
yields the result
    \begin{align} \label{prakemap}
        W_{ \varphi(\op{p})\op{O} \varphi^*(\op{p}) } (q,p)  
        &=  \int dw \, {W}_{\op{O}} (q+w,p) W_{\outprod{\varphi}{\varphi}} (w,p)
        \, .
    \end{align}
Here, $\ket{\varphi}$ is defined to be a state whose \emph{momentum} wave function is given by $\varphi(t)$:
    \begin{align} \label{secondWignerstate}
        \ket{\varphi} \coloneqq \int dt \varphi(t) \pket{t} = \sqrt{2\pi} \varphi(\op{p}) \qket{0}\, .
    \end{align}
The Wigner-function maps in Eq.~\eqref{qrakemap} and Eq.~\eqref{prakemap} are the major results of this Appendix. Each describes a ``raking'' (see Sec.~\ref{subsubsection:phasespaceffectsEC}),  where the input Wigner function is raked in either position or momentum by the the state $\ket{\chi}$ or $\ket{\varphi}$, respectively.

Applying the operators $\chi(\op{q})$ and $\chi(\op{p})$ consecutively to $\op{O}$, we have
\begin{widetext}
    \begin{subequations} \label{WignerMapqp}
    \begin{align} 
        W_{ \varphi(\op{p}) \chi(\op{q}) \op{O} \chi^*(\op{q}) \varphi^*(\op{p}) }  (q,p) 
        \blu & \propto  \iint dw'dw'' \,  W_{\outprod{\varphi}{\varphi}} (w',p)W_{\outprod{\chi}{\chi}} (q+w',w'') W_{\op{O}} (q+w',p-w'') \, , \label{WignerMapqpFIRST}
        \\
        W_{ \chi(\op{q}) \varphi(\op{p}) \op{O} \varphi^*(\op{p}) \chi^*(\op{q}) }  (q,p) 
        & \propto  \iint dw'dw'' \,  W_{\outprod{\chi}{\chi}} (q,w'')W_{\outprod{\varphi}{\varphi}} (w',p-w'') W_{\op{O}} (q+w',p-w'') \, . \label{WignerMapqpSECOND}
    \end{align}
    \end{subequations}
\end{widetext}
Note that these two forms are not the same, which is due to the fact that $\chi(\op{q})$ and $\varphi(\op{p})$ do not, in general, commute.
\blk

\subsection{Action of the embedded-error operator} \label{Appendix:embeddederrorop}

We use the general result above to describe the action of the embedded-error operator $\op{\xi} (\mat{\Xi})$, Eq.~\eqref{embeddederrorop}, on an operator $\op{O}$ in phase space,
    \begin{align} \label{bigOprimed}
        \op{\bar{O}}  & = \op{\xi} (\mat{\Xi}) \op{O} \op{\xi} (\mat{\Xi}) 
        \to W_{\bar{O}}(q,p)
        \, .
    \end{align}

In the low-noise limit, $\op{\xi} (\mat{\Xi})$ separates into two exponential operators, Eq.~\eqref{highqualitynoise}, each being diagonal in either $\op{q}$ or $\op{p}$. Following the previous section, we label these operators as $\chi(\op{q})$ and $\varphi(\op{p})$:
    \begin{subequations} \label{exponentialoperators}
    \begin{align}
         \chi(\op{q})    &= e^{-\frac{1}{2}\kappa^{2}\op{q}^2} \, , \\*
         \varphi(\op{p}) &= e^{-\frac{1}{2}\Delta^{2}\op{p}^2}
         \, ,
    \end{align}
    \end{subequations}
also noting that $\chi(\op{q}) = [\chi(\op{q})]^\dagger$ and $\varphi(\op{p}) = [\varphi(\op{p})]^\dagger$.
Their Wigner functions are univariate,
    \begin{align}
         W_{\chi(\op{q})} (q,p) &= \frac{1}{2\pi} \chi(q) =  \frac{1}{2\pi} e^{-\frac{1}{2}\kappa^{2} q^{2}} \, ,\\
         W_{\varphi(\op{p})} (q,p) &= \frac{1}{2\pi} \varphi(p)=\frac{1}{2\pi} e^{-\frac{1}{2}\Delta^{2} p^{2}} \, .
    \end{align}
The states associated with each of these operators, $\ket{\chi}$ [Eq.~\eqref{appendixchi}] and $\ket{\varphi}$ [Eq.~\eqref{secondWignerstate}], have Wigner functions
    \begin{align}
        W_{\outprod{\chi}{\chi}} (q,p) 
        &=\frac{\sqrpi}{\kappa} G_{\frac{1}{2\kappa^{2}}} (q) G_{\frac{\kappa^{2}}{2}} (p)\, ,\\
        W_{\outprod{\varphi}{\varphi}} (q,p) 
        &=\frac{\sqrpi}{\Delta} G_{\frac{\Delta^{2}}{2}} (q) G_{\frac{1}{2\Delta^{2}}} (p) \, ,
    \end{align}
which are found using Eqs.~\eqref{Wignerfromwavefunctions} and \eqref{Wignerfrommomwavefunctions}, respectively.
    
We now find the Wigner function for the transformed operators
    \begin{align}
        \op{\bar{O}}_1 &= 
        \varphi(\op{p}) \chi(\op{q}) \op{O} \chi(\op{q}) \varphi(\op{p}) \label{order1} \, , \\
        \op{\bar{O}}_2 &= 
        \chi(\op{q}) \varphi(\op{p}) \op{O}  \varphi(\op{p}) \chi(\op{q}) \label{order2}
          \, ,
    \end{align}
using Eq.~\eqref{WignerMapqpFIRST}.
The Wigner function for Eq.~\eqref{order1} is  
    \begin{align} 
        W_{\op{\bar{O}}_1}(q,p) &= \frac{\pi}{\Delta\kappa} G_{\frac{1}{2\Delta^{2}}} (p) \int dq'\, G_{\frac{1}{2\kappa^{2}}} (q') G_{\frac{\Delta^{2}}{2}} (q' - q)\nonumber \\
        &\quad \times \int dw\, W_{\op{O}} (q', p - w) G_{\frac{\kappa^{2}}{2}} (w).\label{O1}
    \end{align}
When the operators are applied in the other order, Eq.~\eqref{order2}, the Wigner function is 
    \begin{align} 
        W_{\op{\bar{O}}_2}  &=\frac{\pi}{\Delta\kappa}G_{\frac{1}{2\kappa^{2}}} (q) \int dp'\, G_{\frac{\kappa^{2}}{2}} (p'-p) G_{\frac{1}{2\Delta^{2}}} (p')\nonumber \\
        &\quad \times \int dw'\, W_{\op{O}} (q + w', p') G_{\frac{\Delta^{2}}{2}} (w'). \label{O2}
    \end{align}
As we are are interested in the low-error limit, $\Delta^{2}, \kappa^{2} \ll 1$, where $\op{O}' \approx \op{O}_1' \approx \op{O}_2'$. We apply the approximation in Eq.~\eqref{approxspikeproduct}, with the result that both Eq.~\eqref{O1} and Eq.~\eqref{O2} yield the Wigner function
    \begin{align}
        W_{\op{\bar{O}}}(\vec{x}) 
        & \approx \frac{\pi}{\Delta\kappa} G_{\frac{1}{2}\mat{\Xi}_{0}} (\vec{x}) \int d^{2} \vec{x}' \,     W_{\op{O}} (\vec{x} -  \vec{x}') G_{\frac{1}{2}\mat{\Xi}_{0}^{-1}} (\mat{\Omega} \vec{x}') \\*
        &= \frac{\pi}{\Delta\kappa} G_{\mat{\Sigma}_{\text{env}}} (\vec{x}) \int d^{2} \vec{x}'  \,        W_{\op{O}} (\vec{x} -  \vec{x}') G_{\mat{\Sigma}_{\text{spike}}} (\vec{x}')
        \, ,
        \label{eq:WignerOappE}
    \end{align}
where we have used Eqs.~\eqref{EnvSpikeCovariances} and \eqref{minenvelope}, and we have written the arguments in terms of $\vec{x}$ for convenience. The convolution above can also be written as a blurring operation, \erf{bludefin}, on the Wigner function for $\op{O}$,
    \begin{align} 
        \int d^{2} \vec{x}'   \,      W_{\op{O}} (\vec{x} -  \vec{x}') G_{\mat{\Sigma}_{\text{spike}}} (\vec{x}') = \mathcal{D}_{\mat{\Sigma}_\text{spike}} W_{\op{O}} (\vec{x})
        \, .
    \end{align}

We now consider an operator~$\op{O}_\Sha$ whose Wigner function is a two-dimensional $\Sha$-function, \erf{Osha}. The convolution in \erf{eq:WignerOappE} replaces each $\delta$-function with a Gaussian of covariance $\mat{\Sigma}_\text{spike}$, yielding a two-dimensional $\theta$-function, giving
    \begin{align} \label{WigOshaAppendix}
        W_{ \op{\bar{O}}_\Sha  }(\vec{x})  
        & \approx \frac{\pi}{\Delta\kappa} G_{ \mat{\Sigma}_\text{env} }(\vec{x}) 
            \Rthetafxn[\mat{A}]{\vec{v}_1}{\vec{v}_2} \left( \vec{x}, 2 \pi i \mat{\Sigma}_\text{spike} \right)\, ,
    \end{align}
recognizing that the factor $\pi/(\Delta \kappa)$ arises from the definitions of the exponential operators, Eqs.~\eqref{exponentialoperators}, but is of no particular physical consequence. 
An important example is $\op{O}_\Sha = \op{\rho}_L$, where $\op{\rho}_L$ is an ideal GKP state with Bloch 4-vector $\vec{r}$. The approximate GKP state, $\op{\bar{\rho}}_L$, has the Wigner function
\begin{align} \label{WigStateAppendix}
    \bar{W}_{\vec{r}}(\vec{x}) \approx \frac{1}{\mathcal{N}} G_{\mat{\Sigma}_{\text{env}}} (\vec{x}) \sum^{3}_{\mu=0} \vec{r}_{\mu} \Rthetafxn[\sqrpi\mat{I}]{\nicefrac{\mat{\Omega}\vec{\ell}_{\mu}}{2}}{\nicefrac{\vec{\ell}_{\mu}}{2}} (\vec{x},2\pi i \mat{\Sigma}_{\text{spike}})
    \, ,
\end{align}
with all prefactors absorbed into the normalization $\mathcal{N}$.

Finally, we estimate the order of the approximation by comparing the convolved envelopes to their unconvolved counterparts. We express the error $\epsilon$ as 
    \begin{align}
        \epsilon(x)  = [G_{\Delta^{2}/2}*G_{\kappa^{-2}/2}] (x) - G_{\kappa^{-2}/2} (x),
    \end{align}
whose $L^{2}$ norm,
    \begin{equation}
        \int dx\, [\epsilon(x)]^{2} = \frac{\kappa}{2\sqrpi} \bigg(1 + \frac{1}{\sqrt{1 + \kappa^{2}\Delta^{2}}} -\frac{2\sqrt{2}}{\sqrt{2 + \kappa^{2}\Delta^{2}}}\bigg)
        \, ,
    \end{equation}
is of order $\mathcal{O}(\kappa^{2}\Delta^{2})$ and thus negligible in our approximation.

\vfill

\bibliographystyle{bibstyleNCM_papers}
\bibliography{ReferencesMaster}

\end{document}

%% file: Menicucci_preamble.tex
\allowdisplaybreaks[4]

\graphicspath{{./figures/}}

\DeclareMathOperator{\Tr}{Tr}

\renewcommand{\Re}{\operatorname{Re}}
\renewcommand{\Im}{\operatorname{Im}}

\makeatletter

\newcommand{\ringplus}{\mathbin{\text{\@ringplus}}}

\newcommand{\@ringplus}{%
  \ooalign{\hidewidth\raise1.3ex\hbox{\tiny$\circ$}\hidewidth\cr$\m@th+$\cr}%
}

\newcommand{\ringminus}{\mathbin{\text{\@ringminus}}}

\newcommand{\@ringminus}{%
  \ooalign{\hidewidth\raise0.9ex\hbox{\tiny$\circ$}\hidewidth\cr$\m@th-$\cr}%
}
\makeatother

\newcommand{\tp}[0]{\mathrm{T}}

\DeclareFontFamily{U}{wncy}{}
\DeclareFontShape{U}{wncy}{m}{n}{<->wncyr10}{}
\DeclareSymbolFont{mcy}{U}{wncy}{m}{n}
\DeclareMathSymbol{\Sh}{\mathord}{mcy}{"58}

\newcommand{\negspace}{\!}
\newcommand{\lsub}[2]{{\protect\vphantom{#1}}_{#2} \negspace {#1}}
\newcommand{\rsub}[2]{{#1} \negspace {\protect\vphantom{#1}}_{#2}}
\newcommand{\lrsub}[3]{{\protect\vphantom{#1}}_{#2} \negspace {#1} \negspace {\protect\vphantom{#1}}_{#3}}

\newcommand{\ketsub}[2]{\rsub {\ket{#1}} {#2}}
\newcommand{\brasub}[2]{\lsub {\bra{#1}} {#2}}

\newcommand{\pket}[1]{\ketsub{#1} p}
\newcommand{\qket}[1]{\ketsub{#1} q}

\newcommand{\inprod}[2]{\left\langle {#1} | {#2} \right\rangle}
\newcommand{\inprodsubsub}[4]{\lrsub {\inprod{#1}{#2}} {#3} {#4}}

\newcommand{\outprod}[2]{\ket {#1}\!\bra {#2}}

\newcommand{\outprodsubsub}[4]{\ketsub {#1}{#3} \brasub{#2}{#4}}

\newcommand{\abs}[1]{\left\lvert{#1}\right\rvert}
\newcommand{\abss}[1]{\lvert{#1}\rvert}

\newcommand{\op}[1]{\hat{#1}}

\newcommand{\opvec}[1]{\op{\vec{#1}}}

\newcommand{\id}[0]{I}

\newcommand{\mat}[1]{\bm{\mathrm{#1}}}

\renewcommand{\vec}[1]{\bm{\mathrm{#1}}}

\newcommand{\blu}{\color{blue!70!cyan}}
\newcommand{\blk}{\color{black}}